\documentclass{jfm}

\usepackage{amsmath,amsfonts,amssymb}
\usepackage{graphicx}
\usepackage{epstopdf, epsfig}
\usepackage{xcolor}

\usepackage{tikz}

\usepackage{natbib}

\DeclareRobustCommand{\dashedredline}{\raisebox{2pt}{\tikz{\draw[-,red,dashed,line width = 1.5pt](0,0) -- (10mm,0);}}}
\DeclareRobustCommand{\blackline}{\raisebox{2pt}{\tikz{\draw[-,black,solid,line width = 1.5pt](0,0) -- (10mm,0);}}}
\DeclareRobustCommand{\redline}{\raisebox{2pt}{\tikz{\draw[-,red,solid,line width = 1.5pt](0,0) -- (10mm,0);}}}
\DeclareRobustCommand{\blueline}{\raisebox{2pt}{\tikz{\draw[-,blue,solid,line width = 1.5pt](0,0) -- (10mm,0);}}}
\DeclareRobustCommand{\greenline}{\raisebox{2pt}{\tikz{\draw[-,black!20!green,solid,line width = 1.5pt](0,0) -- (10mm,0);}}}

\DeclareRobustCommand{\orangeline}{\raisebox{2pt}{\tikz{\draw[-,orange,solid,line width = 1.5pt](0,0) -- (10mm,0);}}}

\def\be{\begin{equation}}
\def\bea{\begin{eqnarray}}
\def\ee{\end{equation}}
\def\eea{\end{eqnarray}}

\def\<{\langle}
\def\>{\rangle}

\allowdisplaybreaks

\shorttitle{Variable-density incompressible turbulent mixing layers}
\shortauthor{J. R. Baltzer and D. Livescu}

\title{Variable-density effects in incompressible non-buoyant shear-driven turbulent mixing layers}

\author
 {
 Jon R. Baltzer\aff{1}
   \corresp{\email{jbaltzer@lanl.gov}}
  \and 
  Daniel Livescu\aff{1},
  }

\affiliation
{
\aff{1}
CCS-2, Los Alamos National Laboratory, Los Alamos, NM 87545
}

\begin{document}

\maketitle

\begin{abstract}
The asymmetries that arise when a mixing layer involves two miscible fluids of differing
densities are investigated using incompressible (low-speed) direct numerical simulations.
The simulations are performed in the temporal configuration with very large domain sizes, to allow the mixing layers to reach prolonged states of fully-turbulent self-similar growth.
Imposing a mean density variation breaks the mean symmetry relative to the classical single-fluid
temporal mixing layer problem. Unlike prior variable-density mixing layer simulations in which the
streams are composed of the same fluids with dissimilar thermodynamic properties, the
density variations are presently due to compositional differences between the fluid streams, leading to different mixing dynamics. Variable-density (non-Boussinesq) effects introduce strong asymmetries in the flow statistics that can be explained by the strongest turbulence increasingly migrating to the lighter fluid side as free stream density difference increases. Interface thickness growth rates also reduce, with some thickness definitions particularly sensitive to the corresponding changes in alignment between density and streamwise velocity profiles. 
Additional asymmetries in the sense of statistical distributions of densities at a given position within the mixing layer reveal that fine scales of turbulence are preferentially sustained in lighter fluid, which also is where fastest mixing occurs. These effects influence statistics involving density fluctuations, which have important implications for mixing and more complicated phenomena that are sensitive to the mixing dynamics, such as combustion.
\end{abstract}

\begin{keywords}
turbulent mixing, shear layer turbulence, turbulence simulation
\end{keywords}

\section{Introduction}

A wide range of applications include the fundamental phenomenon of turbulence sustained by shear between streams of fluids. Frequently, the streams may have
different densities because they consist of different fluids. Such flows can involve miscible or immiscible fluids; we are here concerned only with the miscible case. 
Miscible applications exist in combustion, industrial chemical mixing, and geophysical flows.
The relevance of mixing layer simulations to combustion is reviewed in \cite{givi1989mfs}, and other complex applications of sheared variable-density flows are summarized in \cite{akula2013esr}.

In many cases, the density differences can be large, producing significant changes to the flow evolution.
\cite{dimotakis2005tm}, in a review of turbulent mixing, classified mixing into three categories based on the complexity (physics coupling)
of the mixing phenomena and the importance of correctly capturing the mixing dynamics to the overall predictions.
In the simplest (Level-1) cases, capturing the turbulence but not the mixing itself is sufficient to predict the flow dynamics.
Level-2 indicates that mixing alters the flow dynamics. Inertial effects of the large density variations of the mixing layers investigated herein place the flow in Level-2 with increased complexity that cannot be captured by extending single-density mixing layer results with passive mixing.

In combustion, very large density variations can exist due to differing fluid compositions and thermodynamic variations.
Combustion is among the most complex mixing flows (classified as Level-3) because the mixing
strongly affects reactions that produce changes in the fluids (including heat release) which then couple back to the mixing dynamics. Capturing the inertial effects associated with compositional variations during the mixing of reactants and reaction products can be a significant component of predicting combustion.
\cite{bilger1976tjd} noted the importance of density
differences in turbulent jet diffusion flames. In configurations such as a jet of hydrogen fuel
released into air, the density differences can be very large simply due to the different molar masses of the fluids.

Several recent incompressible studies have revealed interesting effects on turbulent mixing when density differences are large solely due to differing compositions.
The Atwood number $A$ characterizes the difference in densities between streams of fluids:
\begin{equation}
A = \frac{\rho_2-\rho_1}{\rho_2+\rho_1} \;\;\; \implies \;\;\; \frac{\rho_2}{\rho_1} = \frac{1+A}{1-A},
\end{equation}
where $\rho_1$ and $\rho_2$ are the densities of each pure fluid.
Pure helium mixing with air (or nitrogen) corresponds to an Atwood number of $0.75$, while pure hydrogen mixing with air
corresponds to $A=0.85$.
Studying Rayleigh-Taylor (RT) instability in the classical configuration and a triply-periodic version (i.e. homogeneous buoyancy-driven turbulence), \cite{livescu2008vdm,livescu2009mav} found significant changes in behavior when Atwood number was increased to high values.
Atwood numbers of $A \lesssim 0.05$ are typically considered to be the limit of the Boussinesq approximation \citep{livescu2010npv}.
Flows of sufficiently high Atwood number to vary significantly from the Boussinesq approximation have been termed \emph{variable-density}.
\cite{livescu2010npv} showed that changes in alignment between density gradient and local strain is a variable-density effect
associated with reduced mixing in the heavy fluid regions.
Much of the simulation studies of density effects on mixing have occurred in buoyancy-driven turbulence, such as the small density variation study of \cite{batchelor1992hbg} that was later extended to non-Boussinesq flow by \cite{sandoval1995dvd}. \cite{sandoval1995dvd} also considered decaying isotropic turbulence without buoyancy, which was further studied by \cite{jang2007pns}. \cite{movahed2015mrf} studied variable-density mixing in two fluids with decaying isotropic turbulence initially separated by a planar interface. Notable classical RT studies include \cite{cabot2006rne} and \cite{livescu2009hrn}.

Shear-driven mixing layers have historically received a great deal of attention, but mainly for single-fluid configurations. \cite{rogers1994dss} simulated an incompressible mixing layer in the temporal (streamwise-periodic) configuration to self-similar fully-turbulent growth. A similar configuration was simulated by \cite{balaras2001sss} to study the effects of initial conditions. More powerful computational resources have recently enabled performing spatially-developing simulations, which more closely approximate mixing layer experiments. These require much longer streamwise domains to attain a desired mixing layer thickness since they thicken with downstream distance rather than in time as is the case for temporal simulations. (However, meaningful temporal simulations implicitly require sufficiently large domains to not interfere with the growth of turbulent structures.) \cite{wang2007cfs} designed a spatially-developing mixing layer simulation to be comparable to the temporal mixing layer of \cite{tanahashi2001aas} and observed similar energy dissipation rates but increased turbulent kinetic energy. The DNS of \cite{attili2012sst} advanced spatially-developing mixing layer simulations to a very long domain that enabled attaining a relatively large Reynolds number. During self-similar growth, they found remarkable agreement between their self-similar dissipation values and that of the \cite{rogers1994dss} temporal simulation, as well as close agreement for most other statistics.
Relevant low-speed experimental studies include those of \cite{spencer1971sip}, \cite{bell1990dts} and \cite{loucks2012vvg}. Experiments addressing detailed turbulent structure include those of \cite{olsen2003pvm} (which also contained a weak density difference) and \cite{li2010ese}.
In several studies, mixing properties have been investigated with shear-driven mixing layers, but in the absence of density differences between the participating fluids \citep[e.g.,][]{sharan2019tsl}.

High-speed compressible mixing layers have also received a great deal of attention, particularly due to the strong reduction in mixing layer growth rate that occurs with increasing Mach number.
Though density effects associated with compressibility were once thought to affect growth rate \citep[as discussed in][]{brown1974del}, DNS simulations have clarified how compressibility effects reduce the growth due to decreased turbulent kinetic energy production as compressibility decorrelates the strain and pressure fluctuations \citep{vreman1996cml,sarkar1996odp,freund2000cet,pantano2002sce,livescu2004sss}.
Research has continued on this mechanism in compressible mixing layer experiments \citep[e.g.,][]{barre2015des}.
Recent simulations have further investigated the mixing characteristics of compressible mixing layers \citep[e.g.,][]{jahanbakhshi2015bvg}.

Non-buoyant mixing layers with significant density variations (i.e. density ratios larger than 2) have begun to receive attention.
2D and 3D simulations demonstrated that differing free-stream densities significantly changed the early-time growth and Kelvin-Helmholtz (KH) flow structures \citep{joly2001bfs,joly2002ssv}. The pioneering 3D temporal simulations of \cite{pantano2002sce} included an investigation of different free-stream densities within a broader study of compressible mixing layers. The differing densities were established by varying the temperature for a single fluid. They found that increasing Atwood number decreased the temporal thickness growth rate, though the extent depended on how thickness was defined.
During self-similar growth, the Reynolds shear stress changed little in magnitude but shifted to the light fluid side with increasing Atwood number. They also developed a model characterizing the shift of the mean velocity profile to the light fluid side and the associated decrease in momentum thickness growth rate. Mild compressibility effects were likely present because the convective Mach number was $M_c=0.7$.
More recently, \cite{almagro2017nsv} performed DNS using a low-speed approximation for the flow of \cite{pantano2002sce}. Two streams of a single fluid with different temperatures again create the density difference, but compressibility effects are considered negligible at low speeds. They also developed a semi-empirical model for the reduction in momentum thickness growth rate with density ratio.

Details of mixing layers with variable density due to differing fluid compositions are much less understood. Detailed studies of mixing layers involving two different miscible fluids have been rare, particularly when not complicated by other effects such as buoyancy or compressibility, despite earlier attention.
The historic low-speed experiments of \cite{brown1974del} using two gases with different densities found reductions in the growth rates as large as $50\%$ for density ratios up to $7$. These measurements were limited to mean density and streamwise velocity profiles and no details of the changes to turbulence and mixing properties are available. Our present investigation focuses on this flow but in a temporal configuration. The governing equations for this incompressible flow differ from those for a single fluid with thermal-induced density variations, as used by \cite{pantano2002sce} and, in a low speed limit, by \cite{almagro2017nsv}.
The relationship between the equations governing these flows has been reviewed in detail by \cite{livescu2020tlt}. \cite{baltzer2020lst} focused this analysis on applications to mixing layer simulations and found that mean statistical profiles showed little difference when the density difference between free streams was compositionally-induced or thermally-induced. However, these cases had significant differences in their mixing and density probability density function behaviors.

The present temporal simulations are relevant to understanding variable-density effects on growth in the spatially-developing configuration. 
2D simulations of early-time spatially-developing mixing layers show strong differences in entrainment
depending on whether the low or high speed stream has lower or higher density \citep{reinaud2000aps,joly2002ssv}; we are unaware of any spatial simulations of fully turbulent growth. Based on experiments, \cite{brown1974els} studied the thickness growth rate of variable-density spatially-developing fully-turbulent mixing layers. He assumed that the temporal growth rate (i.e., from a frame of reference moving with the mixing layer convection velocity) is independent of the density difference between the streams, which is contrary to the reductions observed by \cite{pantano2002sce} and \cite{almagro2017nsv}. As discussed in \cite{pantano2002sce}, \cite{brown1974els} combined this with the observation that the convection velocity is closer to the velocity of the high-density stream to propose a formula for growth rate reduction with Atwood number. \cite{dimotakis1984tds} refined the formula to account for asymmetric entrainment that is present only in spatially-developing mixing layers.
\cite{ashurst2005odt} studied variable density effects in temporal and spatial mixing layers using the one-dimensional turbulence stochastic simulation method; they captured many of the effects observed in \cite{pantano2002sce}. 

Other studies have addressed variable-density shear-driven mixing layers with buoyancy or other complicating physics playing a significant role.
\cite{olson2011nec} simulated mixing layers with mixed RT (buoyant) and KH (shear) instability and Atwood numbers ranging up to 0.71 using the same governing equations as for our present study.
They focused on early times when complicated interactions between the instabilities produce complex effects on the growth rate.
The linear stability study of \cite{zhang2005esf} also considered a similar configuration.
\cite{barros2011hin} performed linear stability analysis in a similar configuration representative of some environmental flows and highlighted the importance of the variable-density inertial terms beyond a Boussinesq approximation.
Experimentally, \cite{akula2013esr} studied mixed RT and KH instability with air and air/helium mixture streams
shearing past each other, following a number of water-based experiments (also reviewed therein); buoyancy was the principal density effect and the Atwood numbers were low ($<0.04$).
\cite{gat2017ivd} simulated the mixing of vertical columns of fluid with different densities and perturbed interfaces. Gravity accelerates the perturbed heavy column downward within the triply-periodic domain to induce KH instability.
Their configuration contains some of the same physics (shear aligned with buoyancy) as the more complex configuration of a buoyant jet, which was recently studied experimentally by \cite{charonko2017vdm} and received more detailed analysis of the cascade of energy between scales by \cite{lai2018khm}.
Additional multi-composition variable-density shear studies in the presence of other complicating physics include simulations of hydrogen and air streams to address supersonic turbulent combustion by \cite{obrien2014ssb}, reacting mixing layer simulations by \cite{miller2001dns}, and hybrid motor simulations with oxidizer and gasified fuel by \cite{haapanen2008lsa}.

Our present investigation seeks to elucidate the fundamental changes to the self-similar growth in a free shear flow produced by differences in the density of each stream with differing compositions. We perform direct numerical simulations in the simple incompressible temporally-developing configuration with two miscible fluids. In particular, we seek to quantify the asymmetries that appear in the flow statistics due to variable-density effects (whereas the analogous single-density incompressible temporal mixing layer configuration is statistically symmetrical) and explain their effect on growth characteristics.
The paper is structured as follows: Section~\ref{sec:simappr} describes the simulation approach and governing equations, followed by a description of the initial conditions in Section~\ref{sec:initcond}. Section~\ref{sec:defprop} discusses flow properties that can be adduced from the governing equations and introduces definitions of flow measurements. Section~\ref{sec:basicstat} presents an overview of mean and fluctuation statistics from the simulations and relates growth rates to statistical profiles. Section~\ref{sec:condstat} briefly addresses the local effects of density on velocity-related statistics, leading to the conclusions of Section~\ref{sec:concl}. This is followed by appendices addressing (a) the relationship between density profiles and mean cross-stream velocity and (b) contrasts between the present variable-composition flow and variable-thermodynamic-property flow.

\section{Simulation Approach}
\label{sec:simappr}

\begin{figure}
\centering
\includegraphics[scale=.7]{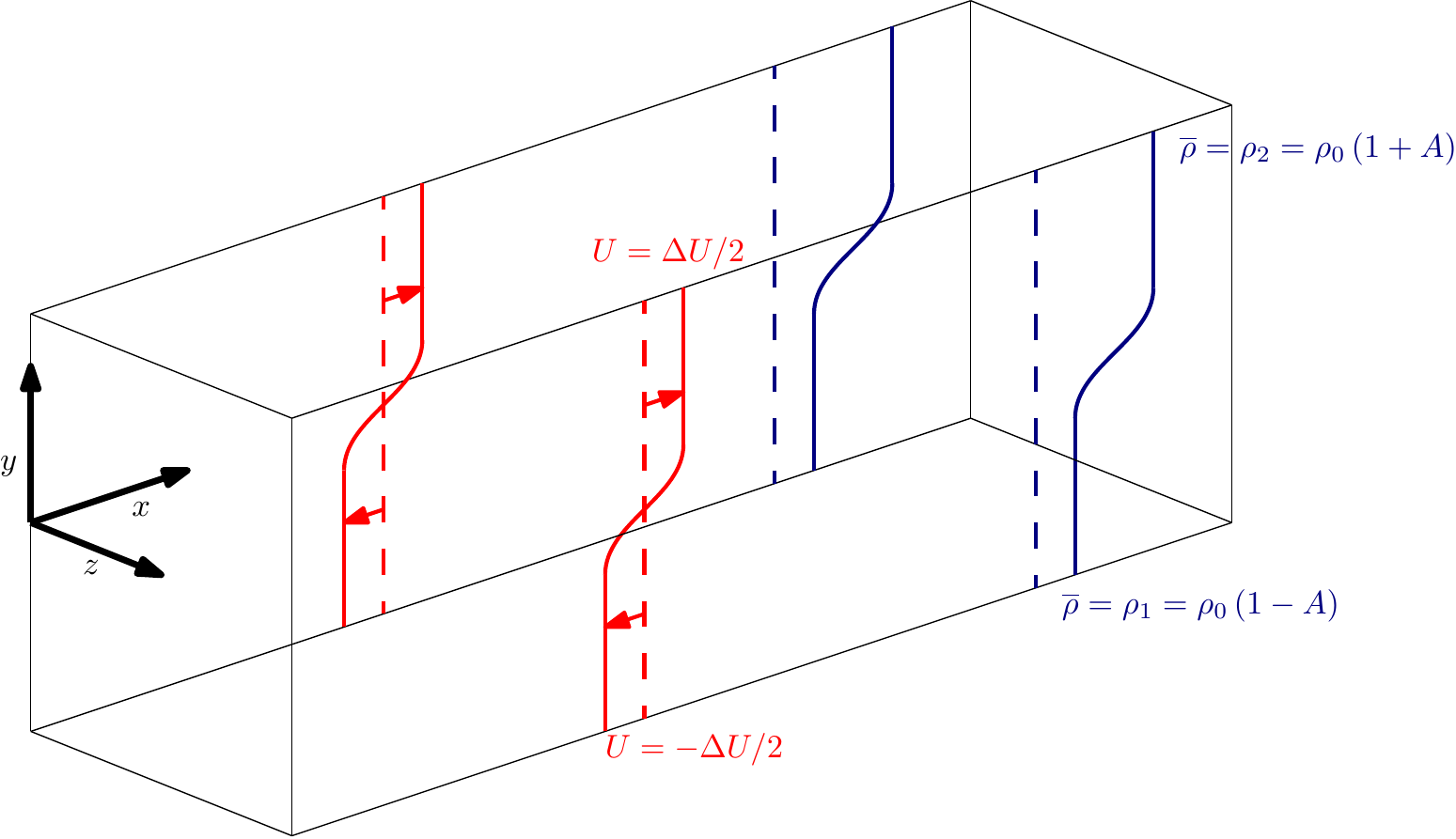}
\caption{Variable-density mixing layer simulation set-up and coordinate system.}
\label{fig:setup}
\end{figure}

The simulations are performed in the canonical temporal configuration, with two velocity streams of equal magnitudes flowing in opposite directions. The temporal configuration can be regarded as the limit of mean convection velocity of a spatial mixing layer approaching zero. In this case, the mixing layer develops with time instead of with spatial position as the flow convects downstream for the latter configuration. By using periodic boundary conditions in the streamwise (and spanwise) directions, the temporal configuration avoids the need for choosing inflow and outflow conditions and focuses on the variable-density effects on mixing in the simplest configuration possible. To our knowledge, this is the first study focusing on variable-density effects due to composition variation without additional effects such as compressibility, reactions, etc.
Following the typical set-up \citep[e.g.,][]{rogers1994dss}, the coordinates are oriented such that $1$ ($x$)
denotes the streamwise direction aligned with the mean velocities, $2$ ($y$) denotes the
cross-stream (transverse) direction normal to the fluid interface, and $3$ ($z$) denotes the spanwise direction (figure \ref{fig:setup}).

\subsection{Governing Equations}
\label{ss:goveqn}

To study incompressible mixing layers involving two fluid streams with strongly differing densities,
the governing equations are formed by considering the full compressible flow equations for a
miscible binary fluid mixture and then obtaining the infinite speed of sound incompressible limit \citep{livescu2013nst}. Gravity is not included here, but otherwise the governing equations are identical to those describing variable-density (non-Boussinesq) RT flow, as simulated by \cite{cook2001tsr}, \cite{livescu2007bdv} and \cite{wei2012lqg}.
To our knowledge, the present study is the first application of these equations to purely shear-driven variable-density fully-turbulent mixing layers.

The equations for the instantaneous variables (with partial derivatives denoted following the comma in the subscript, namely $t$ representing the time variable $t$ and an index $i$ representing the relevant spatial direction $x_i$) are
\begin{align}
\rho_{,t} + \left(\rho u_j\right)_{,j} & = 0\label{eq:cont}\\
\left( \rho u_i\right)_{,t} + \left( \rho u_i u_j \right)_{,j} & = -p_{,i} + \tau_{ij,j}\label{eq:mom}\\
u_{j,j} & = -\mathcal{D} \left(\ln \rho \right)_{,jj}\label{eq:divg},
\end{align}
where the viscous stress, assumed to be Newtonian, is
\begin{equation}
\tau_{ij} = \mu \left[u_{i,j} + u_{j,i} - \frac{2}{3} u_{k,k} \delta_{ij} \right]\label{eq:viscstr}.
\end{equation}
The governing equations are supplemented by slip boundary conditions in the $y$ direction and periodic boundary conditions in $x$ and $z$ directions. 

Equation (\ref{eq:divg}) represents the nonzero divergence of velocity that occurs due to the change in volume
during mixing (while the flow is incompressible). The Fickian form with diffusion coefficient $\mathcal{D}$ represents the infinite 
sound speed limit of the full multicomponent diffusion operator \citep{livescu2013nst}. Equation 
(\ref{eq:divg}) can be derived from the mixture rule $\rho=1/\left(Y_1/\rho_1 + Y_2/\rho_2\right)$ (where $Y_1$ and $Y_2$ are species mass fractions of pure fluids with constant densities $\rho_1$ and $\rho_2$, respectively) and species mass fraction transport equations for each species $\left(\rho Y_m\right)_{,t} + \left(\rho Y_m u_j\right)_{,j} = \mathcal{D}\left(\rho Y_{m,j}\right)_{,j}$ \citep{sandoval1995dvd,cook2001tsr,livescu2007bdv}. The mixture rule can also be connected to the infinite speed of sound limit of the ideal gas mixture equation of state. Alternately, the same divergence relation can be derived as the infinite sound speed limit of the energy transport equation, which demonstrates the consistency of the VD governing equations \citep{livescu2013nst}.
The dynamic viscosity of mixed fluid obeys a relation analogous to the density: $\mu=1/\left(Y_1/\mu_1 + Y_2/\mu_2\right)$,
where $\mu_1$ is the viscosity of the pure fluid with density $\rho_1$ and $\mu_2$ is the viscosity of the pure fluid with density $\rho_2$, which ensures a uniform Schmidt number, $Sc=\mu/(\rho D)$, throughout the mixture.

\subsection{Notations}
Many of the statistics are based on averages, which are indicated by the symbol $\langle \rangle$. Generically, the Reynolds mean of a quantity $q$ is denoted by $\langle q \rangle$ and Reynolds fluctuation is $q' = q - \langle q \rangle$.
For simple expressions, the Reynolds mean will also be indicated by an overbar, i.e. $\bar{q}$, which is equal to $\langle q \rangle$.
As is typical for compressible flows, Favre averaging is employed for the mean governing equations to account for density variations. The Favre mean of a velocity component, $u_i$, is denoted by $\tilde{U}_i = \langle \rho u_i \rangle / \langle \rho \rangle$
and the Favre fluctuation is $u_i'' = u_i - \tilde{U}_i$, in contrast to the Reynolds mean, $\bar{U}_i=\langle u_i \rangle$, and fluctuation, $u_i' = u_i - \bar{U}_i$.

Numerical quantities presented in sections below are obtained from averages computed based on homogeneities present within the flows. Since the flow is periodic and homogeneous in the streamwise and spanwise coordinates $x$ and $z$, area averages are computed across $y$-normal planes.
Self-similar statistics will also be considered in which profiles should not change with time (except for noise due to lack of statistical convergence) when the $y$ coordinate is scaled by an appropriate length scale. For these statistics, time averaging is also performed over the self-similar growth duration to improve statistical convergence (\S\ref{ss:indselfsim}). The averages computed to obtain Reynolds ($\bar{U}_i$) and Favre ($\tilde{U}_i$) averages are $x-z$ area averages only when the statistic is a function of time or not in self-similar coordinates, but time averages are taken of the area averages when self-similar statistics are presented and the same set of notations is used for the averaged quantities.

\subsection{Numerical Approach}
\label{ss:numapprch}

The governing equations (\ref{eq:cont}--\ref{eq:viscstr}) are solved numerically using a pseudo-spectral scheme for spatial discretization in the periodic (streamwise and spanwise) directions and a compact difference scheme for the inhomogeneous (cross-stream) direction of the flow. The algorithm and code are slightly modified from those employed and described by \cite{wei2012lqg,livescu2010npv,livescu2011dns} for variable-density RT simulations; the equations solved are the same except non-zero mean streamwise velocity is present in the mixing layer.

The cross-stream (normal) velocities at the lower and upper slip wall boundaries are maintained at zero, and this is consistent with the governing equations for this temporal mixing layer. Averaging the divergence equation (\ref{eq:divg}) with diffusivity $\mathcal{D}$ assumed constant, and then omitting the terms of the summed indices that vanish due to the homogeneities present in the flow results in
\begin{equation}
    \langle u_2 \rangle_{,2} = - \mathcal{D} \langle \ln \rho \rangle_{,22}.
\end{equation}
Integrating across the $y$ domain, this expression becomes $u_2(y_\mathrm{max})-u_2(y_\mathrm{min}) = -\mathcal{D}\left\{[\ln{\rho}]_{,2}(y_\mathrm{max}) - [\ln{\rho}]_{,2}(y_\mathrm{min})\right\}$. Since density remains constant at the free streams existing at the upper and lower walls, it follows that $u_2(y_\mathrm{max})-u_2(y_\mathrm{min}) = 0$. Thus, the variable-density equations are consistent with the boundary conditions $u_2(y_\mathrm{min}) = u_2(y_\mathrm{max}) = 0$.
This argument also holds for thermally-induced single-fluid variable-density mixing layers \citep[for which the governing equations are summarized and contrasted with the present equations in][]{livescu2020tlt,baltzer2020lst} if the heat conduction coefficient is constant.
More complicated cases such as heat release with
chemical reaction necessitates nonzero normal velocity at the boundaries, e.g., \cite{higuera1994ech}.
Spatially-developing mixing layers also include streamwise gradients in the streamwise velocity, leading to another term remaining in the left-hand side of the divergence equation, which leads to cross-stream velocities at the upper and lower domain velocities associated with entrainment in even the single-density case.

The third-order accurate variable time stepping Adams-Bashforth-Moulton method is used for time integration, coupled with the usual fractional step method. This is adapted for the pressure equation with variable coefficients due to non-zero velocity divergence associated with the variable-density equations. Fourier representations in the periodic coordinate directions allow the variable coefficient Poisson equation for pressure to reduce to an ordinary differential equation in the inhomogeneous direction. Taking advantage of the structure of the compact derivative, direct solvers can be employed for constant coefficient Poisson equations. 
The algorithm was initially devised for triply-periodic buoyant turbulence simulations by \cite{livescu2007bdv} to provide an exact
divergence of momentum and thus avoid degrading the overall order of accuracy. This was an advancement from the algorithm used by \cite{sandoval1995dvd} that required an extrapolation of velocity in time in order to determine the divergence of momentum but could degrade the overall temporal order of accuracy from second-order.

The variable coefficient Poisson equation for pressure is decomposed into the form $\nabla p/\rho^{(n+1)}=\nabla q + \nabla \times \vec{A} + \langle \vec{L} \rangle$, which results in a constant coefficient equation corresponding to the dilatational (curl-free) component, $\nabla q$, and implicit equations for the curl (divergence free), $\nabla \times \vec{A}$, and mean components. The implicit equations are solved iteratively, using the direct Poisson solvers at each step. Due to the periodic boundary conditions, the mean term $\langle \vec{L} \rangle$ is non-zero only in $y$ direction. Differences compared to the RT algorithm appear in the mean term for the mixing layer because of the mean flow in the streamwise direction. For the RT case, the mean velocity is zero in both (periodic) horizontal directions, while for the mixing layer case, it is zero only in the (periodic) spanwise direction.

This algorithm avoids introducing additional errors that could affect mass conservation or degrade the accuracy from the time stepping method. The dilatational component of $\nabla p/\rho$ is related to mass conservation, which is enforced to machine precision due to the direct solvers involved. The curl component, $\nabla \times \vec{A}$, is related to the baroclinic production of vorticity. The iterative procedure is performed until the maximum $x$--$z$ planar average squared change in $\nabla \times \vec{A}$ relative to the previous iteration value reduces to 0.01 times the squared value of $\nabla \times \vec{A}$ averaged within the plane, for each component $\alpha$:
\begin{align}
\max_{\substack{j \in \left\{1, \ldots, N_y\right\} \\ \alpha \in \left\{1, 2, 3\right\}}}{\frac{\sum_{k=1}^{N_z}\sum_{i=1}^{N_x}\left[(\nabla \times \vec{A})_\alpha^{(n)}(x_i,y_j,z_k)-(\nabla \times \vec{A})_\alpha^{(n-1)}(x_i,y_j,z_k)\right]^2}{\sum_{k=1}^{N_z}\sum_{i=1}^{N_x}\left[(\nabla \times \vec{A})_\alpha^{(n)}(x_i,y_j,z_k)\right]^2}}<0.01,
\end{align}
where $n$ denotes the iteration number. This tolerance ensures small differences compared to convergence to machine precision. Note that each step of the iterative procedure is based on a direct Poisson solver.  

No filtering was used in the simulations, so that the small scales are not affected by numerical artifacts. The spatial resolutions were determined by the requirement that the Kolmogorov scale is well resolved and a series of lower resolution, early time mesh convergence studies. The higher Atwood number cases have more stringent spatial resolution requirements, but for consistency, the same resolution was used for all simulations with Atwood number of $0.75$ or below. Therefore, the lowest Atwood number simulations are over-resolved but should yield very high-quality vorticity and velocity gradient statistics. As described below in the discussion of self-similarity, at late times the peak local dissipation decays linearly with time, so the simulations require the finest resolution during the initial growth stage.

\cite{moin1998dns} note that the Kolmogorov length scale is often cited as the smallest scale that needs to be resolved,
but suggest that this requirement is more stringent than necessary for reliable first- and second-order statistics. For spectral methods, resolution is often expressed as $k_\mathrm{max} \eta$, where
$\eta$ is the average Kolmogorov length scale $(\nu^3/\epsilon)^{1/4}$ and $k_\mathrm{max}=a (2\pi/L)$ for a spectral representation
of $N$ grid points in a domain of length $L$. The leading coefficient of the $k_\mathrm{max}$ definition depends on the dealiasing employed, up to a maximum of $N/2$ if no truncation is used. The present simulations calculate the advective terms in skew-symmetric form to reduce the aliasing errors for cubic terms \citep{blaisdell1991nsc}.
In DNS intended to maximize Reynolds number, typical values are $1 \le k_\mathrm{max}\eta \le 2$ \citep{gotoh2013pst},
with $k_\mathrm{max}\eta \approx 1.5$ typical for adequately-resolved DNS of isotropic turbulence \citep{petersen2010fss,pope2000tf}.
Greater resolution may be required when special attention is focused on certain features, such as fine scale structure associated with
stretched spiral vortices in isotropic turbulence that requires $k_\mathrm{max}\eta \gtrsim 4$ \citep{horiuti2008mms}
or the alignment of strain rate and vorticity \citep{hamlington2008lns}.

In the present mixing layer at negligible Atwood number, $k_\mathrm{max} \eta$ for the Fourier
spectral representation of each homogeneous direction reaches a minimum of $\approx 1.7$ at early times at the centerline (where turbulence is most developed) and continuously increases thereafter.
\cite{pantano2002sce} report final values of $k_\mathrm{max} \eta \approx 1.0$, and they rely on spatial filtering
that was shown to produce a relatively small amount of nonphysical dissipation to improve stability in their simulations.
Resolution can also be be quantified in terms of grid spacing relative to the average Kolmogorov scale.
\cite{almagro2017nsv} reported horizontal grid spacing finer than $1.8\eta$ during the self-similar growth, whereas
the corresponding values in \cite{pantano2002sce} are $3$--$4\eta$. In the present low $A$ simulation, the
horizontal grid spacing ($\Delta x$ and $\Delta z$) peaks at $1.8\eta$ during the early-time transition and reduces to $1.0\eta$ during self-similar growth.
Since the mixing layer is inhomogeneous and the Kolmogorov microscales shown above calculated from the dissipation at the
peak $y$ position does not account for inhomogeneities in the flow scales, these values merely represent a guideline.

For the present high Atwood number simulations, resolutions can be similarly estimated using the isotropic turbulence formula for $\eta$ that does not address how scales may vary with local density variations. For the present $A=0.75$ simulation, which has the same grid spacing as the $A=0.001$ simulation, $k_\mathrm{max} \eta$ attains a minimum value of 1.8 at early times and is 3.2 to 3.7 during the self-similar growth (which is similar to the values attained in the $A=0.001$ case). For $A=0.75$, the horizontal grid spacing corresponds to a maximum of $1.8\eta$ at early time and decreases to $1.0\eta$ by the end of self-similar growth. For $A=0.87$, the simulation requires a greater number of grid points for the same physical domain size to maintain numerical stability. The calculated $k_\mathrm{max} \eta$ reaches a minimum value of 2.7 at early times but remains between 4.4 and 5.3 during the identified self-similar growth interval. The horizontal grid spacing corresponds to a maximum of $1.2\eta$ at early time and decreases to $0.6\eta$ by the end of self-similar growth for $A=0.87$. Nonetheless, these values based on isotropic turbulence $\eta$ are not sensitive to localized steep velocity and density gradients at increased Atwood number that are hypothesized to necessitate greater resolution for numerical stability.

The compact finite difference scheme used for the cross-stream ($y$) direction is 6th order accurate for both the momentum and pressure equations. The uniform grid spacing is finer (reduced to a factor of 0.8: $\Delta_y=0.8\Delta_x=0.8\Delta_z$) in the inhomogeneous direction, in order to compensate for the lower accuracy relative to the Fourier directions.
Modified wavenumber analysis for 6th order compact difference equations indicates errors in differentiating modes
become larger at higher wavenumbers \citep{petersen2010fss}.
Since differentiation with the Fourier method is exact up to its highest resolved wavenumber, the Fourier method has no
error until the Nyquist frequency. This corresponds to a grid spacing of $2 \eta$ if $k_\mathrm{max} \eta$=1.5.
Requiring the compact difference method to produce less than 25\% error in differentiating a mode with this same wavelength
dictates that the grid spacing must be refined relative to that of the spectral method by a factor of 0.8.
Note that the vast majority of the energy in the flow is at longer wavelengths that have negligible error, according to the
modified wavenumber analysis: the lowest $3/4$ of the wavenumbers have errors of less than 3.5\%.

The pressure determined by the fractional step method restores the velocity field divergence to be consistent with (\ref{eq:divg}); however, it represents the average pressure over the time step.  To recover the instantaneous pressure for calculating budgets and other statistics, the Poisson equation resulting from obtaining the divergence of (\ref{eq:mom}) is computed as a
post-processing step after the flow has been advanced in time by the fractional step method.
The numerical algorithm has been verified to accurately satisfy the governing equations by comparing the time derivatives calculated for various quantity budgets that appear throughout this paper with the appropriate budget right-hand sides.

\subsection{Domain Size}

The domain lengths in the homogeneous streamwise and spanwise directions $L_x$ and $L_z$ are directly related to the convergence of statistical
quantities obtained by planar averaging. In addition, these dimensions potentially affect the sizes of structures that grow within the domain. Convergence can be improved either by enlarging the domain size or by using an ensemble of smaller domain simulations.
However, a sufficiently large domain is necessary to achieve correct structure growth and interactions.

Several domain sizes were tested and the final dimensions used were found to have minimal evidence of structure growth restriction compared to smaller sizes.
From the perspective of initial KH rollup structures with an assumed streamwise wavelength of the most unstable linear instability mode $\lambda_{ls}$,
the present mixing layer domain accommodates $64\lambda_{ls}$ in the streamwise direction. This corresponds to 6 successive mergers; \cite{vreman1997les} found that lengths of $8\lambda_{ls}$ (i.e., three successive mergers) were required to reach reasonable self-similarity.
In shear flows, the longest scales are oriented along the streamwise direction.
The domain therefore has a $L_x/L_z$ ratio of 4,
which was adopted by a number of previous temporal mixing layer simulations \citep[e.g.,][]{rogers1994dss,obrien2014ssb}.

The cross-stream domain size, $L_y$, must also be sufficiently large that the mixing layer evolves freely without the
slip walls at the $y$ domain boundaries influencing the growth. A series of simulations with different thicknesses
has been performed to ensure the statistics are not influenced by the walls for the self-similar time of interest.
The initial interface is positioned so that it is nearer the heavy-fluid wall than the light fluid wall in proportion to the Atwood number,
since the mean velocity neutral point (interface center) and the most intense turbulence drift to the light fluid side as the flow
develops (\S\ref{ss:profilesssmean}). The interface is centered within the domain for the $A=0.001$ case (as this effect is negligible at low density ratios).
The domain sizes are summarized in Table~\ref{tab:domsize}. Although initial momentum thickness $\delta_{m,0}$ (defined below) is somewhat ill-defined for making comparisons, comparing $L_x/\delta_{m,0}$ suggests that the domain lengths are approximately 10 times those of \cite{pantano2002sce} and 3.9 times those of \cite{almagro2017nsv}.

\begin{table}
  \begin{center}
\def~{\hphantom{0}}
  \begin{tabular}{lcccccccc}
      $A$& $L_x/\delta_{m,0}$ & $L_y/\delta_{m,0}$ & $L_z/\delta_{m,0}$ & $y_{min}/\delta_{m,0}$ & $y_{max}/\delta_{m,0}$ & $n_x$ & $n_y$ & $n_z$\\[3pt]
  $0.001$ & $1803.2$ & 1105.56 & 450.8 & $-552.78$ & $552.78$ & $4096$ & $3072$ & $1024$\\
  $0.25$ & $1803.2$ & 1105.56 & 450.8 & $-594.18$ & $511.38$ & $4096$ & $3072$ & $1024$\\
  $0.50$ & $1803.2$ & 1105.56 & 450.8 & $-594.18$ & $511.38$ & $4096$ & $3072$ & $1024$\\
  $0.75$ & $1803.2$ & 1105.56 & 450.8 & $-626.22$ & $479.34$ & $4096$ & $3072$ & $1024$\\
  $0.87$ & $1803.2$ & 480.60 & 450.8 & $-337.50$ & $143.10$ & $6144$ & $2048$ & $1536$\\
  \end{tabular}
  \caption{Summary of simulation domain parameters. The initial interfaces of both velocity and density are each positioned at $y=0$.}
  \label{tab:domsize}
  \end{center}
\end{table}

\section{Initial Conditions}
\label{sec:initcond}

Mixing layer simulations are typically designed either to approximate a physical mixing layer experiment or
to be in a generic configuration commencing from a simple disturbance. The latter approach is here adopted for generality and
to promote quickly reaching self similarity without artifacts from the initial condition.
Nonetheless, parameters are broadly within the range of those found in experiments.

\subsection{Mean Velocity and Density Profiles}
\label{ss:initcondprofiles}

The initial mean velocity profile that approaches the free-stream velocities of $\pm \Delta U/2$ at the
$y$ boundaries is specified as
\begin{align}
\bar{U}_1(y) = \frac{\Delta U}{2} \tanh\left(\frac{y}{2 \delta_{m,0}}\right),
\end{align} 
where the momentum thickness $\delta_{m,0}$ specifies the initial thickness of the interface.
The hyperbolic tangent profile is commonly used in a wide range of mixing layer simulations,
such as \cite{riley1986dns,pantano2002sce,olson2011nec,obrien2014ssb,almagro2017nsv}.

An initial density profile is prescribed to specify the differing compositions (and thus densities) of the fluid streams.
The simulations focus on the simplest case of two separate streams of different velocities and
densities meeting at a thin interface, so the initial density profiles are aligned with and of the same thickness as the velocity profiles.
Thus, the initial density profile is
\begin{align}
\overline{\rho}(y) = \rho_0 + \frac{\Delta \rho}{2}\tanh \left(\frac{y}{2\delta_{\rho,0}}\right)
\end{align}
with density profile thickness $\delta_{\rho,0}$ chosen to equal $\delta_{m,0}$.
This specification of aligned tanh profiles of density and velocity is similar to the approach of \cite{pantano2002sce}
and \cite{almagro2017nsv}, though their density variations were attained by varying the thermodynamic properties for a single fluid.
In either approach, the mean density of the lower and upper streams of fluid $\rho_0 = \left(\rho_1 + \rho_2\right)/2$ is matched
between all of the simulations within the set. The desired Atwood numbers $A$ are then attained by specifying free-stream densities
$\rho_1 = \rho_0 - \Delta \rho/2$ and $\rho_2 = \rho_0 + \Delta \rho/2$, where $\Delta \rho = \rho_2 - \rho_1 = 2A\rho_0$.
Symmetries present in the temporal mixing layer (but not the spatially-developing case) result in the flow behaviors being equivalent whether
the negative mean streamwise velocity is associated with the light fluid and the positive velocity is associated with the heavy fluid
or vice versa, as also noted by \cite{pantano2002sce}. Thus, results from a different profile convention can be compared by
selecting coordinates to match density profiles and then changing the sign of the mean streamwise velocity to also match.

\subsection{Initial Disturbance}

Only the velocity field is perturbed relative to the mean profile given above to induce the transition to turbulence. This is appropriate because the velocity field drives the instability and turbulence, as observed in the single-density case; this approach also allows the disturbance to be consistent between Atwood numbers. Different velocity disturbances can produce significantly different
growth rates at early times in mixing layers \citep{fathali2008sai}, but the present goal is to quickly establish self-similar growth and minimize long-lived large-scale structures that are uniquely associated with initial disturbances. To roughly resemble physical experiments, the velocity perturbation is confined to a thin (in $y$) region centered at the mean velocity profile interface.

In the present simulations, this is accomplished by generating a random field (filling the full domain) that is divergence-free and has a 3D energy spectrum obeying a Gaussian behavior at high wavenumbers with $k^4$ behavior at low wavenumbers as $E(k) = \left({k}/{k_0}\right)^4 e^{-2\left({k}/{k_0}\right)^2}$. Here, $k=\sqrt{k_1^2+k_2^2+k_3^2}$ is wavenumber and $k_0$ is the prescribed peak wavenumber.
$k_0$ is selected to be $\lambda_{ls}/4$, where $\lambda_{ls}$ is the streamwise wavelength of the least stable mode calculated from temporal linear stability analysis for the base velocity profile ($\lambda_{ls}=28\delta_{m,0}$ for the present set-up). This places much of the energy at small scales to quickly establish turbulent motions. The disturbance spectrum is that used by \cite{pantano2002sce} and the positions of the peak wavelength (relative to the least stable wavelength) are similar. The field is then tapered to a thin interface region by multiplying by the Gaussian profile in $y$ to obey
$\left\langle u_i' u_i' \right\rangle(y) = A e^{-\frac{1}{2} \left({y}/{\sigma}\right)^2}$, where $\sigma$ is the intensity profile thickness chosen to be $2\delta_m$. This is nearly equivalent to the thickness used in \cite{riley1986dns} simulations based on measurements of the intensity
profile in a mixing layer experiment and to the thickness used by \cite{pantano2002sce}.
The peak amplitude $A$ is specified for peak intensity $\left\langle u_i' u_i' \right\rangle$ of $0.03\Delta U^2$ by prescribing a $0.1\Delta U$ RMS fluctuation for each velocity component. This relatively strong disturbance reduces the time
to reach self-similar growth. The self-similar value of the streamwise turbulent velocity fluctuation intensity reaches
approximately 2.5 times this initial value.

This initial velocity disturbance is similar to those used by \cite{riley1986dns} \citep[further described in][]{riley1979dns} and \cite{pantano2002sce} \citep[further described in][]{pantano2000cet}, but details of the implementations differ.
The present approach of multiplying the field by the $y$-intensity profile produces divergence, which is corrected by applying the pressure step of the projection method to the velocity field. This step slightly weakens the intensity of the $u_2$ velocity component. Alternatives exist \citep[e.g., applying the profile to a
vorticity field, thereby producing a divergence-free velocity field as in][]{pantano2000cet},
but the present method produces an initial velocity field divergence fully consistent with the variable-density incompressible divergence condition (\ref{eq:divg}).
A small mean $u_2$ velocity is also produced by this step, which is consistent with the divergence condition (as further explained in Appendix~\ref{appendix:mean}). This mean velocity is concentrated at the interface and decays toward the $y$ boundaries; the magnitude is also very small ($<1\%$ of $\Delta U$ in all simulations shown).

\subsection{Viscosity and Diffusivity}
\label{ss:viscdiffus}

Momentum thickness Reynolds number, $Re_{m}=\Delta U \delta_m/\nu$, can be maximized during the self-similar stage by either growing to a large final thickness $\delta_m$ or having a small viscosity $\nu$. The initial configuration is chosen to maximize the thickness growth so that the fully-turbulent state is less affected by the initial disturbance. This is achieved by selecting a relatively small initial momentum thickness and appropriate viscosity such that all scales are well resolved and the initial growth is not overly damped. The fundamental velocity scale $\Delta U$ to initialize the simulation is arbitrary and can be scaled out.
In consistent units, $\Delta U=1$ is prescribed with initial momentum thickness of $0.5$ and viscosity of $0.00625$. This initialization results in a Reynolds number $Re_{m}$ of $80$; however, this value has limited meaning before mixing layer evolution sustains the scales of motion.

The Schmidt number $\mathrm{Sc}=\nu/\mathcal{D}$ is chosen to maintain a constant value of $1$ everywhere as the fluids mix.
This is imposed by selecting the same values of kinematic viscosity $\nu=\mu/\rho$ for each of the participating fluids (i.e., $\nu_1=\nu_2$)
with constant diffusivity $\mathcal{D}$. The choice of constant kinematic viscosity to maintain constant Schmidt number of $1$ is frequently
used in other multi-fluid mixing studies \citep[e.g.,][]{sandoval1995dvd,cook2001tsr,livescu2007bdv}, though maintaining $Sc=0.7$ (which is typical for gases)
is also common \citep[e.g.,][]{olson2011nec}. Note that the choice of constant $\nu$ implies that $\mu \sim \rho$, whereas with real fluids there is typically a weaker
dependence on density such as $\mu \sim \sqrt{\rho}$ \citep{livescu2010npv}. 

\section{Basic Definitions and Theoretical Flow Properties}
\label{sec:defprop}

While detailed simulations are necessary to obtain many quantities describing the flow, several
characteristics of the flow can be deduced from the governing equations and flow configuration.
The Favre mean equations obtained from (\ref{eq:cont}--\ref{eq:mom}) are
\begin{align}
\bar{\rho}_{,t} + \left(\bar{\rho} \tilde{U}_j\right)_{,j} & = 0\label{eq:contfa}\\
\left( \bar{\rho} \tilde{U}_i\right)_{,t} + \left( \bar{\rho} \tilde{U}_i \tilde{U}_j \right)_{,j} + 
\left(\bar{\rho}\tilde{R}_{ij}\right)_{,j} & = -\bar{P}_{,i} + \bar{\tau}_{ij,j}\label{eq:momfa},
\end{align}
where the Favre Reynolds stresses are
\begin{align}
\tilde{R}_{ij} = \frac{\left\langle \rho u_i'' u_j'' \right\rangle}{\bar{\rho}}.
\end{align}
These equations apply to incompressible variable-density flows as well as fully compressible flows.

When the equations are applied to the geometry and flow conditions of the temporally-developing mixing layer, many of the terms
vanish due to homogeneity and symmetries of the flow. The expanded equations after these simplifications are
\begin{align}
\bar{\rho}_{,t} + \left(\bar{\rho} \tilde{U}_2\right)_{,2} & = 0\label{eq:contfae}\\
\left( \bar{\rho} \tilde{U}_1\right)_{,t} + \left( \bar{\rho} \tilde{U}_1 \tilde{U}_2 \right)_{,2} + (\bar{\rho}\tilde{R}_{12})_{,2} & = \bar{\tau}_{12,2}\label{eq:mom1fae}\\
\left( \bar{\rho} \tilde{U}_2\right)_{,t} + \left( \bar{\rho} \tilde{U}_2 \tilde{U}_2 \right)_{,2} + (\bar{\rho}\tilde{R}_{22})_{,2} & = -\bar{P}_{,2} + \bar{\tau}_{22,2}\label{eq:mom2fae}.
\end{align}
The slip wall boundary condition in the $y$ direction requires that $\bar{U}_2=\tilde{U}_2=0$, $\tilde{R}_{12}=0$, and $\bar{\tau}_{12,2}=0$ at the boundary. These conditions are consistent with the variations outside the mixing layer, where $\bar{\rho}$ and $\tilde{U}_1$ are constant. As shown in Appendix \ref{appendix:mean}, for the incompressible flow considered here, the mean cross velocity can be expressed solely in terms of density moment statistics and their derivatives; the cross-stream velocity is necessarily zero if the flow contains no density variations.

\subsection{Conservation Properties}

Integrating the mean density conservation equation (\ref{eq:contfae}) over the $y$ domain indicates that $\int_{y_1}^{y_2} \bar{\rho} \; dy$ is constant with respect to time (total mass within the domain is
conserved). The mean momentum equations (\ref{eq:mom1fae})-(\ref{eq:mom2fae}), when similarly integrated over the $y$ domain, show that
$\int_{y_1}^{y_2} \bar{\rho}\tilde{U}_i \; dy$ are also constant with respect to time (total momentum within the domain is conserved),
when the remaining terms vanish at the boundaries. This is approximately satisfied for (\ref{eq:mom1fae}) and (\ref{eq:mom2fae})
throughout the duration of the simulation, since the velocity fluctuations remain at low values near the slip walls, and therefore the advective term and Reynolds stress are negligible at the $y$ domain boundaries, while the mean pressure gradients and viscous stresses have relatively little effect.

\subsection{Self-Similarity}
\label{ss:selfsim}

Another property expected of mixing layers is attaining states of self-similar growth.
For the temporal configuration, the statistics are functions only of time and the inhomogeneous $y$ position. Assuming self-similarity and that both mean density and velocity profiles are initially centered at $y=0$ (so that no other length scale is introduced in the problem), the time- and $y$-dependencies are eliminated by introducing a new variable $\eta=y/h$, where for the present purpose, $h$ generically represents a length scale that characterizes the $y$-thickness of the mixing layer and grows with time. Specific choices for defining this thickness are discussed below. The scaled coordinate $\eta$ defined here is separate from the Kolmogorov length scale $\eta$ of \S\ref{ss:numapprch}.

As described in Appendix~\ref{appendix:selfsim}, the mean mass conservation equation (\ref{eq:contfa}) and Favre mean streamwise momentum equation (\ref{eq:momfa}) are satisfied for self-similar growth when the growth rate $dh/dt$ is constant and the mean variables are non-dimensionalized as
\begin{align}
\bar{\rho}(y,t) & = \rho_0 \hat{\rho} (\eta)\label{eq:ssscalingrho}\\
\tilde{U}_1(y,t) & = (\Delta U) \hat{U}_1 (\eta)\label{eq:ssscalingU1}\\
\tilde{U}_2(y,t) & = \left(dh/dt \right) \hat{U}_2(\eta)\label{eq:ssscalingU2}\\
\tilde{R}_{12}(y,t) & = (\Delta U) \left( dh/dt\right) \hat{R}_{12}(\eta)\label{eq:ssscalingR12}.
\end{align}

Analyzing the resulting self-similar mass conservation and streamwise momentum equations (Appendix~\ref{appendix:selfsim}) reveals relations between the scaled $y$ positions at which features in the statistical profiles occur. Let $\eta_2$ be defined as the $\eta$ point where the Favre cross-stream velocity inflection point occurs [$d\hat{U}_2/{d\eta}(\eta_2)=0$] and $\eta_{12}$ as the point where Favre shear stress has its inflection [$d\hat{R}_{12}/{d\eta}(\eta_{12})=0$]. Then the self-similar analysis proves that $\eta_{12} < \eta_{2} < 0$.
That is, the Reynolds stress peak is located further in the light fluid than the peak of mean cross-stream velocity. This analysis does not determine the position $\eta_1$ of the zero-crossing of Favre streamwise velocity [$\hat{U}_1(\eta_1)=0$], but this can be empirically investigated in the simulations.

The above analysis and arguments reach similar conclusions to those presented by \cite{pantano2002sce} after developing the self-similar analysis framework while analyzing their variable-density flow. It should be noted that these self-similar equations and results pertain to \emph{any} variable-density mixing layer that obeys the compressible mass conservation and streamwise momentum equations. Specifying particular cases of the flow (in this case, incompressible binary mixing of species, as opposed to thermodynamic variations or high-speed flow) influences the specific forms of the self-similar quantities (assuming states of self-similar growth are reached).

\subsection{Thickness Definitions}
\label{ss:thickdefn}

The thicknesses of the density and streamwise velocity profiles are among the most basic global quantities characterizing mixing layers growth. Though the density and velocity mean profiles initially coincide, they need not grow identically as the flow evolves, so various thickness measurements are defined based on both profiles.

Thickness of a mixing layer is traditionally quantified based on the mean streamwise momentum profile,
which has a clear connection to the momentum equation (\ref{eq:momfa}).
Momentum thickness is defined as:
\begin{align}
\delta_m(t) & = \frac{1}{\rho_0 \Delta U^2} \int_{-\infty}^{\infty} \bar{\rho} \left[ \tilde{U}_1(y,t) - U_{-} \right] \left[U_{+} - \tilde{U}_1(y,t) \right] dy\nonumber\\
& = \int_{-\infty}^{\infty}  \frac{\bar{\rho}}{\rho_0} \left( \frac{1}{4} - \frac{\tilde{U}_1^2}{\Delta U^2} \right) dy\label{eq:momthickdefn}
\end{align}
As the first form emphasizes, this corresponds to the integral of the product representing deficits relative to
free streams, which have streamwise velocities of $U_{-}=-\Delta U/2$ and $U_{+}=\Delta U/2$.
An analogous thickness could also be defined on a per-mass basis to depend only upon the mean velocity profile:
\begin{align}
\delta_{m,pm} (t) & = \frac{1}{\Delta U^2} \int_{-\infty}^{\infty} \left[ \bar{U}_1(y,t) - U_{-} \right] \left[ U_{+} - \bar{U}_1(y,t) \right] dy\nonumber\\
& = \int_{-\infty}^{\infty}  \left( \frac{1}{4} - \frac{\bar{U}_1^2}{\Delta U^2} \right) dy
\label{eq:momthickpmdefn}
\end{align}
This definition uses Reynolds-averaged streamwise velocity rather than Favre-averaged to avoid any explicit dependence on the density field.
For single-density mixing layers, (\ref{eq:momthickpmdefn}) is commonly given as the definition of the momentum thickness because (\ref{eq:momthickdefn}) reduces to this when density is constant, though (\ref{eq:momthickdefn}) is the most formal definition.

Several other quantities also are commonly used to characterize mixing layer thickness based on the mean velocity profile, but these
are generally less smooth (i.e., more sensitive to lack of statistical convergence) than the integral thicknesses defined above. These other measurements include lengths based on gradients of profiles. Vorticity thickness is obtained from gradients of the Reynolds mean velocity profiles as
\begin{align}
\delta_\omega (t) = \frac{\Delta U}{\max(|d\bar{U}_1/dy|)},
\end{align}
as the vorticity magnitude reduces to $|d\bar{U}/dy|$ in the absence of a mean streamwise gradient in cross-stream velocity. This measure based on only a small portion of the mixing layer (where the mean gradient is steepest) has the potential to produce a misleading representation of the thickness of the layer when significant asymmetries are present.

The distance between
positions at which the mean velocity reaches specific percents (e.g., 10\%) of the difference $\Delta U$ between its
free-steam values $U_{-}$ and $U_{+}$ (which are associated with fluids having densities $\rho_1$ and $\rho_2$, respectively):
\begin{align}
h_{0.1} (t) = y_{\left[\tilde{U}_1=U_{+}-0.1*\Delta U\right]} - y_{\left[\tilde{U}_1=U_{-}+0.1*\Delta U\right]} = y_{\left[\tilde{U}_1=0.4*\Delta U\right]} - y_{\left[\tilde{U}_1=-0.4*\Delta U\right]}.
\label{eq:hdefn}
\end{align}
While momentum thickness
and vorticity thickness have been the most commonly used thickness measurements in the historic mixing layer
literature, \cite{pope2000tf} adopts $h_{0.1}$ in treating planar mixing layers, and it has also been recently used by \cite{schwarzkopf2016tls}, for example.
For brevity, $h$ will be used herein to indicate $h_{0.1}$. This choice of velocity percent produces measurements that are
smoother and less sensitive to statistical fluctuations than selecting a smaller fraction (e.g., $h_{0.01}$) that would yield thicknesses based on
the flow far out in the intermittently turbulent / non-turbulent interface.
Favre-averaged velocity is used for $h$, though it could alternatively be based on Reynolds-averaged velocity, as could any of the other thickness quantities. For even the highest Atwood numbers, the effect of averaging type on the calculated thickness is negligible: using Reynolds averages instead of Favre averages for $A=0.87$ produces about $1\%$ larger values for $h$ and $5\%$ larger values for $\delta_m$. Favre averaging is used for all of the velocity-based thicknesses shown except for $\delta_{m,pm}$ and $\delta_\omega$.

For variable-density mixing layers, similar thicknesses may be defined based instead on the density profiles as
\begin{align}
\delta_{\rho} (t) & = \frac{1}{\Delta \rho^2} \int_{-\infty}^{\infty} \left[ \left(\rho_0 - \frac{\Delta \rho}{2}\right) - \bar{\rho}(y,t)\right] \left[ \bar{\rho}(y,t) - \left( \rho_0 + \frac{\Delta \rho}{2}\right)\right] dy\\
\delta_{d\rho/dy} (t) & = \frac{\Delta \rho}{\max(|d\rho/dy|)}\\
h_{\rho,0.1} (t) & = y_{\left[\bar{\rho}=\rho_{2}-0.1*\Delta \rho\right]} - y_{\left[\bar{\rho}=\rho_{1}+0.1*\Delta \rho\right]} = y_{\left[\bar{\rho}=\rho_{0}-0.4*\Delta \rho\right]} - y_{\left[\bar{\rho}=\rho_{0}+0.4*\Delta \rho\right]}.
\label{eq:delrhody}
\end{align}
Note that $\delta_\rho$ is equivalent to the width measurement introduced by \cite{youngs1991tdn,youngs2009ami} and also used by \cite{livescu2010npv}; it
is typically written as $W = \int_{-\infty}^{\infty} F_1 F_2 \; dy$, defined based on the mean volume fractions of each species
$F_1 = \left(\bar{\rho} - \rho_1\right)/\left(\rho_2 - \rho_1\right)$ and $F_2 = \left(\rho_2 - \bar{\rho}\right)/\left(\rho_2 - \rho_1\right)$.
Typically, a scaling constant $\beta$ is used with $W$ to approximate bubble height in RT flows as $h_b^*=\beta W$;
$\beta$ depends on Atwood number in order to represent asymmetries that develop in RT flow structure as Atwood number increases
\citep{youngs2013drd,livescu2010npv}. For brevity, $h_\rho$ will be used herein to indicate $h_{\rho,0.1}$. As shown below, the mean density profiles develop significant asymmetries at high Atwood numbers, which implies that (\ref{eq:delrhody}) can only accurately represent the layer thickness at very low density ratios.

Additional width quantities commonly used for
variable-density flows \citep[particularly RT instabilities, e.g.,][]{livescu2009hrn, zhou2019tds} are also relevant.
One such quantity, used by \cite{cook2001tsr,livescu2008vdm,livescu2010npv}, is $h_{X_\rho} = \int_{-\infty}^{\infty} X_{P}(\bar{\rho}) \; dy$, where $X_{P}$ represents the amount of product
in a hypothetical fast reaction between the two species 
\begin{equation}
X_{P}(\rho) = \begin{cases} 
      2\frac{\rho - \rho_1}{\rho_2 - \rho_1} & \rho\leq \frac{\rho_1 + \rho_2}{2} \\
      2\frac{\rho_2 - \rho}{\rho_2 - \rho_1} & \rho\geq \frac{\rho_1 + \rho_2}{2}.
   \end{cases}
\end{equation}
$X_P(\bar{\rho})$ corresponds to the mole fraction of fluid fully mixed to the mean density. Physically,
$h_{X_\rho}$ is the thickness of mixed fluid that would result if the two fluids were perfectly homogenized within the mixing layer.

\section{Basic Statistics}
\label{sec:basicstat}

\subsection{Time Evolution of Mean Profiles and Thickness Growth}
\label{ss:thickgrowth}

\begin{figure}
\centering
\includegraphics[scale=0.85]{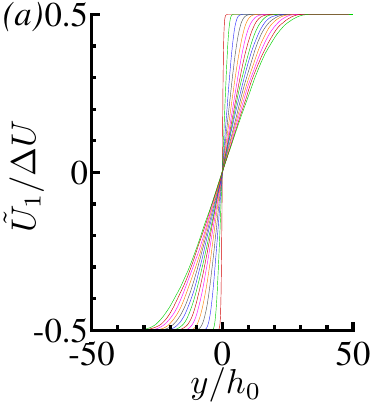}
\includegraphics[scale=0.85]{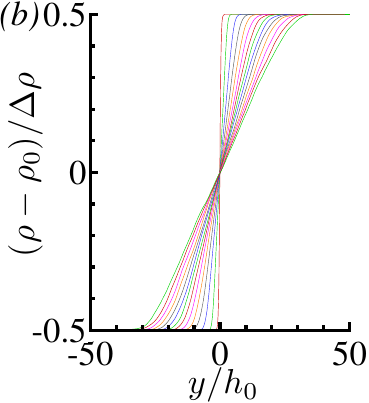}
\includegraphics[scale=0.85]{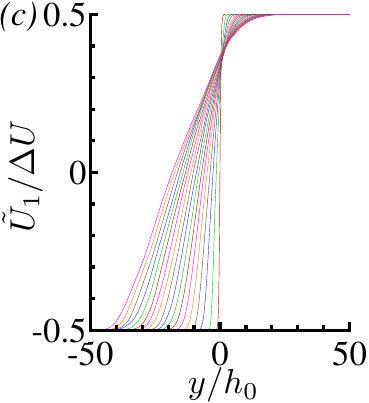}
\includegraphics[scale=0.85]{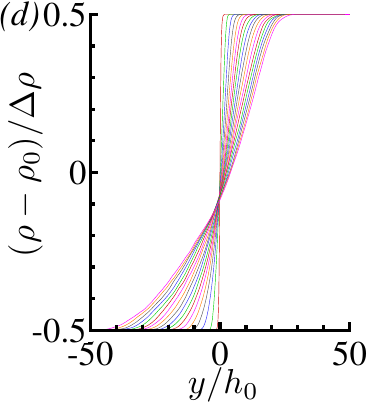}
\caption{Area-averaged mean profiles throughout the simulation run for $A=0.001$ (a--b) and $A=0.75$ (c--d). For Favre mean streamwise velocity $\tilde{U}_1$ (a, c) and scaled density (b, d), the cross-stream coordinate is scaled by the initial thickness $h_0$ and the profiles demonstrate the interfaces thickening with time. The lack of symmetry about $y=0$ that develops with increased Atwood number is apparent.}
\label{fig:profilesmeanall}
\end{figure}

Area-averaged mean profiles of streamwise velocity and density illustrate the basic properties of the mixing layers' evolution with respect to time. These profiles are shown for two representative Atwood numbers (almost single-density and strongly variable density) in Figure~\ref{fig:profilesmeanall}. These profiles form the basis for the thickness scales defined in \S\ref{ss:thickdefn}.

\begin{figure}
\centering
\includegraphics[scale=.73]{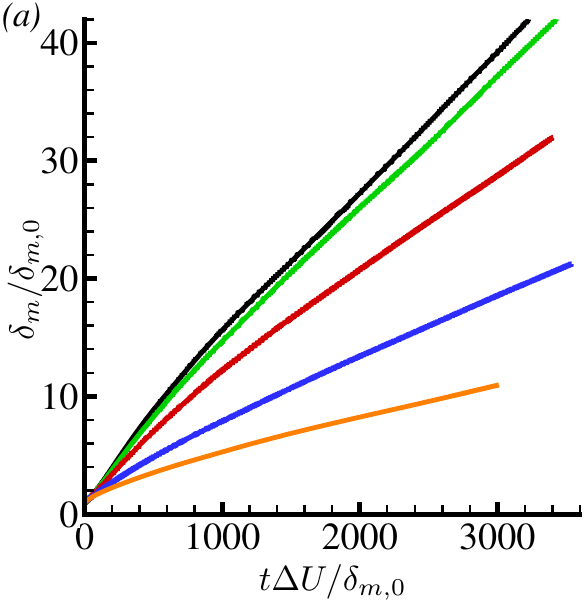}
\includegraphics[scale=.73]{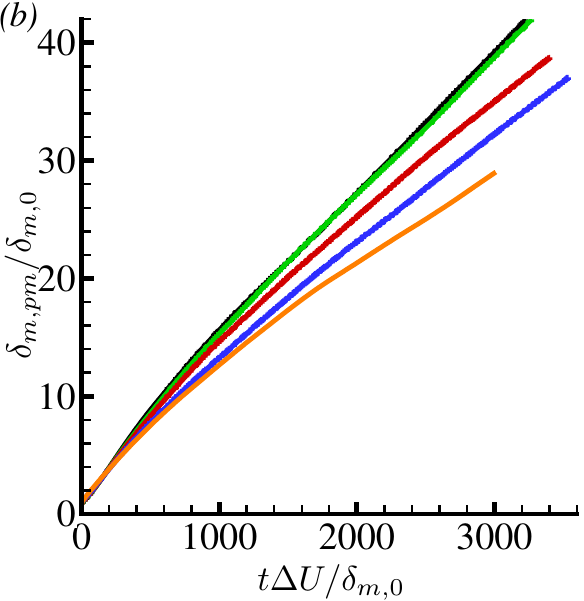}
\includegraphics[scale=.73]{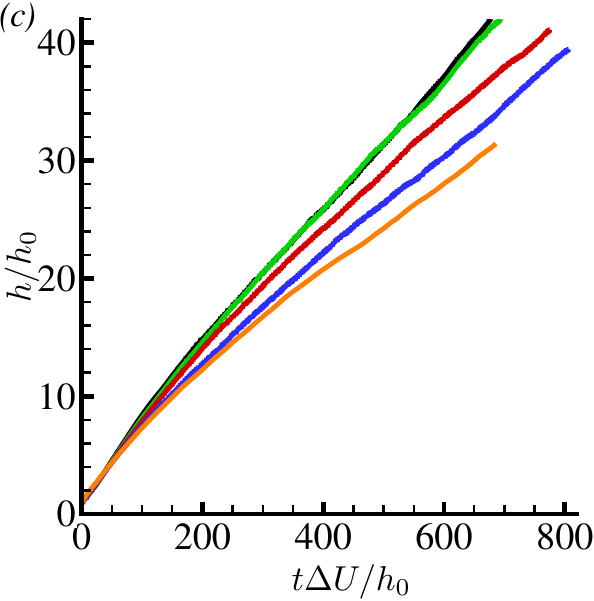}
\caption{Mixing layer widths: (a) momentum thickness $\delta_m$, (b) momentum thickness per unit mass $\delta_{m,pm}$ and
(c) mean velocity thickness $h$ time evolution for each Atwood number. Lines are colored by Atwood number: $A=0.001$ (\blackline); $A=0.25$ (\greenline); $A=0.50$ (\redline); $A=0.75$ (\blueline); $A=0.87$ (\orangeline).}
\label{fig:thicknessvst}
\end{figure}

Figure~\ref{fig:thicknessvst} displays the time evolution of thickness by several definitions involving the above profiles. All measurements indicate
that simulations for each Atwood number approach linear thickness growth with respect to time at late times. Regardless of the specific
thickness definition, thickness growth is retarded with increasing Atwood number.
The momentum thickness (a) indicates a strong reduction in growth with Atwood number, whereas the
momentum thickness per mass (b) and $h$ (c) quantities both indicate weaker growth reduction,
as does vorticity thickness (not shown).
The thickness evolutions also highlight that the mixing layers grow to many times their initial thicknesses, as desired to reach self-similar growth.

It should be noted that $\delta_{m,0}$, used for nondimensionalization, is based on initially aligned profiles at $t=0$, before the shifts of mean streamwise velocity relative to mean density have developed. Thus, $\delta_m$ and $\delta_{m,pm}$ are initially essentially equal but evolve differently as the profile shifts develop.
The correspondence between initial $\delta_m$ and other initial length scales is $h_0 = 4.39 \delta_{m,0}$ and $\delta_{\omega,0} = 4 \delta_{m,0}$; similar
relations apply to the analogous initial density thicknesses $\delta_\rho$, $h_\rho$, and $\delta_{d\rho/dy}$ as well. However, as the profile shapes evolve in transition and turbulent flow, these relations no longer apply.

\begin{figure}
\centering
\includegraphics[scale=.73]{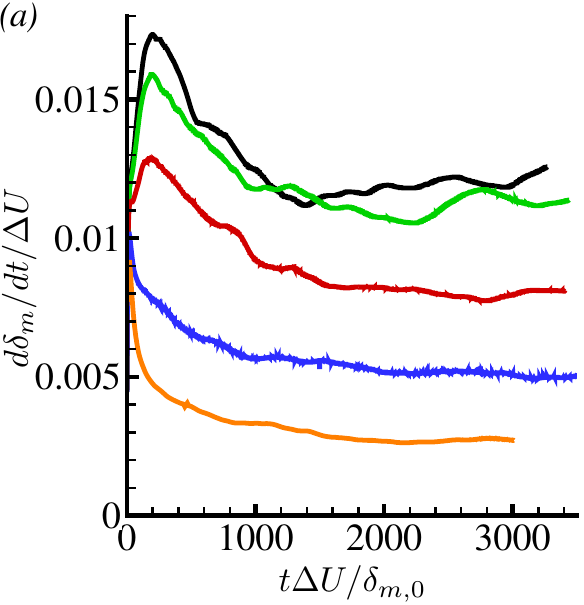}
\includegraphics[scale=.73]{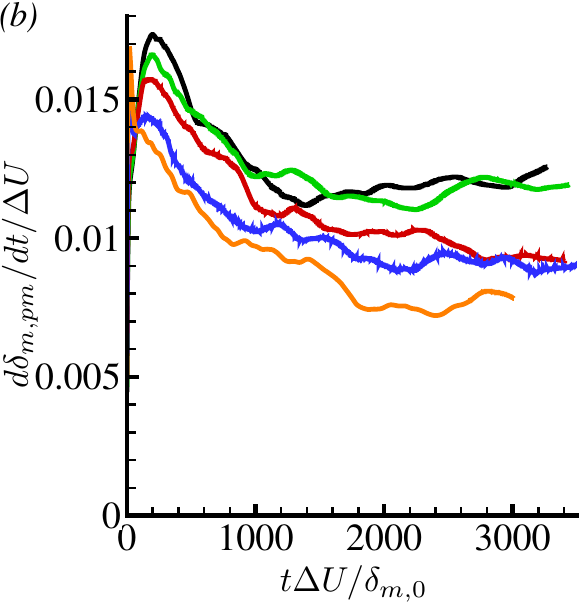}
\includegraphics[scale=.73]{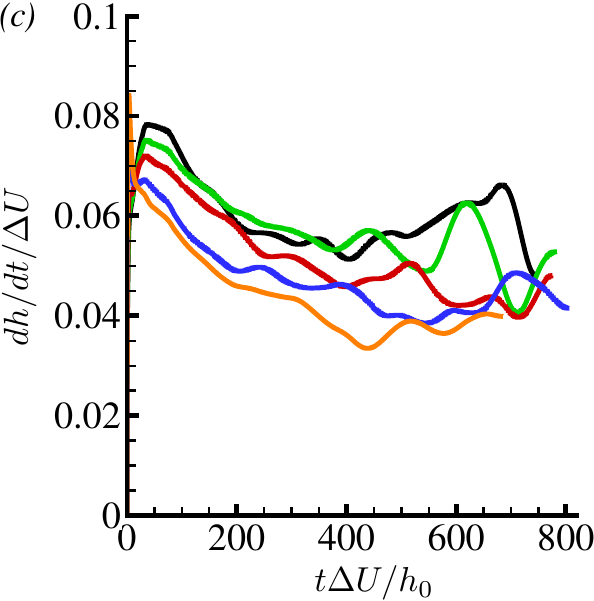}
\caption{Time evolution of mixing layer thickness growth rates based on (a) momentum thickness $\delta_m$,
(b) momentum thickness per unit mass $\delta_{m,pm}$ and (c) mean velocity thickness $h$. These correspond to the time derivatives of the thickness evolutions shown in Figure~\ref{fig:thicknessvst}.}
\label{fig:dthicknessdtvst}
\end{figure}

To evaluate whether constant values for self-similar temporal growth rates are reached, the time derivatives of thicknesses
are shown as functions of time for each Atwood number in Figure~\ref{fig:dthicknessdtvst}.
Thicknesses based on integral measures produce relatively smooth growth rates that in each case asymptote to constant values at late time.
Growth rates based on $h$ contain more noise than the rates based on the integral quantities, but applying a Hann filter to smooth the thickness vs. time
functions produces the result shown in Figure~\ref{fig:dthicknessdtvst}c. These results are also consistent with asymptoting growth rate (though statistical fluctuations are present).
For vorticity and density gradient thicknesses, calculating the gradient of a mean profile and then extracting its $y$-maximum makes these measurements
more sensitive to noise associated with lack of statistical convergence.
The sensitivity of the gradients to small-scale noise dictates that a small amount of spatial smoothing (via a Hann filter) first be applied to the instantaneous mean profiles to remove the finest scales of noise before calculating peak gradients.

\subsection{Determining the Time Interval of Self-Similar Growth}
\label{ss:indselfsim}

In addition to constant growth rate, another consequence of self-similar growth is the statistical profiles collapsing when appropriately scaled. For example, the mean streamwise velocity and density profiles would collapse to single curves for all times during self-similar growth when $y$ is scaled by thickness (e.g., $\delta_m$ or $h$).
As observed by \cite{rogers1994dss}, mean velocity
profiles are relatively insensitive to deviations from self-similar growth. However, fluctuation intensity profiles generally continue to converge after the mean velocities reach their self-similar profiles. Statistical profiles for many quantities are expected to have constant peak values and thus linearly increasing integral
values as thickness grows linearly with time. Directly evaluating the time histories of statistics' peak values comprises a more stringent test of
self-similarity, but evaluating their corresponding integral quantities instead is less sensitive to noise.

\begin{figure}
\centering
\includegraphics[scale=0.7]{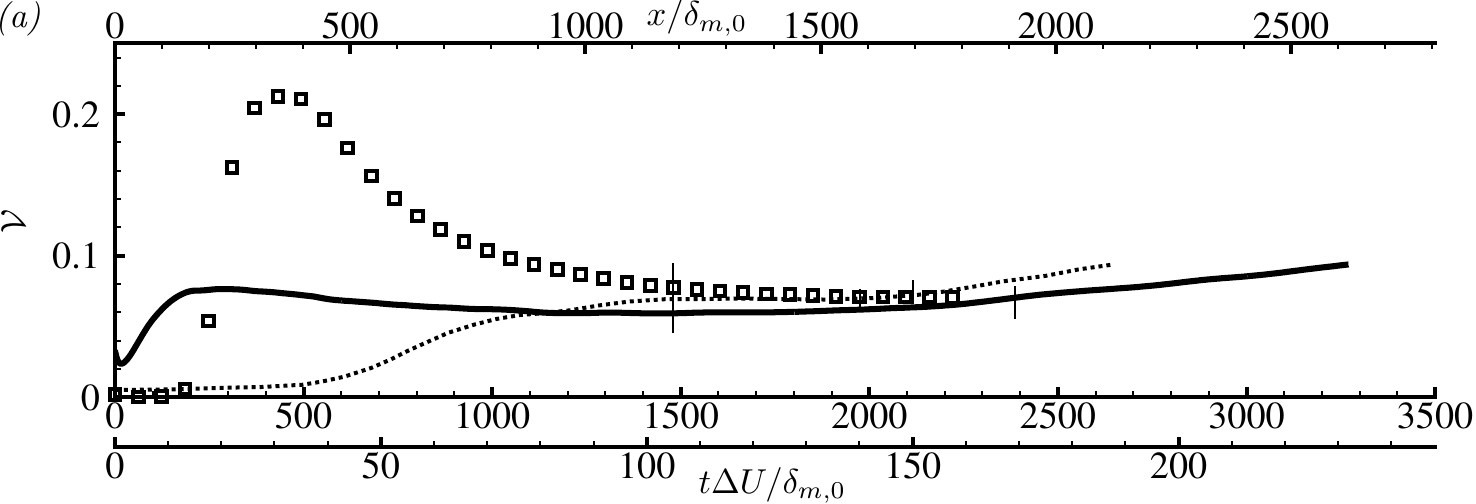}
\includegraphics[scale=0.7]{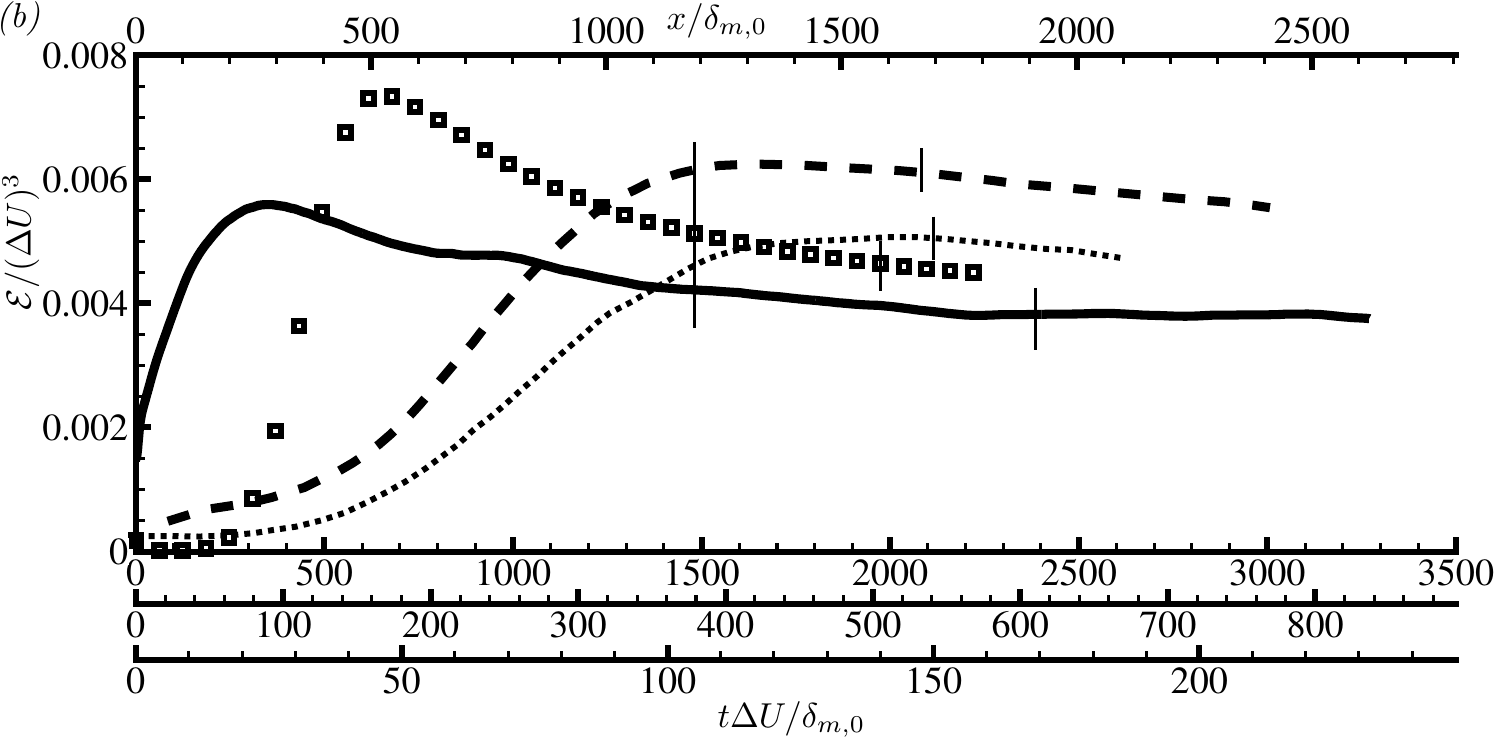}
\caption{Evolutions of $y$-integrated planar-average (a) cross-stream component velocity fluctuation
and (b) dissipation, two indicators of self-similar growth. Self-similar growth periods are indicated between the vertical lines; the
horizontal axes are scaled to match the beginnings such that the tall left-most vertical line applies to all of the flows. The right vertical lines mark
the end of self-similar growth for each flow individually. Note that this scaling is not intended to quantify the relative durations
of self-similar growth between simulations.
Of the lower horizontal axes, the upper-most corresponds to the present $A=0.001$ mixing layer (---$\!$---),
the middle in (b) only corresponds to the density ratio $s=1$ simulation of \cite{almagro2017nsv} (- - -), and the
lower-most corresponds to the simulation of \cite{rogers1994dss} ($\cdots$\,$\cdots$).
The upper horizontal axis corresponds to the spatially-developing simulation of \cite{attili2013fps} ($\square$).}
\label{fig:integralstatselfsim}
\end{figure}

One statistic that is meaningful for evaluating self-similar growth is integral of cross-stream velocity fluctuation intensity:
\begin{align}
\mathcal{V} = \frac{1}{\Delta U^2 \delta_m} \int_{-\infty}^{\infty} \langle u_2' u_2' \rangle \; dy.
\label{eq:vvint}
\end{align}
In earlier simulations emphasizing roll-ups of KH vortex structures and their subsequent mergers, \cite{moser1993tde} showed that
large values of $\mathcal{V}$ are associated with these features. Conversely, when \cite{rogers1994dss} began a mixing layer simulation
from a fully turbulent field, no large values were attained but instead $\mathcal{V}$ slowly increased and then asymptoted to the self-similar value.
\cite{attili2013fps} examined $\mathcal{V}$ for their spatially-developing mixing layer beginning from a thin disturbance
(similar to that for the present simulations). It overshot the self-similar growth value when the vortices played an important role at early time,
but decreased and asymptoted thereafter as the mixing layer reached a self-similar growth regime. This behavior is compared to that of
the present simulation with negligible Atwood number in Figure~\ref{fig:integralstatselfsim}a. The present simulation produces a much
weaker peak in $\mathcal{V}$ than the \cite{attili2013fps} simulation. Despite the weaker peak, the present simulation follows similar behavior of approaching
self-similarity after the peak.
This behavior contrasts with the asymptoting from below that appears to occur for the fully-turbulent initial condition of \cite{rogers1994dss}.
All of the simulations shown in Figure~\ref{fig:integralstatselfsim} display $\mathcal{V}$ values remaining approximately constant throughout their respective self-similar growth periods, and these values are in good agreement between the simulations.
In the present simulations, similar behavior also occurs at increased Atwood numbers.

An important indication of self-similarity employed by \cite{rogers1994dss} is total dissipation of turbulent kinetic energy (TKE), which is planar-averaged dissipation $\varepsilon=- \langle \tau_{ij}' u_{i,j}' \rangle$ (from the TKE budget equation)
integrated across the entire mixing layer:
\begin{align}
\mathcal{E} = \int_{-\infty}^{\infty} \varepsilon \; dy.
\end{align}
The rate at which TKE ultimately is dissipated is set by the large-scale motions that scale (in magnitude) with the
velocity difference between streams $\Delta U$. Since $\mathcal{E}$ has units of velocity cubed, it can be argued on dimensional grounds
that $\mathcal{E}$ scales with $\Delta U$ only and therefore is constant with respect to time during self-similar growth \citep{rogers1994dss}. 
Unlike the velocity fluctuation intensities, the dissipation peak value does not remain constant with respect to time but instead decays in
magnitude proportionally with the mixing layer thickness.
Thus, its integral over the increasing width as the mixing layer thickens remains constant.

For the essentially single-density case, the dissipation evolution is compared with those of other mixing layer simulations
in Figure~\ref{fig:integralstatselfsim}b. The self-similar growth durations are marked as identified in each corresponding reference.
Depending on the route of transition, the peak dissipation may also correspond to an overshoot in dissipation prior to
self-similar growth or to part of the self-similar growth regime. The former scenario applies to the simulation of \cite{attili2013fps}
that begin from a thin disturbance. The latter applies to the simulation of \cite{rogers1994dss} that begins from a field containing fully-turbulent fluid 
and slowly approaches the self-similar state from below (in terms of dissipation). \cite{attili2013fps} discuss these differences and
the role of KH structures in the transition in further detail. The present flow corresponds to the former scenario,
beginning from a thin disturbance leading to structures that cause dissipation to overshoot, though this is weaker than in \cite{attili2013fps} likely due to the
form of the disturbance and the temporally-developing nature of the flow.

Compared to the close agreement of self-similar $\mathcal{V}$ value with the other simulations in the literature, there is significantly more
variation among the self-similar integrated dissipation values. However, the \cite{attili2013fps} mixing layer appears to be asymptoting to
a value near that observed in the present $A=0.001$ simulation. The self-similar time interval shown for this present simulation
(for which $\mathcal{E}$ is one of the determining considerations) maintains $\mathcal{E}$ to a nearly constant value.

\begin{figure}
\centering
\includegraphics[scale=0.8]{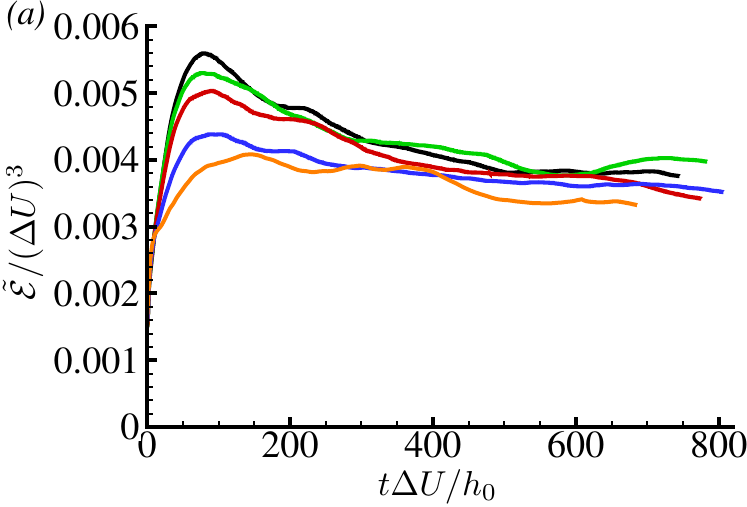}
\includegraphics[scale=0.8]{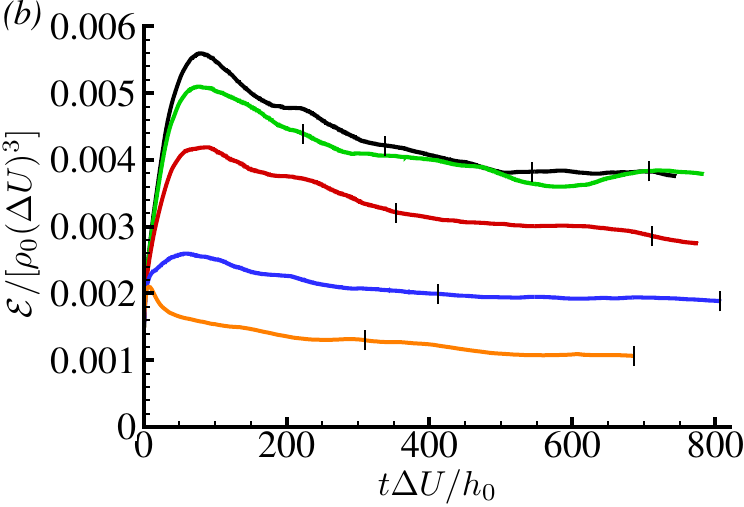}
\caption{The time histories of $y$-integrated planar-average dissipation (a) divided by local mean density and
(b) divided by the average density of the two streams. Both quantities asymptote to constant values
for every Atwood number, which is consistent with self-similar growth. The self-similar time periods are marked by
vertical lines in (b) and are based on time histories of dissipation as well as other quantities reaching values that are constant within a specified tolerance. Line colors for each Atwood number are as for Figure~\ref{fig:thicknessvst}.}
\label{fig:dissipationvstime}
\end{figure}

The dimensional argument described above for constant $\mathcal{E}$ in self-similar growth holds for the variable-density mixing layers as well.
For variable-density mixing layers, the TKE budget equation terms are often defined to include density \citep[e.g.,][]{livescu2009hrn},
unlike the typical budgets written for single-fluid incompressible mixing layers \citep[e.g.,][]{rogers1994dss}. Therefore, the integrated dissipation
must be divided by density to have the units of $(\Delta U)^3$.
One option is to nondimensionalize by $\rho_0$, the average of the two streams. However, the most typical treatment is to divide
$\varepsilon$ by the mean density $\bar{\rho}$, in analogy to Favre averaging other quantities:
\begin{align}
\tilde{\mathcal{E}} = \int_{-\infty}^{\infty} \frac{\varepsilon}{\bar{\rho}} \; dy.
\label{eq:Epstilde}
\end{align}
Figure~\ref{fig:dissipationvstime} demonstrates that $\mathcal{E}$ and $\tilde{\mathcal{E}}$ become constant in self-similar growth for each Atwood number.
The values for $\mathcal{E}$ scaled by $\rho_0$ and ${\Delta U}^3$ decrease strongly with increasing Atwood number, while $\tilde{\mathcal{E}}$ scaled by
${\Delta U}^3$ displays a much weaker dependence.

\begin{figure}
\centering
\includegraphics[scale=0.75]{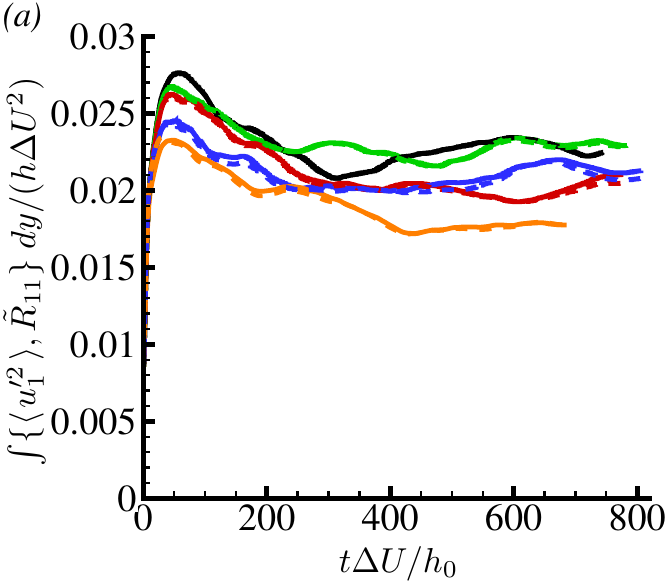}
\includegraphics[scale=0.75]{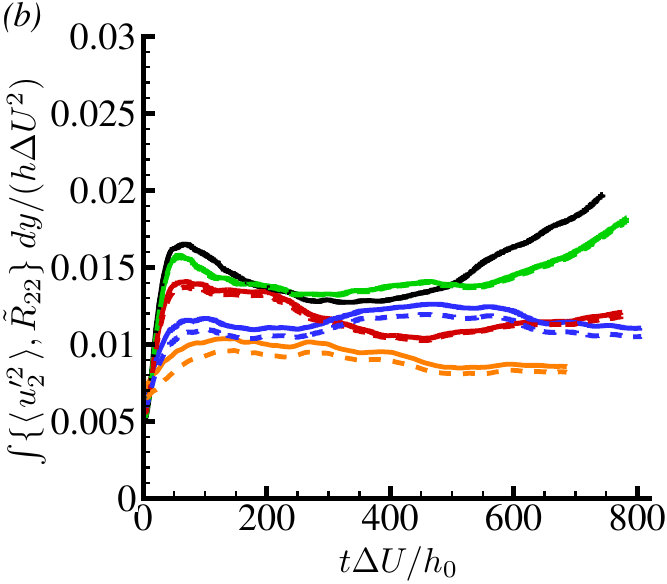}
\includegraphics[scale=0.75]{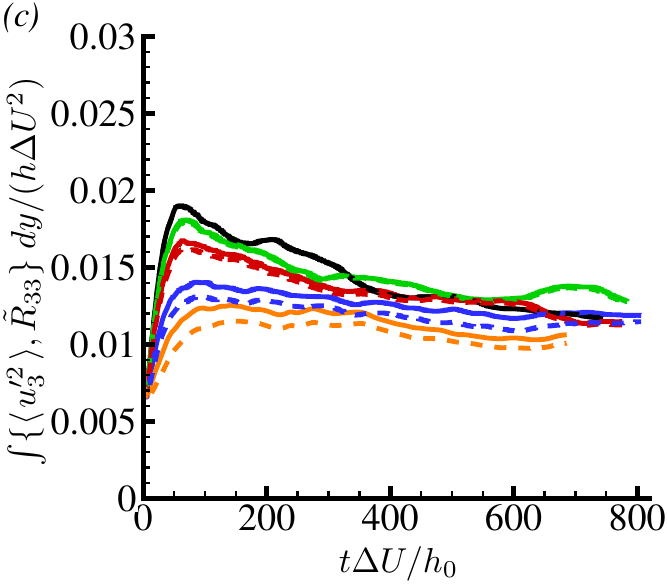}
\includegraphics[scale=0.75]{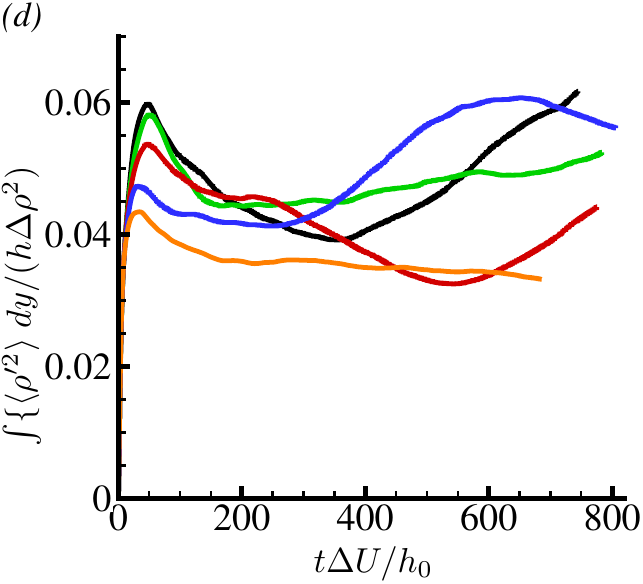}
\caption{(a--c) Time histories of $y$-integrated planar-average turbulence intensities $\langle u_i' u_i'\rangle$ (---$\!$---) and corresponding Favre-average
Reynolds stresses (- - -), each divided by Favre mean streamwise velocity thickness $h$ for the (a) streamwise, (b) cross-stream, and (c) spanwise components.
(d) Time histories of $y$-integrated planar-average density fluctuation intensities. Line colors for each Atwood number are as for Figure~\ref{fig:thicknessvst}.}
\label{fig:intvelmaxrhovarselfsim}
\end{figure}

While linear growth of thickness and constant integrated dissipation are key indicators of self-similar growth
\citep[which have been long been employed, e.g.,][]{rogers1994dss}, comparing additional flow statistics profiles
produces further useful indications. This was recognized by \cite{vreman1997les}, who determined mixing layer growth to be self-similar when
``the development of the shear layer thickness is linear in time and profiles of normalized statistical quantities at different times coincide.''
The time evolutions of profiles can be evaluated by monitoring the peak values of these statistics or examining their integrals in $y$ divided by the thickness
(as with $\mathcal{V}$). This latter approach is less sensitive to statistical variability than the peaks.
A number of profile quantities are considered in determining the self-similar growth time interval; integral velocity variances and Reynolds stresses are shown in
Figure~\ref{fig:intvelmaxrhovarselfsim}(a--c), while additional profiles (e.g., cross-correlations between velocity and density) are considered but not shown for brevity.
For each Atwood number, the integral turbulence intensities match very closely with the corresponding integral Favre-averaged Reynolds stresses and are
nearly identical for $A=0.5$ and below. Comparing between Atwood numbers, there is a consistent trend to lower intensities with increasing $A$ during transition
(when the values peak); during self-similar growth, the trend is weak and easily obscured by statistical variability. The $y$-integrated values shown may conceal some of the complexity in weakly changing profile shapes.
For the cross-stream component (Figure~\ref{fig:intvelmaxrhovarselfsim}b), the intensity increasing at late time is hypothesized to be associated the turbulent fluctuations reaching and accumulating near the slip walls to affect the interior of the mixing layer. This is expected to occur soonest for the lowest Atwood numbers because they experience the fastest growth. The self-similar time interval is determined to end before this phenomenon affects the flow.

Variable-density mixing layers introduce additional quantities to be considered for self-similarity, most importantly the density fluctuation intensity $\langle \rho' \rho' \rangle$. The integral values of this planar-mean quantity are shown for each Atwood number in Figure~\ref{fig:intvelmaxrhovarselfsim}d.
$\langle \rho' \rho' \rangle$ can remain within a tolerance of a constant value later than other statistics and thus determine
when the self-similar interval begins. These profiles are related to the mixing of the two streams, which is dependent on
how fluid is transported into the cores of the mixing layers.
Despite the complex mixing behavior, the simulations indicate that the density fluctuation intensity profiles for each Atwood number approach a unique self-similar scaled profile that remains approximately constant with respect to time.

The integral $\langle \rho' \rho' \rangle$ for $A=0.75$
(blue curve) is suggestive of reaching self-similar growth at particularly late time, with a leveling occurring at earlier time before it again increases and levels off.
It appears that the flow configuration changes during the second period of rapid increase. This behavior is responsible for the late starting time of the self-similar period.
This increase in maximum density fluctuation intensity also appears to be associated with a smaller increase in integral cross-stream component velocity fluctuation
intensity, as shown in Figure~\ref{fig:intvelmaxrhovarselfsim}b.

In summary, the self-similar periods are determined by seeking constant thickness growth rates, constant values of integrated dissipation, and statistical
profiles that remain constant when the cross-stream coordinate is self-similarly scaled. In addition to the velocity intensity profiles, density fluctuation intensity profiles must also be considered for variable-density mixing layers.
To identify self-similar growth periods in a consistent manner for all Atwood numbers, these conditions are approximated by requiring that thickness growth rates as well as integrals across the cross-stream domain of dissipation, velocity fluctuation intensity $\langle u_i' u_j' \rangle$, and density fluctuation intensity $\langle \rho'^2 \rangle$ be constant to within a specified threshold. The integrals of fluctuation intensity profiles are scaled by thickness ($h$) to attain constant values (or equivalently are integrated with respect to $y/h$), since the integrals would grow proportional to thickness if the self-similar scaled profiles remain constant.
Mean profile convergence is accomplished by ensuring the more sensitive fluctuation intensity profiles are converged.
This algorithm is consistently applied by determining the longest time interval that each of the quantities specified above remains within 10\% of any value and then retaining the intersection of these time intervals as the self-similar time interval.
The very large simulations produce satisfactory adherence to a relatively stringent set of criteria that must be simultaneously satisfied, as indicated by the self-similar periods marked in Figure~\ref{fig:dissipationvstime}.
The self-similar periods for other simulations compared in Figure~\ref{fig:integralstatselfsim} are taken from their respective publications. Due to the effects of differing initial momentum thicknesses (and how they
relate to the disturbances), the scaled times $t\Delta U/\delta_{m,0}$ (or scaled downstream position $x/\delta_{m,0}$ for the spatial-developing case) in this comparison
cannot be meaningfully related between simulations. The significantly smaller domains that
were feasible for many previous studies could contribute to the difficulties reported in reaching self-similarity \citep[e.g.,][]{vreman1996cml,vreman1997les,pantano2002sce}.
In general, questions remain about the universality of the self-similar state \citep[e.g., ][]{dimotakis1976mlh,rogers1994dss,vreman1997les}.
However, the thin and broadband disturbance is intended to reduce idiosyncratic large-scale vortices that persist after transition as a result
of the initial condition so the present simulations reach generic self-similar states.

Another consideration relevant to the self-similar growth regime is flow Reynolds number.
For the flow statistics to be representative of the fully turbulent mixing in practical applications, the Reynolds numbers must be sufficiently large throughout the averaging time duration. In general, significant changes in mixing behavior have been observed to occur at a Reynolds number threshold
\citep[i.e., the mixing transition,][]{dimotakis2000mtt}. Relevant Reynolds numbers are typically defined using the mixing layer thickness or the Taylor microscale. Both scales continuously grow as the mixing layers thicken with time.
According to \cite{dimotakis2000mtt}, general necessary conditions for passing the mixing transition for turbulent flows
are that the outer-flow Reynolds number exceeds $Re \approx 1$--$2 \times 10^4$ and that Taylor Reynolds number exceeds $Re_\lambda \approx 100$--$140$.
Dimotakis defines the former Reynolds number using a visual thickness scale $\delta_{sh}$ that is used in experiments; it has been estimated
as $\delta_{sh} \approx 2 \delta_{\omega}$ for numerical simulations \citep[e.g.,][]{rogers1994dss}.
This criterion corresponds to attaining $Re_{\omega} \approx 0.5$--$1 \times 10^4$.
Table~\ref{tab:renumcomp} confirms that this condition is satisfied for the self-similar growth statistical averaging periods. The decrease of $Re_m$ values with Atwood number is a consequence of $\delta_m$ decreasing as the velocity profiles shift into lighter density fluid. This complicates interpreting $Re_m$ in variable-density mixing layers.

Though Taylor microscale is anisotropic in its most fundamental definition, it is estimated using a relation that strictly only applies to homogeneous isotropic turbulence, $\lambda_g = \sqrt{10 \tilde{k} \frac{\nu}{\tilde{\varepsilon}}}$. (Averaging the homogeneous-coordinate components of the fundamental Taylor microscale shows good agreement with this estimate for the present mixing layers.) The velocity scale is also taken as $u'_\mathrm{rms} \approx \sqrt{\left(\langle u_1'u_1' \rangle + \langle u_2'u_2' \rangle + \langle u_3'u_3' \rangle\right)/3} = \sqrt{2k/3}$. Using the turbulent kinetic energy and dissipation at the $y$ position of most intense turbulence, the estimate of Taylor microscale Reynolds number is $Re_\lambda=\tilde{k}\sqrt{\frac{20}{3}\frac{1}{\tilde{\varepsilon}\nu}}$.
Using $u'_\mathrm{rms}$ produces consistency with the velocity scale used in the $\lambda_g$ definition
as well as consistency between the turbulent kinetic energy and dissipation included in turbulent kinetic energy budget (in analogy to isotropic turbulence).
Though similar definitions are also used for other relevant flows \citep[e.g.,][]{sekimoto2016dns}, mixing layer literature often uses
$\sqrt{2k}$ as the velocity scale (rather than $u'_\mathrm{rms}=\sqrt{2k/3})$ to form $Re_\lambda=(2k)\lambda_g/\nu=k \sqrt{20/(\varepsilon \nu)}$
\citep[e.g.,][]{pantano2002sce,obrien2014ssb,almagro2017nsv}. Renormalized to the present convention, the $Re_\lambda$ range during self-similar growth for the single-density mixing layer of \cite{rogers1994dss} is $84$--$99$ and for \cite{almagro2017nsv} is $81$--$87$, for example.
The present simulations generally satisfy the $Re_\lambda \approx 100$ (with $Re_\lambda$ is defined in this way) mixing transition guideline given by \cite{dimotakis2000mtt} before their self-similar growth periods end.
The consistency of the statistics within the self-similar growth periods suggests the turbulence is well-developed throughout. The initial condition that produces rapid transition is expected to lead to this state more quickly than the large-scale features that persist through other mixing layers' transitions.

\begin{table}
  \begin{center}
\def~{\hphantom{0}}
  \begin{tabular}{lccc}
      Simulation& $Re_\lambda$ & $Re_m$ & $Re_\omega$\\[3pt]
  $A=0.001$ & $82$--$108$ & $1700$--$2550$ & $8800$--$12700$\\
  $A=0.25$ & $72$--$128$ & 1150--3070 & 6100--15900\\
  $A=0.50$ & $80$--$104$ & 1360--2380 & 8600--14600\\
  $A=0.75$ & $81$--$106$ & 990--1700 & 8500--15500\\
  $A=0.87$ & $70$--$92$ & 510--880 & 6400--10900\\
  \end{tabular}
  \caption{Reynolds numbers during self-similar growth for the present simulations.}
  \label{tab:renumcomp}
  \end{center}
\end{table}

\subsection{Time-Averaged Self-Similar Statistical Profiles}
\label{ss:profilesssmean}

\begin{figure}
\centering
\includegraphics[scale=0.85]{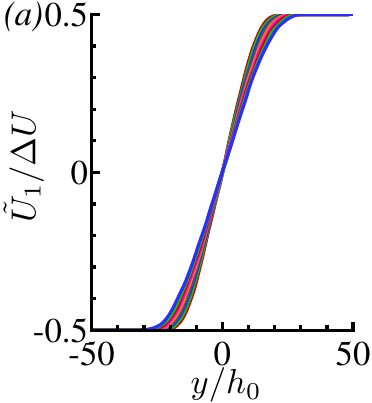}
\includegraphics[scale=0.85]{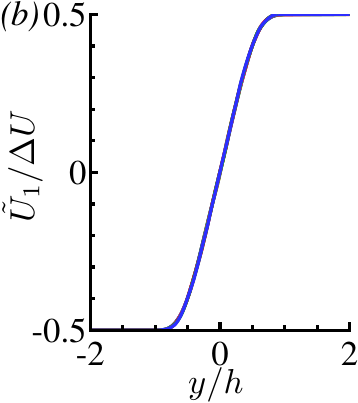}
\includegraphics[scale=0.85]{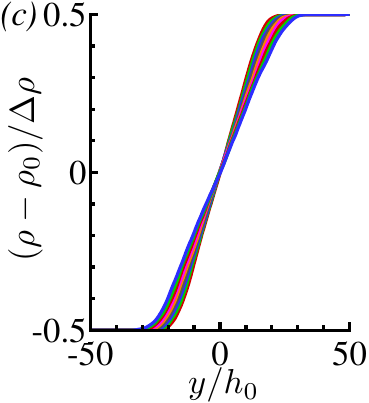}
\includegraphics[scale=0.85]{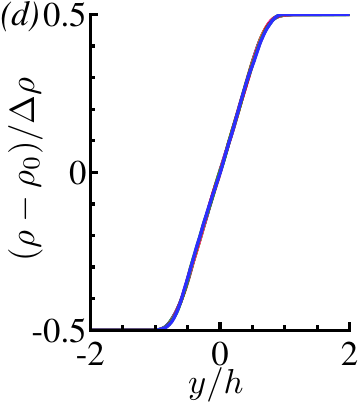}
\includegraphics[scale=0.85]{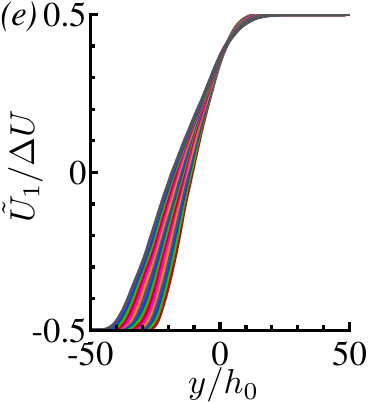}
\includegraphics[scale=0.85]{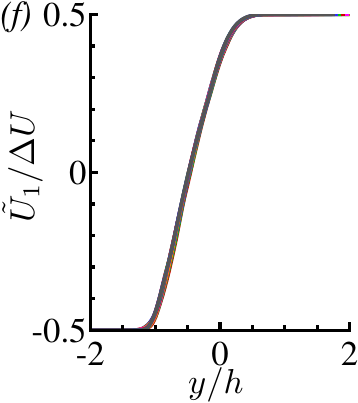}
\includegraphics[scale=0.85]{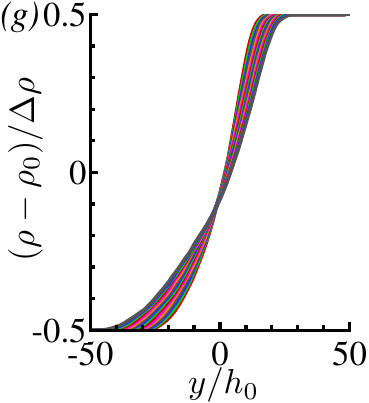}
\includegraphics[scale=0.85]{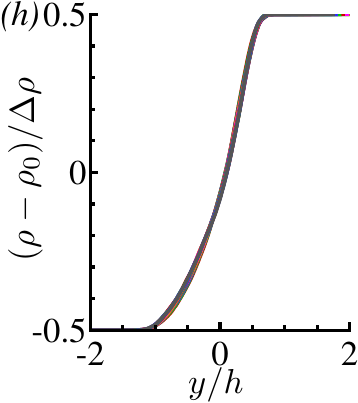}
\caption{Mean profiles during self-similar growth for $A=0.001$ (a--d) and $A=0.75$ (e--h). For Favre mean streamwise velocity $\tilde{U}_1$ (a--b, e--f) and scaled density (c--d, g--h), scaling the cross-stream coordinate by the initial thickness $h_0$ shows only the self-similar time interval curves from Figure~\ref{fig:profilesmeanall} and demonstrates the growth of the mixing layer (a,c,e,g), whereas scaling instead by each curve's thickness $h$ causes this same time series to collapse to a single curve for each simulation (b,d,f,h).}
\label{fig:profilesmeancollapse}
\end{figure}

Figure~\ref{fig:profilesmeancollapse}.
The times included in the plots correspond to the self-similar growth regimes, for which the determination is explained below (\S\ref{ss:indselfsim}). Figure~\ref{fig:profilesmeancollapse} demonstrates that the time series of mean streamwise velocity and density profiles collapse to single curves when the cross-stream coordinate is scaled by the thickness measurement $h$. Similar collapse is also observed when the cross-stream coordinate is instead scaled by $\delta_m$, $\delta_{m,pm}$, or $\delta_\omega$. While $\delta_m$ was used as the thickness length scale in the discussions above to allow comparison with other studies, scaling statistics in terms of the $h$ scale offers interpretive advantages in variable-density flow. For consistency, $h$ will be used as the thickness scale henceforth, except for when making certain comparisons with other studies. The collapse of mean profiles is one indication that self-similar growth is achieved. During self-similar growth, it is thus appropriate to time-average the scaled profiles to improve statistical convergence. This averaging is also applied to all of the other scaled statistics presented below.

\begin{figure}
\centering
\includegraphics[scale=0.7]{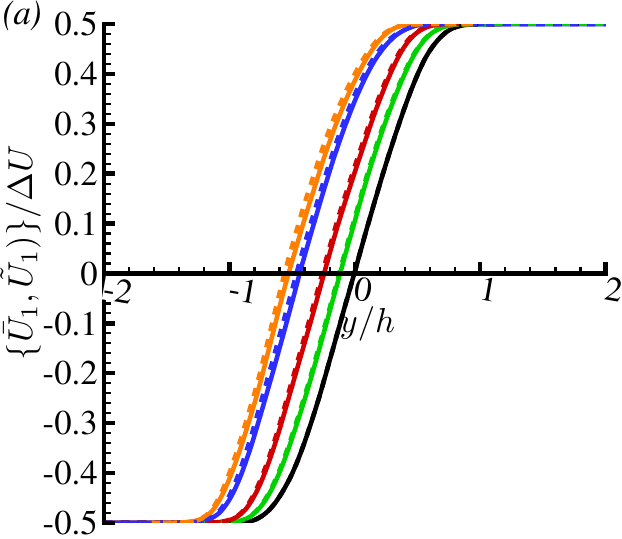}
\includegraphics[scale=0.7]{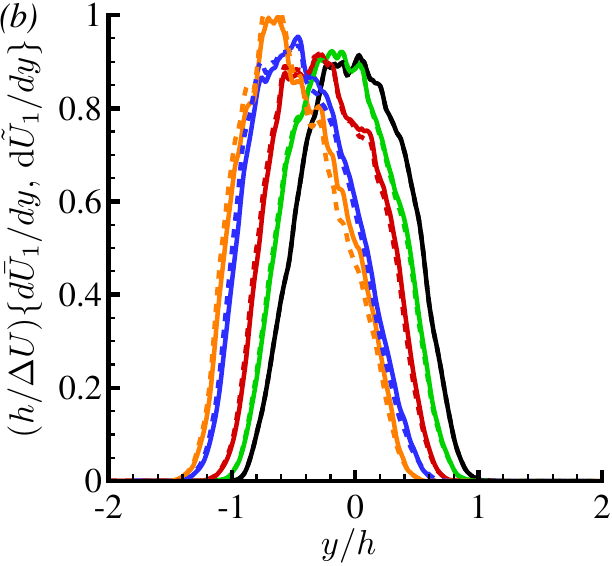}
\includegraphics[scale=0.7]{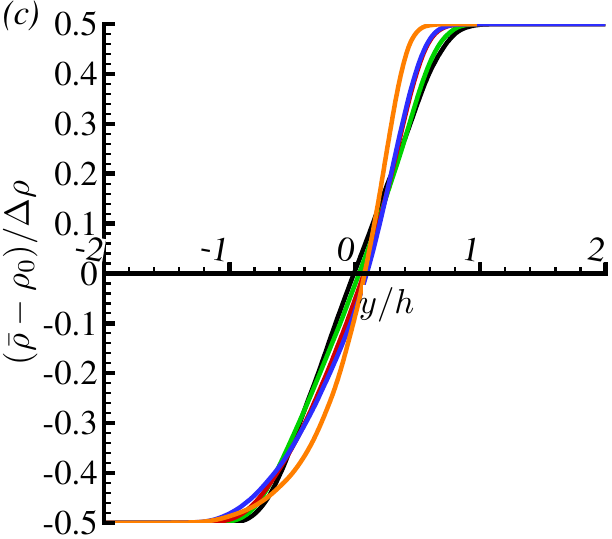}
\caption{Self-similar time-averaged profiles for all Atwood numbers showing (a) mean streamwise velocity, (b) its $y$ gradient, and (c) scaled mean density. In (a--b), the solid line represents Reynolds mean, while the dashed line represents Favre mean. Lines are colored by Atwood number as in Figure~\ref{fig:thicknessvst}.}
\label{fig:profilesmean}
\end{figure}

\begin{figure}
\centering
\includegraphics[scale=0.7]{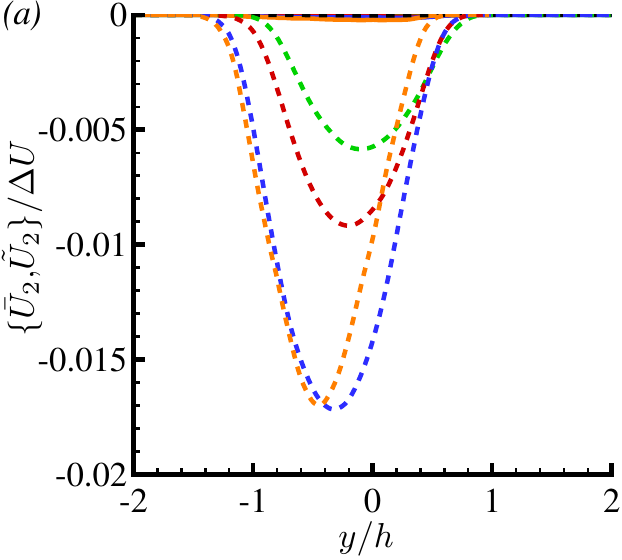}
\includegraphics[scale=0.7]{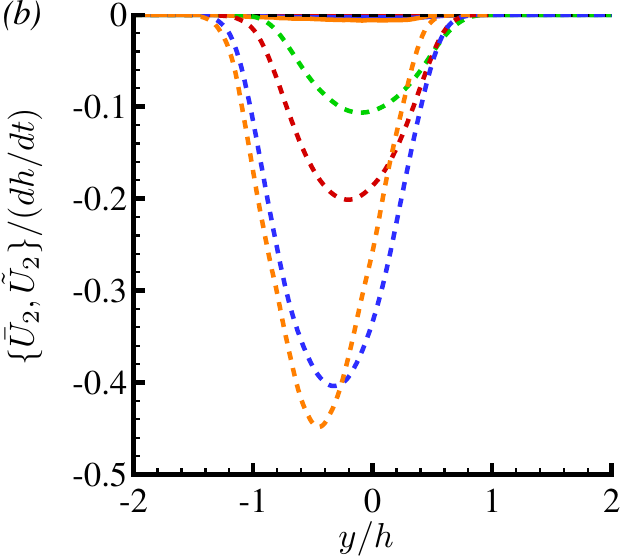}
\includegraphics[scale=0.7]{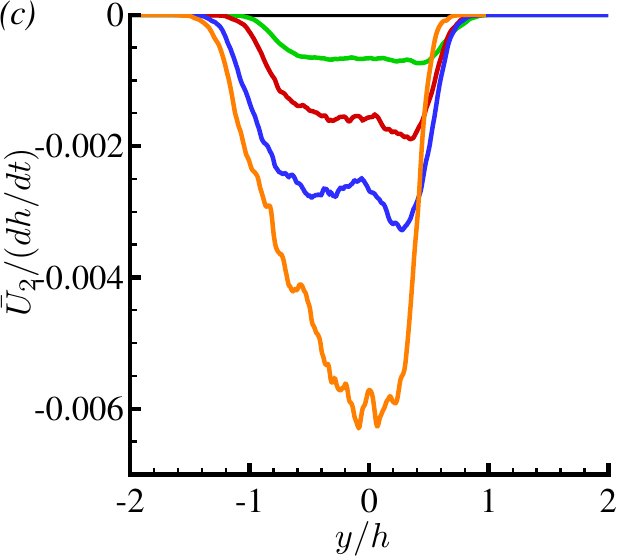}
\caption{Self-similar time-averaged profiles of cross-stream velocity for all Atwood numbers. Solid lines represent Reynolds mean, while the dashed lines represent Favre mean. Line colors are by Atwood number as in Figure~\ref{fig:thicknessvst}. The velocities are scaled using (a) mean streamwise velocity and (b-c) growth rate as suggested by the self-similar analysis (\S\ref{ss:selfsim}). (c) is scaled to show the Reynolds mean, which is negligible compared to the Favre mean in (b); the Reynolds means should be understood to pertain only to their respective averaging times, since they do not become constant in time. No scaling by Atwood number is applied, so the magnitudes for the $A=0.001$ (black line) are small in comparison to the other cases and appear near $\tilde{U}_2/\Delta U=0$ on this vertical scale.}
\label{fig:profilesmeanV}
\end{figure}

Comparing the self-similar scaled profiles among Atwood numbers (Figure~\ref{fig:profilesmean}) illustrates several basic changes that occur as the density difference between streams increases. For $A=0.001$, the mean streamwise velocity and mean density profiles are essentially centered at $y=0$ and symmetric about that point. A shift in the mean streamwise velocity profiles to the light fluid side (i.e., $\eta_1<0$) that increases in magnitude with increasing Atwood number is apparent. With increasing Atwood number, the shapes of these velocity profiles remain generally similar as they shift to the light fluid side, but the asymmetry in their gradients (Figure~\ref{fig:profilesmean}b) reveals an additional steepening on the light fluid side and shallowing on the heavy fluid side. Conversely, the neutral points of the density profiles (where $\bar{\rho}(y)=\rho_0$) remain relatively stationary while the density profiles steepen on the heavy fluid side but become shallower on the light fluid side with increasing Atwood number.

Figure~\ref{fig:profilesmeanV} displays the corresponding profiles for the cross-stream mean velocity component. The magnitudes are much smaller than those of the streamwise velocity. However, as the self-similar analysis indicates, the cross-stream velocity has an important relationship with mass conservation and mixing layer growth in variable-density mixing layers. In Figure~\ref{fig:profilesmeanV}(b-c), these velocity profiles are shown with the scaling suggested by the self-similar analysis, using $h$ based on the Favre mean streamwise velocity for the thickness scaling. The Reynolds averaged cross stream velocity is much smaller in magnitude than the corresponding Favre average quantity. In addition, $V$ can be shown to strongly depend on the mean density gradient (Appendix~\ref{appendix:mean}) and therefore not reach a time-constant magnitude during self-similar growth; the averages in Figure~\ref{fig:profilesmeanV}(c) should be understood to pertain only to their particular averaging time periods. It is shown below (\S\ref{ss:densqty}) that $\tilde{V}$ is dominated by the turbulent mass flux, which does approach a constant value during self-similar growth. 

\begin{figure}
\centering
\includegraphics[scale=0.78]{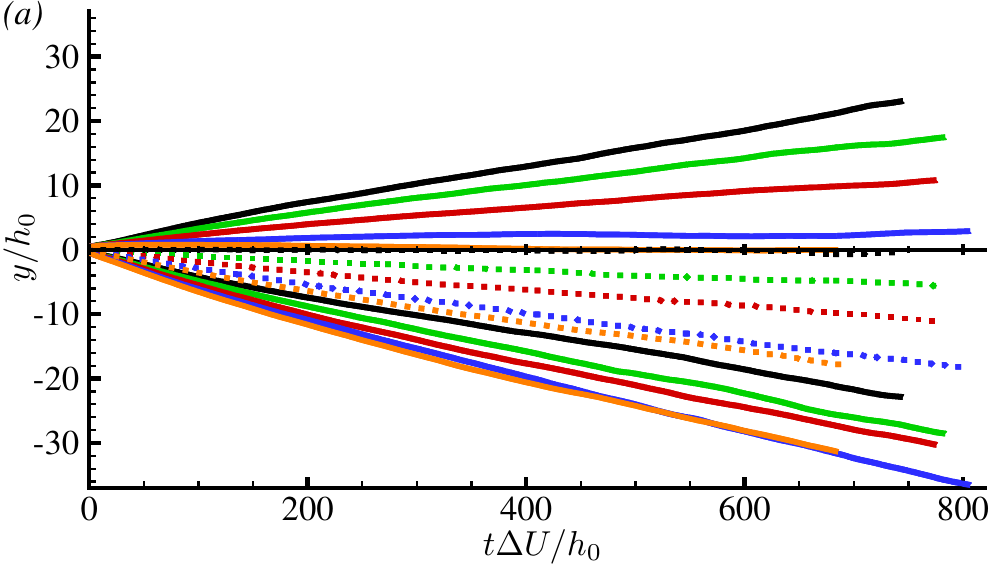}
\includegraphics[scale=0.80]{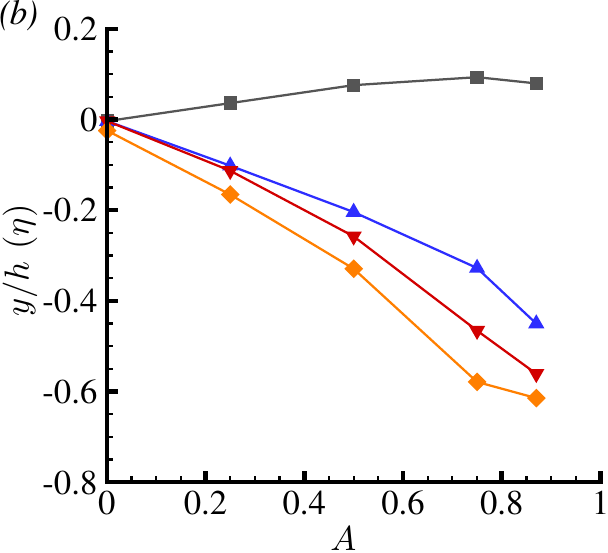}
\caption{(a) Favre-mean streamwise velocity profile edge position (10\%, as defined in (\ref{eq:hdefn})) evolutions for the range of Atwood numbers (colored as in Figure~\ref{fig:thicknessvst}) are shown by solid lines. Their neutral positions (i.e., $\tilde{U}_1=0$) are also shown by dotted lines. (b) The self-similar positions are compared as a function of Atwood number for density neutral point (${\color{darkgray}\blacksquare}$), peak cross-stream velocity $\eta_2$ (${\color{blue}\blacktriangle}$), streamwise velocity neutral point $\eta_1$ (${\color{red}\blacktriangledown}$), and peak Reynolds stress $\eta_{12}$ (${\color{orange}\blacklozenge}$).}
\label{fig:spreadpositions}
\end{figure}

The positions of the neutral points (i.e., $\rho=\rho_0$ and $\tilde{U}_1=0$) and positions of extrema for various statistical quantities (e.g., $\min(\tilde{U}_2)$) are important in characterizing the shape of the mixing layer during the self-similar regime. The mixing layer growth and its asymmetry can be summarized by tracking the points at which the mean streamwise velocity is equal to 10 and 90 percent of the free-stream difference $\Delta U$: $y_{\left[\tilde{U}_1=U_{-}+0.1*\Delta U\right]}$ and $y_{\left[\tilde{U}_1=U_{+}-0.1*\Delta U\right]}$. These are the points whose separation define $h$ in (\ref{eq:hdefn}). In Figure~\ref{fig:spreadpositions}a, the linear growth of these positions (scaled by initial thickness) with respect to time, approximately extending from $y=0$ at $t=0$, is consistent with the positions collapsing to fixed self-similar scaled (e.g., $y/h$) values. The $\tilde{U}_1$-based positions also evolve linearly and likewise collapse to fixed $y/h$ values. Plotting the scaled positions of these points as a function of Atwood number (Figure~\ref{fig:spreadpositions}b) highlights the prominent features observed in Figure~\ref{fig:profilesmean}: an increasing drift of the mean streamwise velocity profile to the light fluid side with increasing Atwood number, while the density profile remains approximately centered at the initial interface. In addition, Figure~\ref{fig:spreadpositions}b indicates that the mean cross-stream velocity $\tilde{U}_2$ peak similarly drifts to the light fluid side, as well as the peak Reynolds stress $\tilde{R}_{12}$ (\S\ref{ss:profilesfluc}). The relative magnitudes of the drifts confirm the predictions of the self-similar analysis (\S\ref{ss:selfsim}) and are consistent with previous simulations of other variable-density mixing layers \citep[e.g.,][]{pantano2002sce,almagro2017nsv}. For the range of Atwood numbers simulated, $\eta_{12} < \eta_{1} < \eta_{2} < 0$: the Reynolds stress peak is located further in the light fluid than the neutral point of mean streamwise velocity, which itself is further than the peak of mean cross-stream velocity.

\subsection{Velocity Fluctuation Intensity Profiles}
\label{ss:profilesfluc}

\begin{figure}
\centering
\includegraphics[scale=0.7]{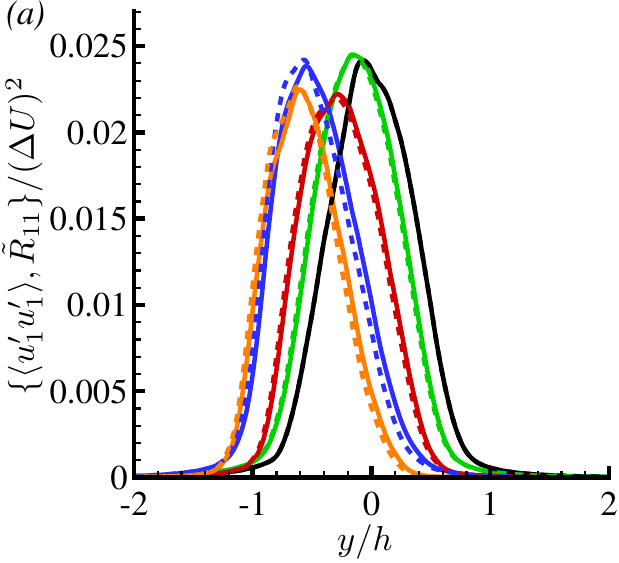}
\includegraphics[scale=0.7]{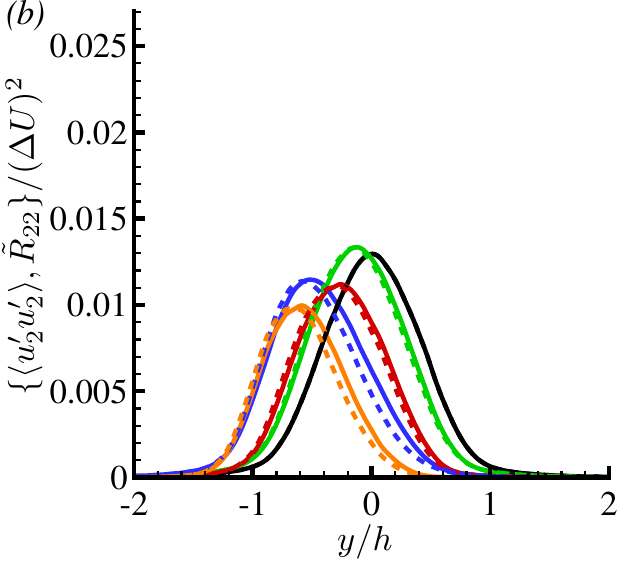}
\includegraphics[scale=0.7]{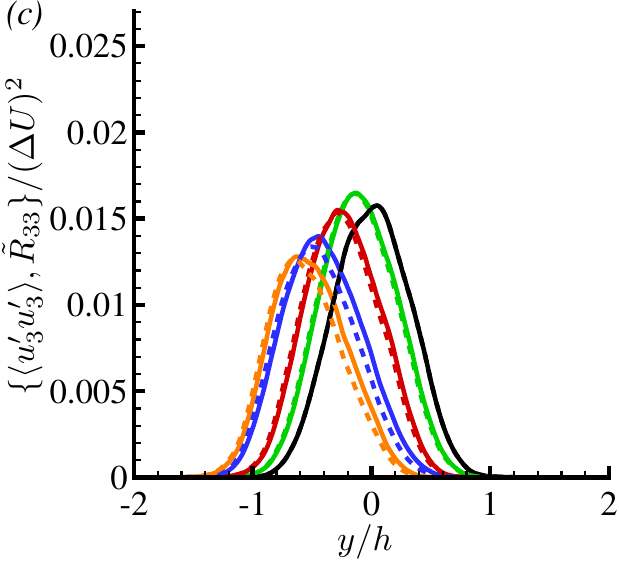}
\includegraphics[scale=0.7]{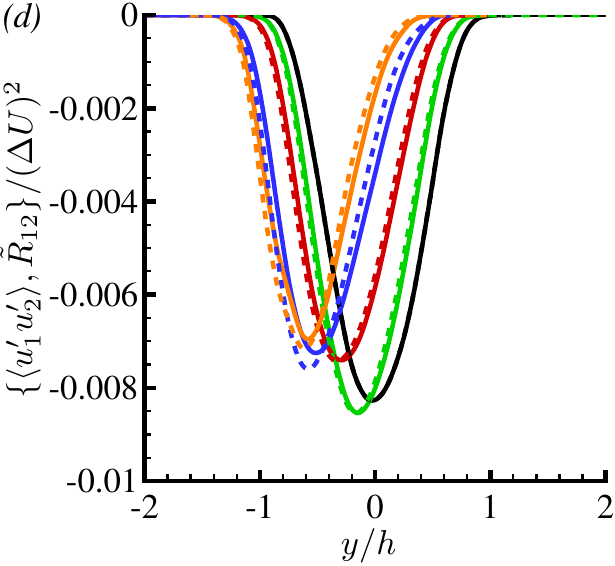}
\includegraphics[scale=0.7]{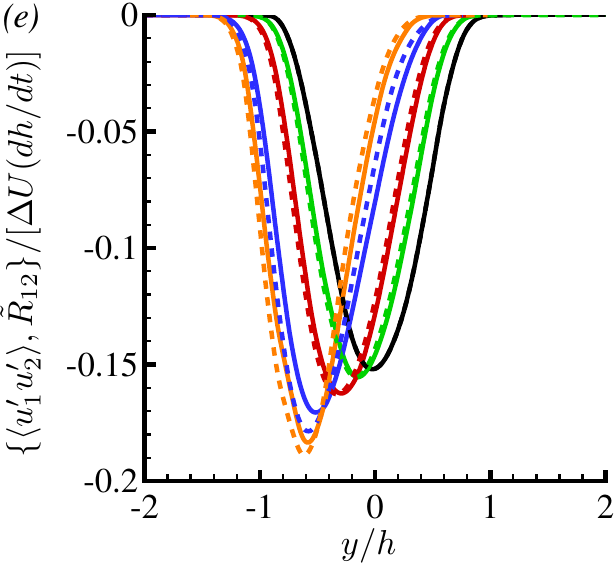}
\includegraphics[scale=0.7]{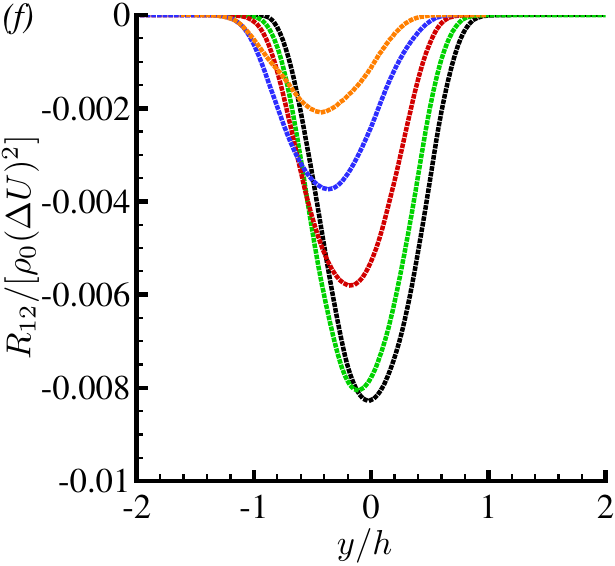}
\caption{Self-similar profiles for all Atwood numbers (colored as in Figure~\ref{fig:thicknessvst}) showing velocity variances and Reynolds stresses. In (a--d), the solid line represents the velocity variance $\langle u_i' u_j' \rangle$ while the dashed line represents Favre-averaged Reynolds stress $\tilde{R}_{ij}=\langle \rho u_i'' u_j'' \rangle/\bar{\rho}$. (e) $\tilde{R}_{12}$ as in (d) but scaled by $\Delta U (dh/dt)$ as suggested by the self-similar analysis. (f) compares the Favre shear stress of (d) not scaled by local average density but $R_{ij}=\langle \rho u_i'' u_j'' \rangle$ scaled by the average of the free-stream densities $\rho_0$.}
\label{fig:profilesfluc}
\end{figure}

Statistical profiles for velocity fluctuations are similarly obtained using self-similar scaling applied to the $y$ coordinate. It has also been verified that these profiles collapse over the self-similar growth time period (apart from a small amount of statistical variability) when scaled in this manner. These time-averaged profiles are compared among Atwood numbers in Figure~\ref{fig:profilesfluc}.

Overall, the behaviors of the velocity variances for the low Atwood number case agree well with other published single-density mixing layer simulations. However, there can be significant differences in the magnitudes.
The peak variance magnitudes of \cite{rogers1994dss} are 23\% larger than those of the present $A=0.001$ simulation. The peak magnitudes of the density ratio $1$ simulation of \cite{almagro2017nsv} are on average 52\% larger than those of the present simulation.  
The magnitudes for the Reynolds stress $\tilde{R}_{12}$ peak likewise differ between the simulations by similar amounts.
The spatially developing mixing layer simulations of \cite{attili2012sst} that reach relatively high Reynolds numbers have peak magnitudes on average 19\% greater than the present results.

One factor likely contributing to the differences of intensity magnitude is the determination of self-similar averaging time. With the present initial disturbance, an overshoot in the turbulence intensities occurs, and after a significant period of time the overshoot decays and asymptotes to the final self-similar growth state as the mixing layer thickens. Other simulations approach self-similar growth differently, and the self-similar period may be determined differently. Despite the difference of the spatial vs.~temporally developing configuration, the \cite{attili2012sst} intensity profiles appear to agree most closely with the present simulation. Their simulation attains higher Reynolds number and greater thickness growth than the other temporal simulations cited. Differences in simulation domain sizes could potentially alter the turbulence dynamics by restricting structure growth and thereby affect fluctuation intensities. An additional factor may be persisting effects of the differing initial disturbances. Among experiments, there is significant scatter in the intensity magnitudes, e.g., the differences between \cite{bell1990dts} and \cite{spencer1971sip} as shown in \cite{almagro2017nsv}. \cite{rogers1994dss} summarized the wide range of magnitudes for streamwise velocity variances measured in experiments (as well as the mixing layer growth rates, which are closely related to $\langle u_1' u_2' \rangle$). They also noted the perspective of \cite{dimotakis1976mlh} that persisting influence of the initial conditions may be responsible.

When Atwood number is increased, the behavior of the intensities and Reynolds stresses remain similar to the $A=0.001$ case, except they shift to the light fluid side and generally decay slightly in magnitude.
As shear moves to the light fluid side with increasing Atwood number, the turbulence intensity peak moves to the light fluid side as well.
(The close relation between mean shear and the production of turbulent kinetic energy $\tilde{k} = \tilde{R}_{ii}/2$ is apparent from the shear production term $- \bar{\rho} \tilde{R}_{12} \tilde{U}_{1,2}$ that dominates the budget for $\bar{\rho} \tilde{k}$.)
Velocity variances $\langle u_i' u_j' \rangle$ and Reynolds stresses $\tilde{R}_{ij}$ [Figure~\ref{fig:profilesfluc}(a--d)] both increasingly shift to the light fluid side with increasing Atwood number; this applies to the on-diagonal ($i=j$) elements as well as the streamwise-cross-stream ($i=1$, $j=2$).

Figure~\ref{fig:profilesfluc}(a--c) suggests that there is only a weak reduction in peak turbulent kinetic energy with increasing Atwood number. The reduction in peak $\tilde{R}_{12}$ Reynolds stress (or $\langle u_1' u_2' \rangle$) is as strong as that experienced by any of the on-diagonal turbulent kinetic energy contributions, yet it is reduced by no more than about 30\% from $A=0.001$ to $A=0.87$. When $\tilde{R}_{12}$ Reynolds stress is scaled by $\Delta U$ and $dh/dt$ as suggested by the self-similar analysis, rather than by $\Delta U^2$ as is typically reported, the peak magnitudes weakly increase with increasing Atwood number (Figure~\ref{fig:profilesfluc}e).

If Reynolds stress is scaled using the average density of the two free streams ($\rho_0$) rather than the local mean density, the reduction in peak value with Atwood number is enormous (Figure~\ref{fig:profilesfluc}f). This is further confirmation that the intense turbulent motions move to (and are sustained in) light density fluid. 
$\left\langle u_i u_j \right\rangle$ and $\tilde{R}_{ij}= \left\langle \rho u_i'' u_j'' \right\rangle/\bar{\rho}$ agree very closely
for even the highest Atwood number throughout self-similar growth (while there are significant differences during transition with high $A$). This agreement is remarkable because
these quantities do not agree well with $R_{ij}=\rho \tilde{R}_{ij}$ due to the shift of strong fluctuations to fluid on average lighter than $\rho_0$.
In other words, at elevated $A$, $\bar{\rho}$ is much smaller than $\rho_0$ at the position of peak turbulence intensity, but
$\left\langle \rho u_i'' u_j'' \right\rangle$ is also commensurately smaller so their ratio is nearly the same as for low $A$.
Details of the local density distributions and how they correlate with velocity-based fluctuations will be further considered (\S\ref{sec:condstat}).

\subsection{Analysis of Thickness Growth Rate During Self-Similar Growth}
\label{ss:ssthickgrowthanalysis}

The statistical profiles discussed above can be related to growth rate attained by each mixing layer during its self-similar growth regime. The average growth rates calculated over the self-similar growth time intervals obtained above are first summarized as a function of Atwood number.
At very low Atwood number ($A=0.001$), the momentum thickness growth rate $d\delta_m/dt/\Delta U=0.012$ agrees well with the value of 0.014
reported by \cite{rogers1994dss}. \cite{almagro2017nsv} reports a somewhat higher growth rate of 0.017 when density is constant.
In terms of thickness measured by $h$, the present simulation's growth rate $dh/dt/\Delta U$ of 0.069 is consistent with the 0.062 value for the simulation of \cite{rogers1994dss}, though both of these growth rates are toward the lower end of the $0.06$--$0.11$ values typically observed in experiments \citep{pope2000tf,dimotakis1991tfs}.

\begin{figure}
\centering
\includegraphics[scale=0.8]{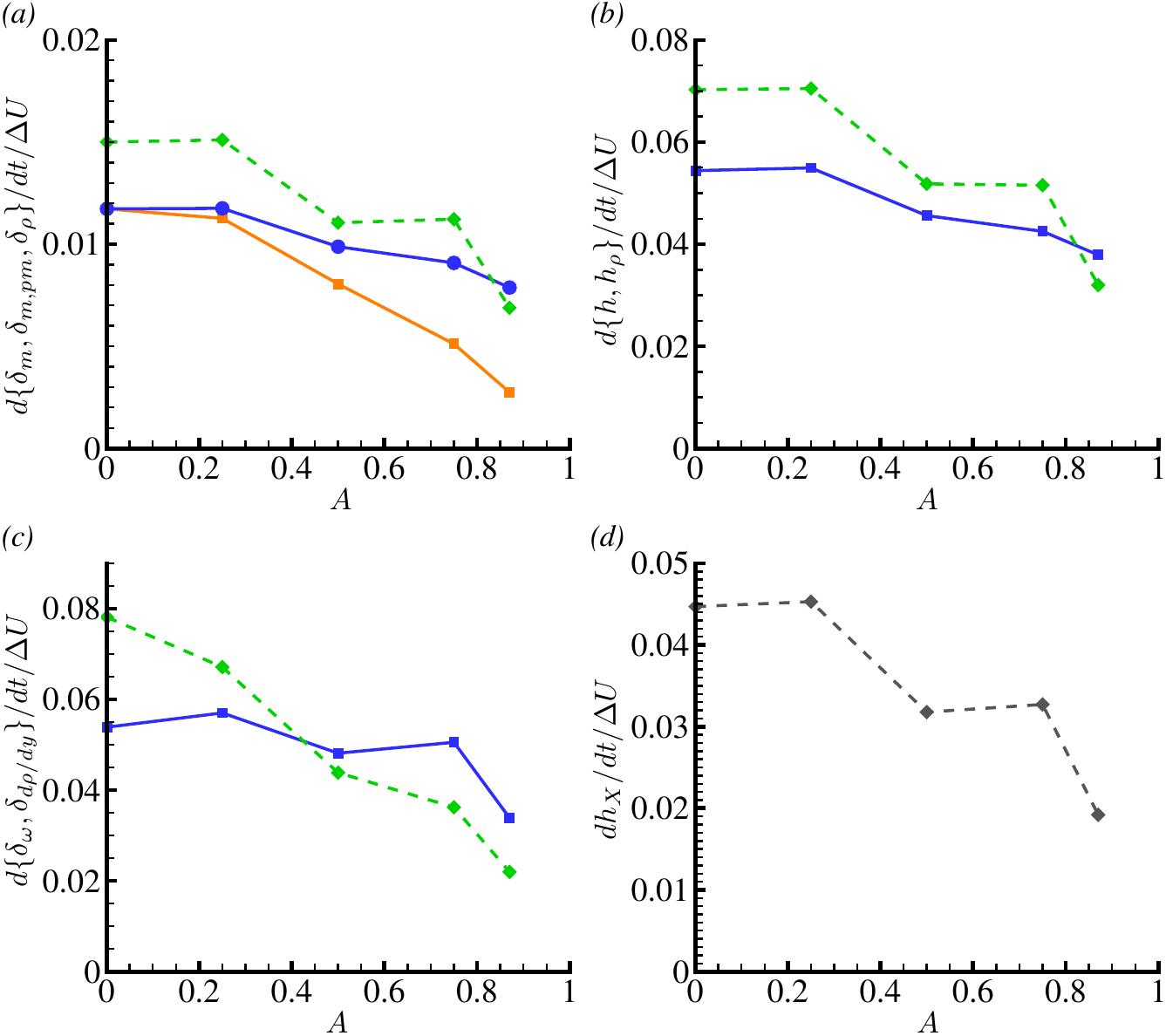}
\caption{The effects of Atwood number on growth rate are displayed for a variety of thickness measurements
based on the mean velocity profile (blue) and the mean density profile (dashed green). 
The momentum thickness calculated from the mean velocity profile and weighted by density (orange) is shown in (a).
The fast reaction product thickness $h_{X_\rho}$ in (d) is based solely on the density profiles.
(a) $\delta_m$ ({\color{orange} ---$\blacksquare$---}), $\delta_{m,pm}$ ({\color{blue} ---$\bullet$---}), $\delta_\rho$ ({\color{green} - - $\blacklozenge$ - -});
(b) $h$ ({\color{blue} ---$\blacksquare$---}), $h_\rho$ ({\color{green} - - $\blacklozenge$ - -});
(c) $\delta_\omega$ ({\color{blue} ---$\blacksquare$---}), $\delta_{d\rho/dy}$ ({\color{green} - - $\blacklozenge$ - -});
(d) $h_{X_p}$ ({\color{gray} - - $\blacklozenge$ - -}).}
\label{fig:growthratevsA}
\end{figure}

Assessing the growth rate reductions as a function of Atwood number across the present simulations,
Figure~\ref{fig:growthratevsA}(a) shows that the momentum thickness growth rate for $A=0.87$ is reduced by 77\% from the
rate for the single-density case, while the momentum thickness per mass growth rate ($\delta_{m,pm}$) and the analogous
integral growth rate for the density profile ($\delta_{\rho}$) experience lesser but nonetheless significant reductions.
The stronger reduction for $\delta_m$ can largely be explained by a misalignment that develops between density and velocity profiles (\S\ref{ss:profilesssmean}).
The reductions in growth rate based on $h$ and $h_{\rho}$ (Figure~\ref{fig:growthratevsA}b) are similar to those of the
$\delta_{m,pm}$ and $\delta_{\rho}$ integral quantities (Figure~\ref{fig:growthratevsA}a); the reductions for $h_{X_p}$ are also similar (Figure~\ref{fig:growthratevsA}d).
The thicknesses derived from mean profile $y$-gradients ($\delta_\omega$ and $\delta_{d\rho/dy}$), however, display less smooth growth rate reduction behavior (Figure~\ref{fig:growthratevsA}c). This could be a consequence of greater sensitivity to noise associated with a lack of statistical convergence when the maximum gradient is calculated on the smoothed profile, but could also appear due to the greater sensitivity of gradient-based measurements to the details of the profile asymmetries that appear with increased Atwood number.
Though $h_{X_p}$ is an integral measurement and therefore lacks the extreme sensitivity to local noise in the mean profile of gradient-based measurements, it appears to display less smooth reductions in growth rate than the other integral thickness quantities. This suggests that some measurements may be particularly sensitive to specific features of the profile shapes.
The close correspondence between most of the growth rates (particularly for $\delta_m$, $\delta_{m,pm}$, and $h$) confirms that any of
the corresponding thickness measurements would be acceptable for scaling the flow statistics profiles during self-similar growth.
Generally, the growths of density thickness quantities (due to mixing of fluids) also behave similarly to the growths of
velocity thickness quantities.

\begin{figure}
\centering
\includegraphics[scale=.95]{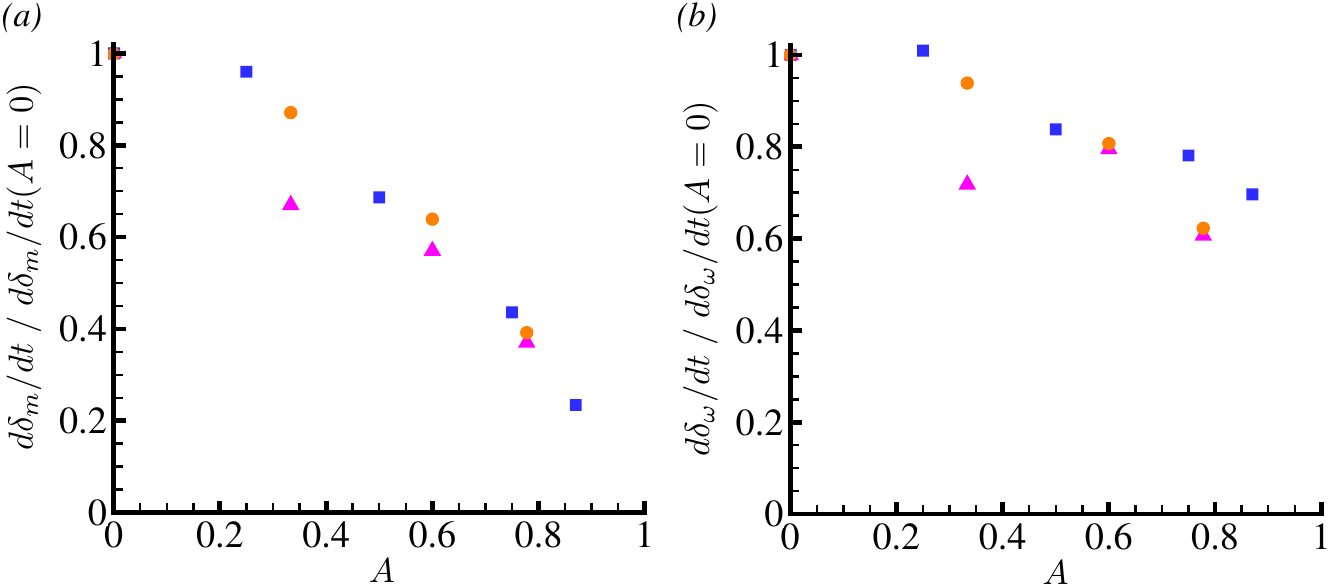}
\caption{Atwood number effects on mixing layer growth rates measured by (a) momentum thickness and (b) vorticity thickness for the (${\color{blue} \blacksquare}$) present incompressible INBM simulations, (${\color{orange} \bullet}$) the LMNOB (thermal variation) simulations of \cite{almagro2017nsv}, and (${\color{magenta} \blacktriangle}$) 0.7 Mach number NOB (thermal variation) simulations of \cite{pantano2002sce}. The growth rates are scaled by the corresponding growth rate with $A\approx 0$ for each data set.}
\label{fig:growthratevsAps}
\end{figure}

To compare the growth rate effects of Atwood numbers for other variable-density mixing layers, Figure~\ref{fig:growthratevsAps} includes single-species variable-temperature simulations of \cite{almagro2017nsv} (low-speed limit) as well as \cite{pantano2002sce} (moderately compressible 0.7 convective Mach number). In contrast to the present species mixing governed by simplified INBM (incompressible non-Boussinesq mixing) equations, the latter cases with thermal variations are governed by the LMNOB (low-Mach number non-Oberbeck-Boussinesq) and fully compressible non-Oberbeck-Boussinesq (NOB) equations \citep{livescu2020tlt}. 
A detailed comparison between simplified INBM equations and LMNOB equations has been made at $A=0.75$ by \cite{baltzer2020lst}.
Atwood number for the wide range shown in Figure~\ref{fig:growthratevsAps} affects both momentum and vorticity thickness growth rates relatively similarly between density variation mechanisms (Figure~\ref{fig:growthratevsAps}), particularly given the differences between simulations in addition to species vs.~thermal transport, e.g.~domain sizes, initial disturbance, determination of self-similar growth period, etc.
The growth rates are normalized by the $A \approx 0$ growth rate for each configuration (simulation set), and doing so conceals important physical differences and their effects between cases: the growth rate of the $A=0$ \cite{pantano2002sce} mixing layer had reduced by $40\%$ relative to their low-speed simulation solely due to compressibility effects. Their simulations investigating the range of Atwood numbers are only available at a Mach number of 0.7.
Since vorticity thickness is based on the maximum gradient of the mean streamwise velocity profile, more scatter in these growth rates is to be expected as they are sensitive both to the shape of the profile and noise in this quantity (smoothing was applied when calculating for the present simulations).

The influence of Atwood number on growth rate can be further explained based on the behavior of the statistical profiles.
One useful property of the momentum thickness definitions (\ref{eq:momthickdefn}--\ref{eq:momthickpmdefn}) is that they
straightforwardly lead to relations between growth rates and flow statistical quantities. This was explored by \cite{vreman1996cml}, who showed
that an informative relation can be formed based on the time derivative of (\ref{eq:momthickdefn}) that yields
\begin{align}
\frac{d\delta_m}{dt} = -\frac{1}{\rho_0 \Delta U^2} \int \frac{d}{dt} \left( \bar{\rho} \tilde{U}_1 \tilde{U}_1 \right) \; dy \label{eq:momthicktd1}
\end{align}
as other terms vanish using the $y$-integrated averaged continuity and momentum equations.
Multiplying the Favre mean momentum equation (\ref{eq:momfa}) for $i=1$ by $\tilde{U}_1$ produces an equation relating
the time derivative of mean kinetic energy to Reynolds stress as
\begin{align}
\frac{d}{dt} \left( \bar{\rho} \tilde{U}_1 \tilde{U}_1 \right) = \left(\bar{\rho} \tilde{U}_1 \tilde{U}_1 \tilde{U}_2 + 2\bar{\tau}_{12}\tilde{U}_1 - 2\tilde{U}_1\bar{\rho}\tilde{R}_{12}\right)_{,2} + 2\tilde{U}_{1,2}\bar{\rho}\tilde{R}_{12} - 2\bar{\tau}_{12}\tilde{U}_{1,2}.
\end{align}
Many terms are cast in conservative forms, so they vanish when integrated over the $y$ domain in (\ref{eq:momthicktd1}).
Then, (\ref{eq:momthicktd1}) becomes
\begin{align}
\frac{d\delta_m}{dt} \approx -\frac{2}{\rho_0 \Delta U^2} \int_{-\infty}^{\infty} \bar{\rho} \tilde{R}_{12} \tilde{U}_{1,2} \; dy \label{eq:momthicktd2}
\end{align}
after neglecting the mean viscous term, which is consistent with the self-similar analysis (\S\ref{ss:selfsim}) and has been shown to be small by scaling arguments of \cite{pantano2002sce}. The growth rates calculated from this equation using the self-similar averaged profiles 
match those directly measured in the flow to within several percent.

The above relation shows that the momentum thickness growth rate depends on the density, mean streamwise velocity, and Reynolds shear stress profiles.
As shown in Figure~\ref{fig:profilesfluc}(d), there is relatively little change in Favre-averaged Reynolds shear stress with the density
normalized by the local mean.
However, when normalized by the average of the two free stream densities $\rho_0$, the $\bar{\rho} \tilde{R}_{12}$ quantity that appears in (\ref{eq:momthicktd2}) reduces strongly with increasing Atwood number, as shown in Figure~\ref{fig:profilesfluc}(f). 
The simulation suite was designed such that the average of the free-stream densities $\rho_0$ remains consistent across all Atwood numbers. A consequence is that, if the fluids were fully mixed in equal proportion in the core of the mixing layer, the density there would be the same for every Atwood number. Therefore, reductions as observed in the aforementioned quantity normalized by $\rho_0$ are indicative of $\tilde{R}_{12}$ having its strongest magnitude increasingly further in the light fluid. Furthermore, this density becomes smaller relative to $\rho_0$ as Atwood number increases.
In (\ref{eq:momthicktd2}), the growth rate is the product of $\bar{\rho} \tilde{R}_{12}$ with the mean streamwise velocity gradient, which weakly changes
in magnitude with Atwood number (Figure~\ref{fig:profilesmean}b). This density weighting reflects the dependence on density in the momentum thickness definition (\ref{eq:momthickdefn}). Thus, the principal cause of growth rate reduction for $\delta_m$ is the turbulent motions and the associated momentum deficit shifting to the light fluid side.

\begin{figure}
\centering
\includegraphics[scale=0.8]{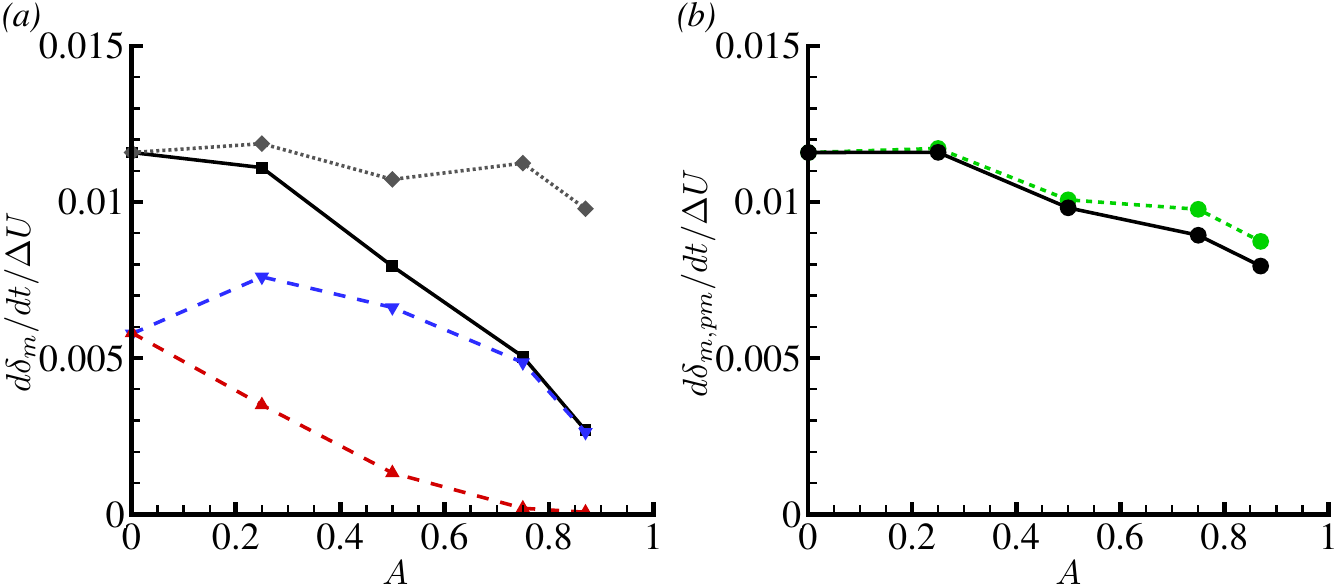}
\caption{(a) Momentum thickness growth rate predicted from self-similar averaged statistical profiles using (\ref{eq:momthicktd2}).
The total momentum growth rate ({---$\blacksquare$---}) is decomposed into light fluid ({\color{blue} - - $\blacktriangledown$ - -}) and
heavy-fluid ({\color{red} - - $\blacktriangle$ - -}) contributions according to (\ref{eq:momthicktd2split}).
The hypothetical growth rate predicted by (\ref{eq:momthicktd2}) if the profiles' drifts were artificially removed ({\color{gray} $\cdots$$\blacklozenge$$\cdots$)})
highlights that much of the growth rate reduction is associated with the intense turbulence and shear shifting to the light fluid.
(b) Momentum thickness per mass growth rate predicted from (\ref{eqn:deltampmgrowthratesfinal}) ({---$\bullet$---}).
The mean shear-Reynolds stress term ({\color{green}- -$\bullet$- -}) has the same form as the production for TKE and dominates the other growth rate terms.
However, variable-density terms also appreciably reduce the growth rate at high Atwood numbers, as shown by the distance between the curves.}
\label{fig:growthrateequations}
\end{figure}

The dominance of the profile shifting effect is demonstrated by artificially realigning the profiles such that the peak in $\tilde{R}_{12}$ and inflection point in $\tilde{U}_1$ are returned to the point where $\bar{\rho} = \rho_0$ (Figure~\ref{fig:growthrateequations}). Eliminating the shifts in this way significantly weakens the growth rate reduction effect. The magnitudes of growth rate reductions after these artificial shifts are approximately those observed for the
growth rate of momentum thickness per mass (which lacks the density weighting).
For other turbulent variable-density mixing layers, \cite{pantano2002sce} and \cite{almagro2017nsv} have developed semi-empirical
formulas that estimate momentum thickness growth rate reductions as functions of density ratio (or Atwood number) largely based on the shifts that develop
between the mean streamwise and mean density profiles.

Equation (\ref{eq:momthicktd2}) also shows that the mixing layer growth rate is the integral over the entire width of the mixing layer of a term that is closely related with the production of TKE (through the shear production mechanism, $- \bar{\rho} \tilde{R}_{12} \tilde{U}_{1,2}$).
It is informative to split this integral into contributions from the light fluid and heavy fluid sides of the domain:
\begin{align}
\frac{d\delta_m}{dt} \approx -\frac{2}{\rho_0 \Delta U^2} \int_{-\infty}^{y_{\rho_0}} \bar{\rho} \tilde{R}_{12} \tilde{U}_{1,2} \; dy -\frac{2}{\rho_0 \Delta U^2} \int_{y_{\rho_0}}^{\infty} \bar{\rho} \tilde{R}_{12} \tilde{U}_{1,2} \; dy, \label{eq:momthicktd2split}
\end{align}
where $y_{\rho_0}$ is the $y$ value at which $\bar{\rho}(y) = \rho_0$. Since mean density increases monotonically across the interior of the mixing layer, it follows that the first term is the light fluid contribution to the growth rate and the second term is the heavy fluid contribution (in a mean sense).
The $y_{\rho_0}$ position remains at $y=0$ for $A \rightarrow 0$, thus splitting the mixing layer in half. As Atwood number increases, the $y_{\rho_0}$ position moves much more weakly compared to points on the mean velocity profiles or $\tilde{R}_{12}$ (and $y_{\rho_0}$ instead moves toward the heavy fluid side), as shown by Figure~\ref{fig:spreadpositions}(b). 

When the density differences are very weak ($A=0.001$), the contributions are essentially equally split between the light and heavy sides
(Figure~\ref{fig:growthrateequations}a). The heavy-fluid growth rate contribution monotonically decays with Atwood number, which is consistent with the
intense turbulence and momentum deficit (relative to the free streams) drifting to the light fluid side. As Atwood number increases from $A=0.001$ to $0.25$, the light-fluid growth rate contribution weakly
increases, but then decays for higher Atwood numbers.
These growth rate changes can be interpreted in light of the time histories of mean profile positions for various Atwood numbers (Figure~\ref{fig:spreadpositions}).
Beginning from the symmetric growth of the mean streamwise profile edge positions for $A=0.001$ in Figure~\ref{fig:spreadpositions}(a), the edge positions
reduce in their penetration to the heavy fluid side but increase in penetrating the light fluid side for Atwood numbers up to $A=0.75$. The penetrations
into the light fluid sides stagnate as Atwood number increases beyond $0.75$, and by $A=0.87$ the growth rate has decreased so much that
\emph{even the growth into the light fluid has reduced below that for $A=0.75$} while growth into the heavy fluid side is negligible. This reduction
in growth into the light fluid side is manifested in the sharply reducing light fluid growth contribution in Figure~\ref{fig:growthrateequations}(a).

While much of the reduction in momentum thickness growth rate with increasing Atwood number can be explained by the
velocity neutral point moving into the light fluid, weaker though significant reductions in growth rate also occur
in all other thickness measurements, despite their lack of explicit density profile dependencies.
This differs from the assumption of constant temporal growth rates
with Atwood number sometimes used to develop theories addressing growth rate for spatially-developing variable-density mixing layers,
as described in the introduction.
To address these subtler reductions in growth rate with Atwood number, we derive an analogous equation relating the growth rate of momentum thickness
per mass to statistical profiles of the flow. A similar derivation instead beginning from (\ref{eq:momthickpmdefn}) leads to
\begin{align}
\frac{d\delta_{m,pm}}{dt} = \frac{U_{-} + U_{+}}{\Delta U^2} \int \frac{d \bar{U}_1}{dt} \; dy - \frac{1}{\Delta U^2} \int \frac{d}{dt}\left(\bar{U}_1 \bar{U}_1\right) \; dy = -\frac{1}{\Delta U^2} \int \frac{d}{dt}\left(\bar{U}_1 \bar{U}_1\right) \; dy,
\end{align}
since $U_{-} = -U_{+}$ for this flow configuration.
Developing an expression for the time derivative in the integral based on the Reynolds mean momentum equation
leads to
\begin{align}
\left(\bar{U}_1 \bar{U}_{1}\right)_{,t} = & 2 \Big[ \left(-\bar{U}_1 \bar{U}_1 \bar{U}_j - \bar{U}_1 \left\langle u_1' u_j'\right\rangle \right)_{,j} + \bar{U}_{1,j} \bar{U}_1 \bar{U}_j + \bar{U}_{1,j} \left\langle u_1' u_j'\right\rangle \nonumber \\
& - \left\langle \frac{p_{,1}}{\rho} \right\rangle \bar{U}_1 + \left\langle \frac{\tau_{1j,j}}{\rho} \right\rangle \bar{U}_1 + \bar{U}_{j,j} \bar{U}_1 \bar{U}_1 + \left\langle u_{j,j}' u_1'\right\rangle \bar{U}_1 \Big].
\end{align}
Again using the homogeneities of this flow and noting that both $\bar{U}_2$ and $\left\langle u_1' u_j'\right\rangle$ vanish on the boundaries
(so the conservative terms integrate to 0) simplifies the expression to
\begin{align}
\frac{d\delta_{m,pm}}{dt} = & -\frac{2}{\Delta U^2} \int \Big[ \bar{U}_{1,2} \left\langle u_1' u_2'\right\rangle + \bar{U}_{1,2} \bar{U}_1 \bar{U}_2 + \left\langle u_{j,j}' u_1'\right\rangle \bar{U}_1 \nonumber \\
& + \bar{U}_{2,2} \bar{U}_1 \bar{U}_1 - \left\langle \frac{p_{,1}}{\rho} \right\rangle \bar{U}_1 +  \left\langle \frac{\tau_{1j,j}}{\rho} \right\rangle \bar{U}_1 \Big] \; dy \label{eqn:deltampmgrowthrate3}.
\end{align}
It can be shown that the terms based on the Reynolds mean cross-stream velocity decay as the flow evolves (and are of small magnitude, Figure~\ref{fig:profilesmeanV}).
In contrast, the $\left\langle u_{j,j}' u_1'\right\rangle$ and $\left\langle {p_{,1}}/{\rho} \right\rangle$ terms, as well as the dominant $\bar{U}_{1,2} \left\langle u_1' u_2'\right\rangle$ term, can be shown to maintain constant magnitudes with the appropriate self-similar scaling. Retaining only these terms, the relation simplifies to
\begin{align}
\frac{d\delta_{m,pm}}{dt} \approx -\frac{2}{\Delta U^2} \int \Big[ \bar{U}_{1,2} \left\langle u_1' u_2'\right\rangle + \left\langle u_{j,j}' u_1'\right\rangle \bar{U}_1 - \left\langle \frac{p_{,1}}{\rho} \right\rangle \bar{U}_1 \Big] \; dy \label{eqn:deltampmgrowthratesfinal}.
\end{align}
For this particular flow in which kinematic viscosity maintains a constant value everywhere, the mean viscous term simplifies to $\nu \bar{U}_{1,22} \bar{U}_1$. It decays
with time and during the self-similar growth generates a very small effect, so it is neglected.
Of the variable-density terms, $\int \left\langle {p_{,1}}/{\rho} \right\rangle \bar{U}_1 \; dy$ is negative and dominates in magnitude over $\int -\left\langle u_{j,j}' u_1'\right\rangle \bar{U}_1 \; dy$,
which is positive, for all of the Atwood numbers considered. 
Note that in the single-density case, with $\bar{U}_2=0$ and $u_{j,j}$ everywhere zero, this equation simplifies to
${d\delta_{m,pm}}/{dt} = -2/\Delta U^2 \int \bar{U}_{1,2} \left\langle u_1' u_2'\right\rangle \; dy$,
which is consistent with the usual momentum thickness growth equation (\ref{eq:momthicktd2}).

This growth rate equation indicates that variable-density effects can modify the $\delta_{m,pm}$ growth rate from its single-density value both through the additional terms (related to the density variations and corresponding divergence of the velocity field) and through changes in the mean velocity gradient $\bar{U}_{1,2}$ and/or the Reynolds stress $\left\langle u_1' u_2'\right\rangle$ that appear in the mean shear-Reynolds shear stress term. Applying this equation to each Atwood number (Figure~\ref{fig:growthrateequations}b) demonstrates
that the mean shear-Reynolds shear stress term dominates the growth rate equation for all Atwood numbers and significantly decreases in magnitude
for the highest Atwood numbers. To understand the reduction in growth rate associated with the dominant $\bar{U}_{1,2} \left\langle u_1' u_2'\right\rangle$ term,
it can be seen that there is little change in peak $\bar{U}_{1,2}$ value (Figure~\ref{fig:profilesmean}) while peak $\left\langle u_1' u_2'\right\rangle$ magnitude
decreases moderately (Figure~\ref{fig:profilesfluc}) with increasing Atwood number. A combination of lower peak stress magnitude and subtle changes in
the mean streamwise velocity and stress profile shapes account for the reduction in growth rate. These phenomena also contribute to the conventional (density-weighted) momentum thickness growth rate reductions but are weaker than the effect of shifting mean streamwise velocity and density profiles.

\subsection{Profiles Involving Density Fluctuations}
\label{ss:densqty}
Density-velocity correlations are contained in the normalized mass flux quantity $a_i=\langle \rho' u_i' \rangle/\bar{\rho}$. The turbulent mass fluxes also quantify the relationship between Favre and Reynolds averages for the velocity and Reynolds stress quantities considered above.
Relations include $\tilde{U}_i - \bar{U}_i = u_i' - u_i'' = a_i$, $a_i = - \langle u_i'' \rangle$, and $\bar{\rho} \tilde{R}_{ij} = \bar{\rho} \langle u_i' u_j' \rangle - \bar{\rho} a_i a_j + \langle \rho' u_i' u_j' \rangle$ \citep[with additional identities provided by][]{livescu2009hrn}.
As observed in Figure~\ref{fig:profilesmeanV}, the Favre average cross-stream velocity $\tilde{U}_2$ is much larger in magnitude than the Reynolds average cross-stream velocity $\bar{U}_2$. According to the above relations, $\tilde{U}_2 = \bar{U}_2 + a_2$ is dominated by the $a_2$ turbulent mass flux while the Reynolds mean is relatively insignificant and is explained by the relation developed in Appendix~\ref{appendix:mean}. In single-fluid incompressible temporal mixing layers, the mean cross-stream velocity is zero, in order to satisfy the divergence-free condition. Thus, the correlations between the cross-stream velocity and density in these variable-density mixing layers dominate the Favre mean cross-stream velocity.
In addition, density fluctuation properties as revealed by fluctuation intensity are relevant to the structure of the flow.

\begin{figure}
\centering
\includegraphics[scale=0.9]{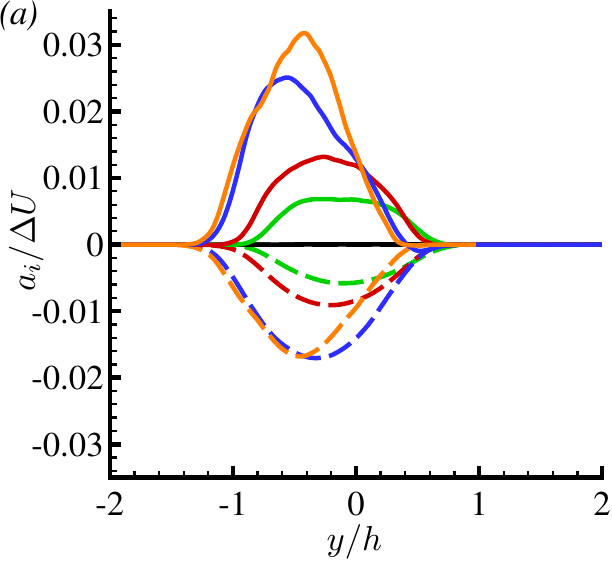}
\includegraphics[scale=0.9]{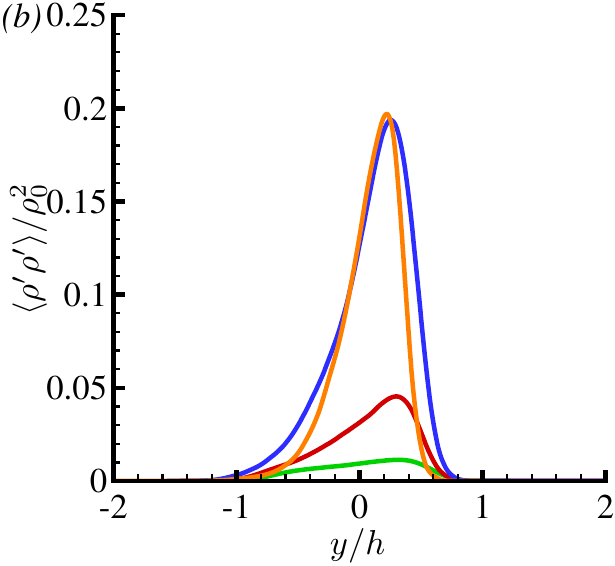}
\includegraphics[scale=0.9]{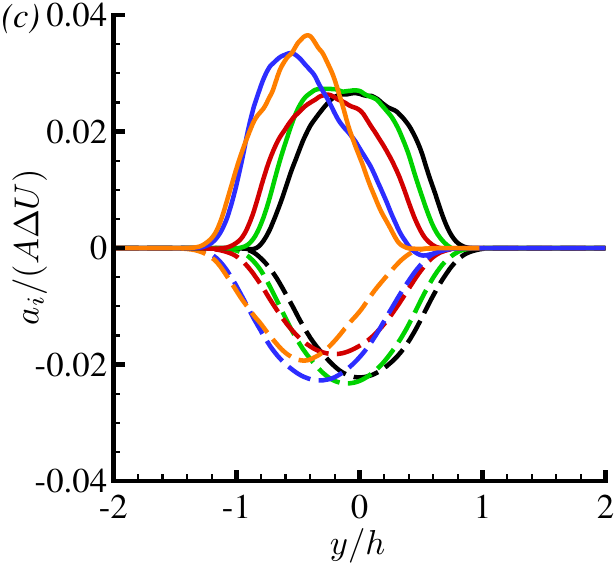}
\includegraphics[scale=0.9]{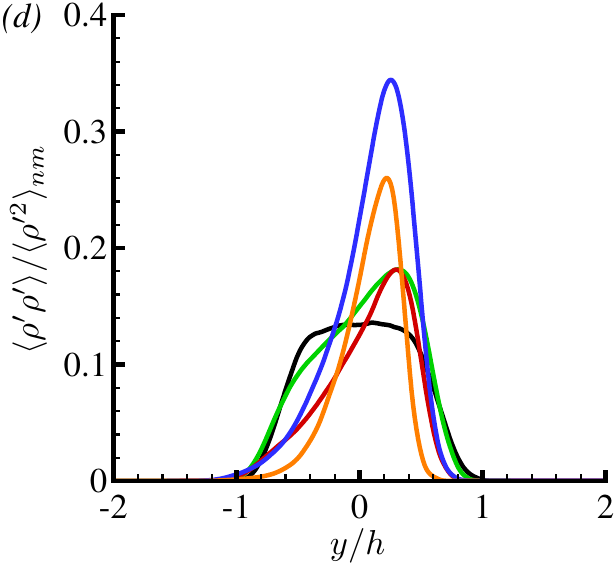}
\caption{Profiles of correlations involving density scaled using $\rho_0$: (a) normalized mass flux $a_i/\Delta U$ with $a_1$ as solid lines and $a_2$ as dashed lines and (b) density variance $\langle \rho'^2 \rangle/\rho_0^2$.
Profiles of correlations involving density scaled using no-mix (nm) intensities based on Atwood number: (c) normalized mass flux $a_i/(A \Delta U)$ with $a_1$ as solid lines and $a_2$ as dashed lines and (d) density variance $\langle \rho'^2 \rangle/\langle \rho'^2 \rangle_{nm}$. Line colors represent the different Atwood number simulations as in Figure~\ref{fig:thicknessvst}.}
\label{fig:profilesaibrho}
\end{figure}

Two types of normalizations for $a_i$ and density fluctuation intensity are considered. The $\bar{\rho}$ denominator included in the $a_i$ definition removes the density dimensional dependency, but a consequence is that $a_i$ generally grows with Atwood number as density fluctuations become more pronounced. Likewise, as Atwood number increases, density fluctuations increase in magnitude. No-mix, denoted by nm, corresponds to the quantities' values in a hypothetical configuration in which the two fluids are distributed without any mixing, and results in the highest possible magnitudes for $\langle \rho'^2 \rangle$. It can be shown that $\langle \rho'^2 \rangle_{nm} = (\Delta \rho/2)^2 = \rho_0^2 A^2$ \citep{livescu2008vdm}. By analogy, $A \Delta U$ is an appropriate scaling for $a_i$ that is equivalent to scaling by $[\Delta \rho/(2 \rho_0)] \Delta U$. Figure~\ref{fig:profilesaibrho}(c--d) shows the self-similar profiles of (a--b) with these scalings applied.

\begin{figure}
\centering
\includegraphics[scale=0.082]{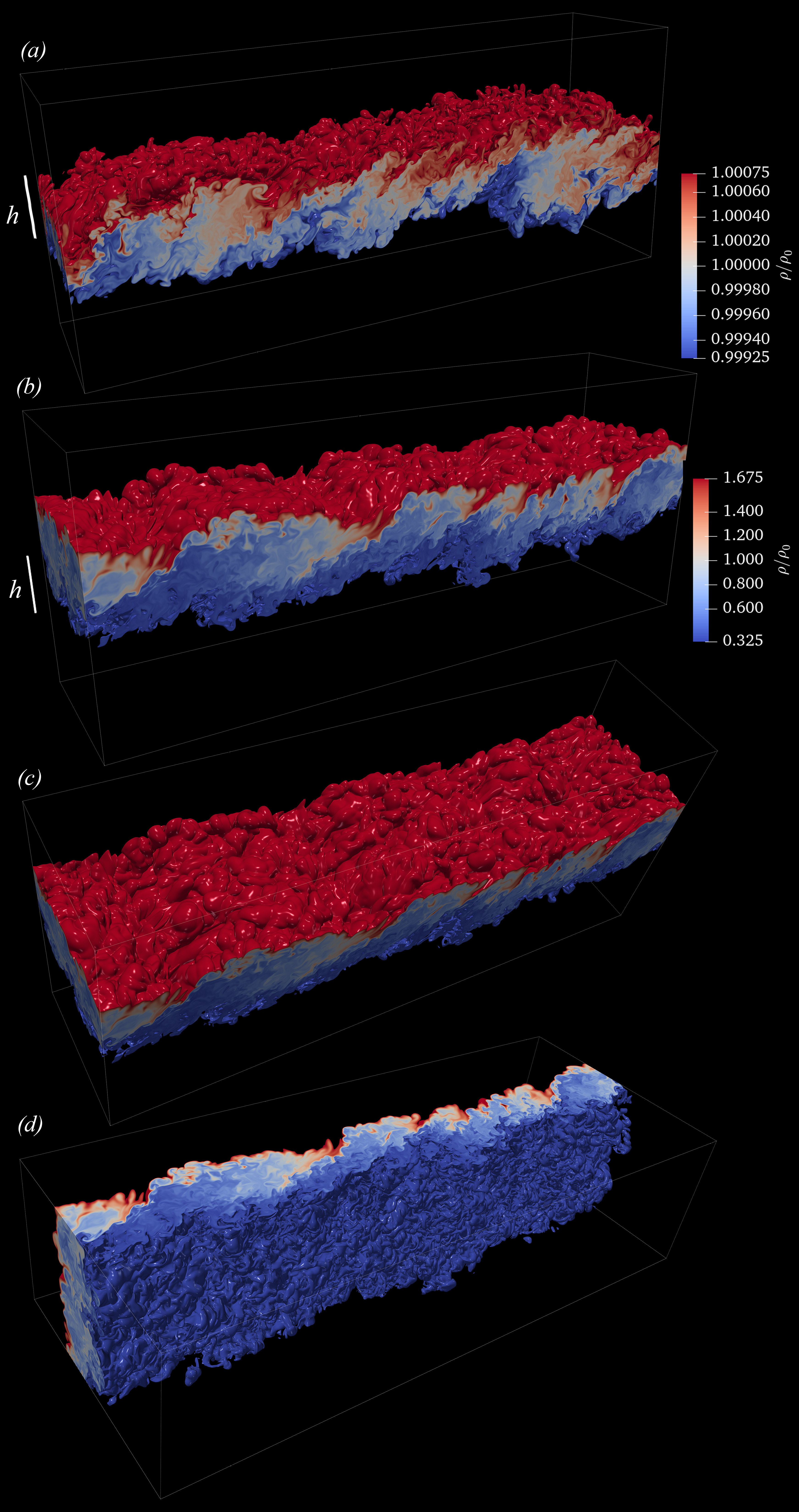}
\caption{Surfaces of density colored from light (blue) to heavy (red) during self-similar growth at approximately $t\Delta U/h_0 = 730$ for (a) $A=0.001$ and (b-d) $A=0.75$. Viewing the $A=0.75$ case from below (d) makes apparent the much finer scales on the light fluid side relative to the heavy fluid side (c). This is consistent with the strongest turbulent vortices being concentrated near the light fluid side. Thickness $h$ as defined based on the velocity field is also included.}
\label{fig:densvis}
\end{figure}

In contrast to the velocity fluctuations' peak magnitudes shift to the light fluid side, the strongest density fluctuations position in the heavy fluid side with increasing Atwood number (Figures~\ref{fig:profilesaibrho}b,d). Fluctuation profiles for these density-based quantities (intensity and $a_i$) are also almost completely symmetric about $y=0$ for the $A=0.001$ case as the intense turbulence remains at the initial interface position. At increased Atwood number, large-scale disturbances (e.g., corrugations) form on the heavy fluid side while the fine scales of motion producing faster mixing are concentrated on the light fluid side.
This behavior can be adduced from the mean density and velocity fluctuation intensity profiles and by density visualizations  (Figure~\ref{fig:densvis}). The smoother yet disturbed heavy fluid side interface suggests the dominance of large-scale structures while small scales concentrate at the light fluid side for high Atwood number. The large displacements of the largely unmixed fluids relative to the background density gradient leads to large density fluctuation magnitudes on the heavy fluid side. In contrast, the greater mixing on the light fluid side produces local densities on average nearer to the mean density $\bar{\rho}$, thus resulting in smaller density fluctuation intensities.

In absolute terms (scaled by $\rho_0$ in Figure~\ref{fig:profilesaibrho}b), the density fluctuation intensity magnitudes increase up to $A=0.75$ but then stagnate for higher Atwood number. Peak intensity positions penetrate less and less deeply into the heavy fluid side for increasing Atwood numbers. These effects are likely a consequence of the less energetic turbulence sustained on the light fluid side reducing in ability to overcome the heavy fluid side's inertia and disturb it. Scaling using the mean density $\rho_0$ of the two streams, which remains constant for all Atwood numbers, reveals the relative magnitudes of the fluctuation intensities.
The no-mix scaled density fluctuation intensity profiles (Figure~\ref{fig:profilesaibrho}d) increase in magnitude up to $A=0.75$ but decay for higher Atwood number. 
No-mix intensity is proportional to the differences in densities between streams $\Delta \rho$, so scaling by this quantity would be expected to scale out the effect of increasing density differences for Boussinesq mixing. Thus, magnitude differences under this scaling reveal non-Boussinesq effects in the fluctuation intensities.

\section{Conditional Statistics}
\label{sec:condstat}

\begin{figure}
\centering
\includegraphics[scale=0.99]{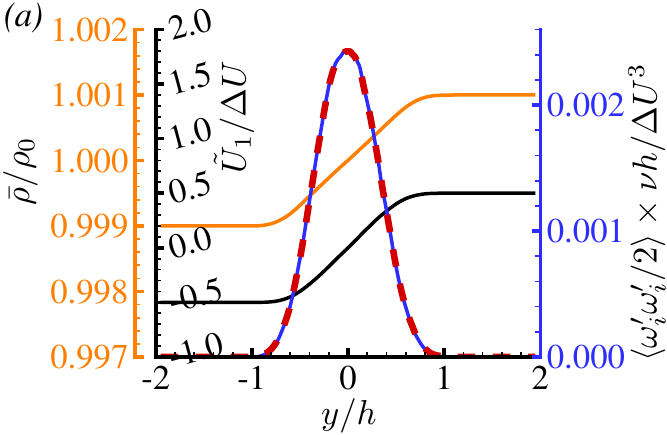}
\includegraphics[scale=0.99]{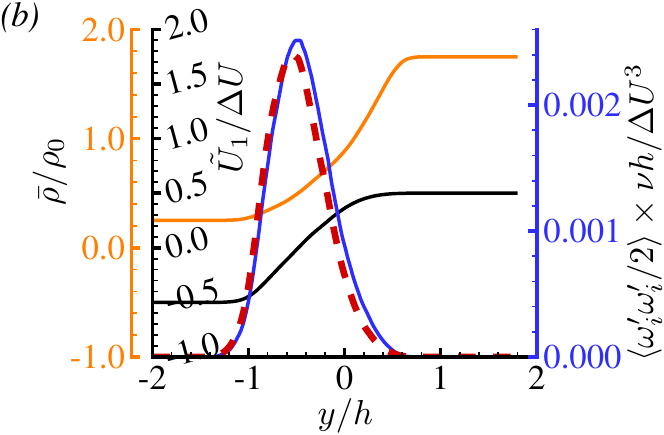}
\includegraphics[scale=0.99]{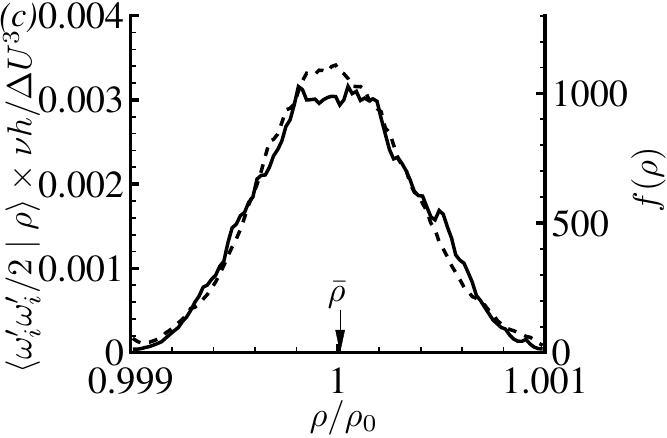}
\includegraphics[scale=0.99]{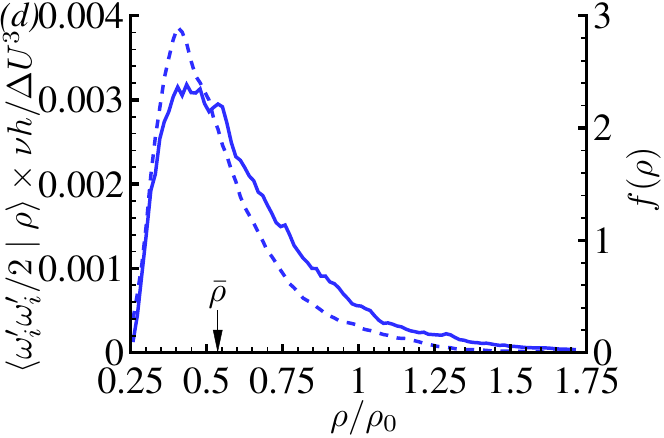}
\caption{(a--b) Mean enstrophy profiles with mean density (\orangeline) and mean streamwise velocity (\blackline) to reveal relative positions. The enstrophy (\blueline) is compared with the scaled dissipation approximation of (\ref{eq:enstrdissrel}) (\dashedredline). (c--d) Conditional enstrophy (solid line) and density pdf (dashed line) indicating the prevalence of fluid as a function of density, both shown at the position of strongest enstrophy identified above from a single field of each simulation with matching thicknesses ($t \Delta U/h_0=380$ for $A=0.001$ and $t \Delta U/h_0=405$ for $A=0.75$). The mean density at the position shown is indicated by an arrow. The conditional averages and pdfs were estimated using 100 discrete $\rho$ bins. (a,c) are for $A=0.001$ and (b,d) are for $A=0.75$.}
\label{fig:enstprofcondenst}
\end{figure}

Conditional statistics expose correlations with local fluid density to further reveal variable-density effects in the flow.
While the unconditional statistical moments above quantify the asymmetries (in a mean sense) with respect to $y$ position, statistics conditioned on density reveal further asymmetries with respect to local density at fixed $y$.
The unconditional statistics have demonstrated that increasing Atwood number concentrates the most intense turbulent motions at descending $y$ positions of mean density progressively lighter than $\rho_0$.
TKE provides one indication of where turbulence is concentrated, while the dissipation term of its budget is associated with intense, small-scale turbulent motions. Enstrophy is another quantity associated with intense vortical motions. Turbulent kinetic energy dissipation per unit volume can be related to fluctuation enstrophy (based on vorticity $\omega_k = \epsilon_{ijk} u_{j,i}$ where $\epsilon_{ijk}$ is the Levi-Civita symbol) as 
\begin{align}
\varepsilon = & \bar{\mu} \langle \omega_i' \omega_i' \rangle - \frac{2}{3} \bar{\mu} \langle d'^2 \rangle + \langle \mu' \omega_i' \omega_i' \rangle - \frac{2}{3} \bar{\mu} \langle \mu' d'^2 \rangle + \bar{U}_{i,j} \langle \mu' u_{i,j}' \rangle + \bar{U}_{j,i} \langle \mu' u_{i,j}' \rangle \nonumber\\
& - \frac{2}{3} \bar{d} \langle \mu' d' \rangle + 2 \bar{\mu} \langle u_{i,j}' u_{j,i}' \rangle + 2 \langle \mu' u_{i,j}' u_{j,i}' \rangle,
\end{align}
where $d=u_{i,i}$ is divergence \citep{morinishi2004dns}.
In constant-viscosity divergence-free incompressible flow, only the first term contributes on the right-hand side. Numerical simulations have indicated that the first term dominated while the third term made a small contribution and the others were negligible, although the detailed study was performed in a wall-bounded configuration \citep{morinishi2004dns}. Therefore, a reasonable approximation to the relation between enstrophy and dissipation is
\begin{align}
\langle \omega_i' \omega_i' \rangle \approx \frac{\varepsilon}{\bar{\mu}} = \frac{\tilde{\varepsilon}}{\nu},
\label{eq:enstrdissrel}
\end{align}
where $\tilde{\varepsilon}$ is defined by $\tilde{\varepsilon} = \varepsilon / \bar{\rho}$ as with (\ref{eq:Epstilde}) and $\nu$ is constant within the present flow (\S\ref{ss:viscdiffus}). The enstrophy and scaled dissipation agree well for both Atwood numbers shown in Figure~\ref{fig:enstprofcondenst}(a--b). The peaks of enstrophy and dissipation are also shown to coincide with steep mean streamwise velocity gradients and, in the case of high Atwood numbers, to be located where mean density is significantly lower than the average of the two free-stream densities ($\rho_0$).
Since the definition of enstrophy is independent of density, it is a useful quantity for investigating density effects on the kinematics of turbulence.

The $y$ plane of peak enstrophy intensity is an informative location for which to study the relationship between local density and intense turbulence.
At this position, the enstrophy conditioned on density is plotted for each Atwood number in Figure~\ref{fig:enstprofcondenst}(c--d).
The mean enstrophy is related to conditional counterpart by
\begin{align}
\langle \omega_i' \omega_i' \rangle = \int_{\rho_1}^{\rho_2} \langle \omega_i' \omega_i' | \rho \rangle f(\rho) \; d\rho,
\end{align}
where $f$ is the $\rho$ probability density function (pdf), which indicates the frequencies with which fluid of a given density exists at this location.
Thus, the peak enstrophy values shown in Figure~\ref{fig:enstprofcondenst}(a--b) can be obtained from the conditional enstrophies shown in Figure~\ref{fig:enstprofcondenst}(c--d) via this relation.

The conditional enstrophy plots reveal that, at low Atwood number, the fluid carrying the greatest amounts of enstrophy has density equal to the mean density at this position, which coincides with the average of the two free streams' densities. Since no pure (free stream) fluid has that density, fluid of such density is a mixture of the pure fluids, and it can be concluded that fluid of this density is associated with strong mixing. As Atwood number becomes large, the mean density at the plane of strongest enstrophy is less than $\rho_0$ and the local fluid densities associated with the strongest enstrophy magnitudes have yet lower values according to the conditional statistics.
Since mixing is associated with the intense small-scale motions, this suggests that the lighter fluid is carrying the turbulence and these motions are also active in mixing with the heavier fluid.
Conversely, the larger scales of turbulent motion lack strong velocity gradients and are thus associated with weaker enstrophy; the conditional enstrophy suggests that the larger scales are thus associated with the denser fluid.
The $\nu h/\Delta U^3$ scaling of the enstrophies shown in Figure~\ref{fig:enstprofcondenst} is based on the arguments given in \cite{rogers1994dss} using the relation between enstrophy and dissipation as well as the property that integrated dissipation remains constant in self-similar growth.

The pdfs also shown in Figure~\ref{fig:enstprofcondenst} demonstrate that the densities of peak conditional average are frequently present for both the lowest $A$ and $A=0.75$ cases. Thus, the densities associated with strong enstrophy magnitudes are representative of fluid parcels playing dominant roles and not merely rare events. The $A=0.001$ (essentially passive scalar) case indicates that fluid near the local mean density $\bar{\rho}$ carries the strongest enstrophy per unit volume and is most prevalent, for the position of strongest enstrophy that occurs near $y=0$. At high Atwood number, fluid lighter than the local mean density carries most of the enstrophy and is also most prevalent, at the peak total enstrophy position that has drifted toward the light-fluid side ($y<0$) relative to the initial interface.

While the conditional enstrophy statistics provides quantitative evidence that vorticity is concentrated in the light fluid, flow visualizations are consistent with this result and illustrate the asymmetries present in the flow. Surfaces indicating scarcely-mixed (nearly free-stream density) fluid are shown in Figure~\ref{fig:densvis}. Since these surfaces were initially parallel planes, the topologies of the surfaces at subsequent times are consequences of the motions that transport the fluid. The surfaces are symmetric in appearance for the lowest Atwood number case (Figure~\ref{fig:densvis}a). For $A=0.75$, however, the higher-density red surface (Figure~\ref{fig:densvis}c) is smoother with larger-scale corrugations compared to the lower-density blue surface (Figure~\ref{fig:densvis}d) that indicates the presence of much finer scales of motion. These fine scales are associated with strong enstrophy. The visualization for elevated Atwood numbers is thus consistent with the the average enstrophy peak existing in the vicinity of $y$ values at which the density is lower than the average of the two streams.

\section{Conclusions}
\label{sec:concl}

The present set of shear-driven variable-density mixing layer DNS spans a wide range of Atwood numbers. Since these simulations have reached self-similar growth at sufficient Reynolds numbers to be past the mixing transition, they form a comprehensive data set for evaluating the variable density effects on late-time turbulence dynamics.
The results demonstrate that, as Atwood number is increased while keeping the average density of the two free streams constant, the most intense turbulence is sustained in lighter-than-average fluid during self-similar growth. This occurs both in the sense of the intense turbulent motions shifting to $y$-positions at which the mean density is lower and also in the sense of the the strongest small-scale turbulent motions preferentially concentrating in fluid of lighter-than-mean density at a given position.

The main Atwood number effects on the most basic statistics, the mean density and velocity profiles, can be explained by self-similar growth properties and flow physics arguments. Self-similar analyses of the mean mass conservation and mean streamwise momentum balance equations have shown that the peak cross-stream velocity occurs in the light fluid side, while the neutral point of streamwise velocity moves further to the same side, and the peak $\tilde{R}_{12}$ stress moves yet further into the light fluid side (\S\ref{ss:selfsim}). The intense turbulent motions occur where production of turbulence is concentrated, which is where the mean velocity profile is steepest and $\tilde{R}_{12}$ magnitude is large. Since the intense turbulent motions are also associated with mixing that smooths the density profile, the mean density profile becomes shallower near the strong small-scale turbulence regions, while thickness growth of both the mean density and mean streamwise velocity interfaces preferentially occurs on the light fluid side. The alignment of the peak enstrophy with the mean streamwise velocity and density profiles confirms this behavior (Figure~\ref{fig:enstprofcondenst}).
The drift of the velocity and Reynolds stress profiles to the light fluid side, as well as the asymmetry of shallower decay on the light fluid side of the mean density profiles, strengthens in degree as Atwood number increases. These effects are robust with respect to statistical noise and occurs in mixing layers with streams of differing density produced by a single fluid with varied thermodynamic properties, both in low-speed \citep{almagro2017nsv} and compressible but subsonic \citep{pantano2002sce} regimes. These profiles for incompressible multi-species and low-speed varied thermodynamic properties cases are directly compared by \cite{baltzer2020lst}.

A prominent variable-density effect is the reduction in growth rates as Atwood number increases. Using the commonly-used momentum thickness quantity to measure growth rate, this reduction has been shown to be primarily associated with the density weighting in the definition. This reflects the momentum deficit (relative to free-stream) growing in progressively lighter density fluid with increasing Atwood number; the thickness growth produces smaller changes in momentum as the density in which the growth occurs decreases. When the thickness is defined in a manner not weighted by the fluid density relative to the average of the pure-fluid densities (such as based on momentum on a per-mass basis or a quantity such as $h$ that considers only the velocity field), the thickness growth rates display much less reduction as Atwood number increases. This is consistent with the mixing layer being sustained in lighter-than-average density fluid and, to a first approximation, behaving as a single-density mixing layer with a smaller nominal density (i.e., $\bar{\rho}$ where the turbulence is strongest rather than $\rho_0$). For an actual single-fluid mixing layer, the growth rate is dependent only on the velocity difference across the free-streams regardless of the density. However, when conventional momentum thickness growth rate is calculated for variable-density mixing layers using $\rho_0$ as the density scaling in the definition, there is a mismatch between $\rho_0$ and the actual density in which the turbulence is sustained. Since turbulence is sustained in increasingly lighter fluid relative to $\rho_0$ with increasing Atwood number, this definition indicates a growth rate reduction with Atwood number. When this dependency is removed, as in the case of the other thickness measures, the growth rate reductions are more modest but nonetheless significant. These latter effects indicate variable-density induced departures from idealized single-density behavior. Such reductions are principally associated with decreases in $\langle u_1' u_2' \rangle$ magnitude, as indicated by (\ref{eqn:deltampmgrowthratesfinal}).

The low-Atwood number limit of variable-density mixing layers captures much of the mass flux (density-velocity correlation) behavior, though the density field approaches passive scalar behavior. At high Atwood number, the mass fluxes become asymmetric and peaked on the light fluid side (particularly for the streamwise component), in addition to shifting to the light fluid side from the origin. Normalizing by only $\Delta U$ while $\rho_0$ remains constant, the turbulent mass flux magnitudes generally increase with Atwood number but appear to decrease past $A=0.75$. This may be due to the greater heavy fluid inertia damping out the turbulence (as suggested by the Reynolds stress magnitudes at increasing Atwood numbers) despite the strengthening maximum density fluctuations.

The shifting of turbulent motions to the light fluid side influences the density field evolution.
The intense turbulent motions progressively shifting toward the lighter fluid stream produces the asymmetry in the mean density profile discussed above. Furthermore, only larger spatial scales of velocity fluctuations are present toward the heavy fluid stream, which explains density fluctuation behavior: While the larger scales of motions near the heavier fluid are much less effective at producing mixing, they are effective at transporting parcels of largely unmixed heavy fluid, thereby producing particularly strong density fluctuations near the heavy fluid side at high Atwood number. These large density contrasts in partially-mixed light fluid and mostly-unmixed heavy fluid also produce large density fluctuation ($\langle \rho'^2 \rangle$) magnitudes there.
Conditional statistics support the picture of turbulence being sustained within relatively light density fluid and penetrating into higher density regions where it is damped by the greater fluid inertia at high Atwood number.

The widespread nature of variable-density multi-fluid mixing motivates further advancements in properly capturing and modeling the variable-density effects on the kinematic structure of turbulence. This is particularly true given these effects' importance for predicting mixing and more complicated phenomena that closely depend on mixing, such as reactions, that appear in a wide range of flows. While many properties of variable-density mixing layers closely resemble those of single-density mixing layers, complex interactions between density field and velocity field must be captured to predict the flow. In particular, the intense turbulence migrates to locally light fluid that interacts through neighboring fluid both through advection and mixing; this process alters the mean velocity and density profile evolutions as well as detailed statistics of mixing. Capturing all of these phenomena in a consistent manner presents significant challenges for RANS- and LES-type models.

\vspace{9pt}

This work has been authored by employees of Triad National Security, LLC which operates Los Alamos National Laboratory under Contract No. 89233218CNA000001 with the U.S. Department of Energy/National Nuclear Security Administration.
Computational resources were provided at Los Alamos National Laboratory through the Institutional Computing (IC) program and
at Lawrence Livermore National Laboratory through the Advanced Simulation and Computation (ASC) program.

\vspace{9pt}

Declaration of Interests. The authors report no conflict of interest.

\appendix
\section{Mean cross-stream velocity}
\label{appendix:mean}
The two-species incompressible fluid mixing equation (\ref{eq:divg}) relates gradients of velocity to density.
Applied to a temporal configuration, the mean velocity and density profiles are inhomogeneous only with respect to the 2 ($y$) direction.
The $\tilde{U}_{2,2}$ gradient of Favre mean velocity can thus be related to the density field.
Multiplying (\ref{eq:divg}) with $\rho$ and then averaging yields
\begin{equation}
\left\langle \rho u_{1,1} \right\rangle + \left\langle \rho u_{2,2} \right\rangle + \left\langle \rho u_{3,3} \right\rangle = \left\langle -\mathcal{D} \rho \left(\ln \rho \right)_{,ii} \right\rangle.
\label{eq:meanveldenrel1}
\end{equation}
Since $\bar{\rho} \tilde{U}_{i,j} = \left\langle \rho u_{i,j} \right\rangle$ and only the $\tilde{U}_{2,2}$ on-diagonal mean velocity gradient is nonzero in the temporal configuration, the left-hand side simplifies to $\bar{\rho} \tilde{U}_{2,2}$. The right-hand side may be decomposed by noting that $\left(\ln \rho\right)_{,ii} = \rho_{,ii}/\rho - {\rho_{,i}}^2/\rho^2$.
If diffusivity is constant, the right-hand side becomes $-\mathcal{D}\left( \langle \rho_{,ii} \rangle - \langle {{\rho_{,i}}^2}/{\rho} \rangle \right)$. For the second term in parentheses, introducing the relevant Reynolds decompositions and writing in terms of specific volume $v \equiv 1/\rho$ and density-specific volume correlation $b \equiv -\langle \rho' v'\rangle$ (both functions of $y$) yields
\begin{equation}
\left\langle v {\rho_{,i}}^2 \right\rangle = \frac{1 + b}{\bar{\rho}} {(\bar{\rho}_{,i})}^2 + \frac{1 + b}{\bar{\rho}} \left\langle {\rho'_{,i}}^2 \right\rangle + 2 \bar{\rho}_{,i} \left\langle v' \rho'_{,i} \right\rangle + \left\langle v' {\rho'_{,i}}^2 \right\rangle.
\label{eq:vrhopricorr}
\end{equation}
Mean specific volume $\bar{v}$ was replaced using the identity $\bar{v} = \left(1 + b\right)/\bar{\rho}$. The complete expression is
\begin{align}
\bar{\rho} \tilde{U}_{2,2} & = -\mathcal{D} \left[ \bar{\rho}_{,ii} - \frac{({\bar{\rho}_{,i})}^2}{\bar{\rho}} - \frac{b}{\bar{\rho}} {(\bar{\rho}_{,i})}^2  - \frac{1 + b}{\bar{\rho}} \left\langle {\rho'_{,i}}^2 \right\rangle - 2\bar{\rho}_{,i} \left\langle v' \rho'_{,i} \right\rangle - \left\langle v' {\rho'_{,i}}^2 \right\rangle \right]\nonumber\\
\rightarrow \;\; \tilde{U}_{2} & = -\mathcal{D} \int_{y_\mathrm{min}}^y
\left[ \left( \ln \bar{\rho} \right)_{,ii} - \frac{b}{\bar{\rho}} {(\bar{\rho}_{,i})}^2  - \frac{1 + b}{\bar{\rho}} \left\langle {\rho'_{,i}}^2 \right\rangle - 2\bar{\rho}_{,i} \left\langle v' \rho'_{,i} \right\rangle - \left\langle v' {\rho'_{,i}}^2 \right\rangle \right] \; dy
\label{eq:meanveldenrel2}
\end{align}

The mean cross-stream velocity is determined by only the mean density profile and the profiles of the density fluctuations (and their correlations); this result does not explicitly depend of the mean streamwise velocity.
If there are no spatial density fluctuations besides the $y$-dependent mean density profile, only the first term on the right-hand side is nonzero and (\ref{eq:meanveldenrel2}) reduces to
$\tilde{U}_{2} = -\mathcal{D} \int_{y_\mathrm{min}}^y \left( \ln \bar{\rho} \right)_{,ii} dy$. This expression is valid when the simulation is initialized.
If density is constant, (\ref{eq:meanveldenrel2}) dictates that $\tilde{U}_2=0$, which is consistent with the
divergence-free nature of incompressible constant-density flow. Away from the interface in the variable-density mixing layers, each free stream has uniform (but different) density, so the equations indicate that $\tilde{U}_2$ will be constant in these regions.
Since each slip wall dictates that $\tilde{U}_2$ will be zero at the wall while these equations indicate that $\tilde{U}_2$ will be constant
approaching each wall, it is concluded that $\tilde{U}_2$ is zero away from the region of mean density gradient or active mixing of unequal-density fluids.

\section{Self-similar analysis}
\label{appendix:selfsim}

Self-similar analysis similar to that performed by \cite{pantano2002sce} is now given for the present flow configuration.
If self-similar growth occurs, each self-similar profile quantity is a function only of $\eta$ (for a given Atwood number/flow configuration) rather than position $y$ and time $t$ (or thickness) independently. Considering first the mean mass conservation equation (\ref{eq:contfa}) and substituting the self-similar variable dependencies yields
\begin{equation}
- \eta [dh/dt] d\bar{\rho}/d\eta +  d(\bar{\rho} \tilde{U}_2)d\eta = 0.
\label{eq:contss1}
\end{equation}
In order to ensure that the terms in this equation depend on $\eta$ only, the mixing layer thickness  
needs to vary linearly in time. Thus, self-similarity requires that
\begin{equation}
dh/dt=C, 
\end{equation}
where $C$ (the growth rate) is a constant. If $\tilde{U}_2$ is scaled by $C$, equation (\ref{eq:contss1}) becomes independent of $C$, which indicates growth rate, rather than $\Delta U$, as the self-similar scaling of $\tilde{U}_2$.

Next, the self-similar variable forms are applied to the Favre mean streamwise momentum equation (\ref{eq:momfa}). When the flow is self-similar, the viscous term has a small effect on the mean momentum, as the mean velocity profile is relatively shallow after the earliest times (whereas the viscous term continues to produce large contributions to the energy balance because these contributions are based on instantaneous gradients within the turbulent motions). In self-similar variables, (\ref{eq:momfa}) then becomes
\begin{equation}
- \eta [dh/dt] d(\bar{\rho} \tilde{U}_1)/d\eta + d(\bar{\rho} \tilde{U}_1 \tilde{U}_2)/d\eta  +  d(\bar{\rho} \tilde{R}_{12})/d\eta = 0.
\label{eq:mom1ss1}
\end{equation}
Assuming that the streamwise mean velocity scale is $\Delta U$, this equation also indicates that 
the self-similar scaling for $\tilde{R}_{12}$ is $\Delta U dh/dt$, since for this scaling the equation becomes independent of the flow.

When both the mean density and velocity profiles are initially specified and centered at $y=0$, the mean density profile is monotonic, with $d\hat{\rho}/d\eta>0$. As the flow evolves, it continues to vary between the same density values for light and heavy fluid. The self-similar growth behavior implies that a non-monotonic density profile would be very unlikely to maintain. The behavior of the density profiles suggests choosing the density scale, $\rho_0$, as the average of the two density extremes, $(\rho_1+\rho_2)/2$, which also corresponds to the initial centerline density. 

Using the scalings identified above, which are summarized in equations~(\ref{eq:ssscalingrho})--(\ref{eq:ssscalingR12}), the self-similar equations then become
\begin{align}
- \eta d\hat{\rho}/d\eta + d(\hat{\rho} \hat{U}_2)/d\eta & =0 \label{eq:contss2}\\
- \eta d(\hat{\rho} \hat{U}_1)/d\eta + d(\hat{\rho} \hat{U}_1 \hat{U}_2)/d\eta  + d(\hat{\rho} \hat{R}_{12})/d\eta & = 0 \label{eq:mom1ss2},
\end{align}
which can be re-written as:
\begin{align}
(\hat{U}_2-\eta)d\hat{\rho}/d\eta+\hat{\rho}d\hat{U}_2/d\eta &=0  \label{eq:contss3}\\
(\hat{U}_2 - \eta) \hat{\rho} d\hat{U}_1/d\eta + d(\hat{\rho} \hat{R}_{12})/d\eta &= 0  \label{eq:mom1ss3}.
\end{align}

From these equations, several conclusions about the behavior of variable-density flow during self-similar growth can be drawn. Equation (\ref{eq:contss2}) shows  that, for $A>0$ (so that $d\hat{\rho}/d\eta>0$), $\hat{\rho} \hat{U}_2$ has a peak at $\eta=0$, which corresponds to the initial centerline ($y=0$). On the other hand, from equation (\ref{eq:contss3}), $d\hat{U}_2/d\eta$ is zero at the location where $\eta  = \hat{U}_2$. Let this location be $\eta=\eta_2$. Equation (\ref{eq:contss3}) also shows that inside the layer
\begin{align}
\eta<\eta_2\, &\rightarrow\ d\hat{U}_2/d\eta<0 \\ 
\eta>\eta_2\, &\rightarrow\ d\hat{U}_2/d\eta>0. 
\end{align}
This implies that $\eta_2$ is unique, otherwise for the region between two $\eta_2$ solutions,
$d\hat{U}_2/d\eta$ needs to be both positive and negative. Since $\hat{U}_2$ is zero outside the layer and its derivative has only one zero inside the layer, it follows that $\hat{U}_2$ has constant sign across the layer. Therefore, $\hat{U}_2<0$ within the layer, otherwise $\hat{U}_2$ cannot become zero outside the layer. As a consequence, at the centerline, $\hat{U}_2$ is strictly negative and $d\hat{U}_2/d\eta>0$, which implies that $\eta_2<0$. Thus, simply from mass conservation considerations for self-similar growth, it is established that the cross-stream velocity peaks on the light fluid side at $\eta  = \hat{U}_2(\eta)$ and that $\hat{U}_2<0$ within the layer. 

Equation (\ref{eq:mom1ss3}) shows that $d(\hat{\rho}\hat{R}_{12})/d\eta$ is also zero at $\eta=\eta_2$. Let then $\eta=\eta_{12}$ be the location where $d\hat{R}_{12}/d\eta$ is zero. Using the mean configuration set-up considered here (such that $d\hat{U}_{1}/d\eta>0$ and $d\hat{\rho}/d\eta>0$), similar arguments as above can be made to show that $\eta_{12}$ is unique within the layer and $\eta_{12}<\eta_2$. Thus, $\hat{R}_{12}<0$ and has its largest magnitude further to the light fluid side than $\hat{U}_2$. Analogous conclusions were previously drawn by \cite{pantano2002sce} from these equations for their configuration.

\bibliographystyle{jfm}

\bibliography{vardenspaper_part1}

\end{document}